\documentclass[twocolumn,american]{revtex4-1}
\usepackage[T1]{fontenc}
\usepackage[latin9]{inputenc}
\usepackage{prettyref}
\usepackage{mathtools}
\usepackage{amsmath}
\usepackage{amssymb}
\usepackage{graphicx}

\makeatletter
\usepackage{xr}
\usepackage{hyperref}
\newrefformat{eq}{\hyperref[#1]{Eq.~(\ref{#1})}}
\newrefformat{cap}{\hyperref[#1]{Fig.~\ref{#1}}}
\newrefformat{fig}{\hyperref[#1]{Fig.~\ref{#1}}}
\newrefformat{tab}{\hyperref[#1]{Table ~\ref{#1}}}
\newrefformat{sec}{\hyperref[#1]{Section~\ref{#1}}}
\newrefformat{sub}{\hyperref[#1]{Section~\ref{#1}}}
\newrefformat{cha}{\hyperref[#1]{Chapter~\ref{#1}}}

\makeatother

\usepackage{babel}
\begin{document}
\title{Hidden Connectivity Structures Control Collective Network Dynamics} \author{Lorenzo Tiberi$^{1,2,3}$} \author{David Dahmen$^{1}$} \author{Moritz Helias$^{1,2,3}$} 
\affiliation{$^{1}$Institute of Neuroscience and Medicine (INM-6) and Institute for Advanced Simulation (IAS-6) and JARA-Institute Brain Structure-Function Relationships (INM-10),    Jülich Research Centre, Jülich, Germany } \affiliation{$^{2}$Institute for Theoretical Solid State Physics, RWTH Aachen University, 52074 Aachen, Germany} \affiliation{$^{3}$Center for Advanced Simulation and Analytics, Forschungszentrum Jülich, 52425 Jülich, Germany}
\begin{abstract}
Many observables of brain dynamics appear to be optimized for computation.
Which connectivity structures underlie this fine-tuning? We propose
that many of these structures are naturally encoded in the space that
more directly relates to network dynamics -- the space of the connectivity
eigenmodes. We develop a mathematical theory to impose eigenmode structures
on connectivity, systematically characterizing their effect on network
dynamics. We find the density of nearly-critical eigenvalues to be
a particularly fundamental structure. It flexibly controls the power-law
scaling of dynamical observables, in analogy with the system's spatial
dimension in classical critical phenomena. This mechanism provides
control over observables which are found to be fine-tuned in brain
networks, but remained so far unexplained by traditionally studied
structures, such as connectivity motifs. Specifically, the slope of
the principal component spectrum of neural activity can be fine-tuned,
as observed in primary visual cortex of mice. Furthermore, a novel
transition between high and low dimensional activity allows for a
wide and flexible tuning of dimensionality, as observed throughout
cortex. The here discovered structures thus largely complement motifs.
In fact, they are of a different, collective nature: they are not
reflected by any local motif configuration. This result shows that
many functionally relevant structures can remain hidden within the
apparent randomness of highly heterogeneous cortical circuits. Our
methods enable revealing these structures and investigate their effect
on network dynamics.
\end{abstract}
\maketitle

\section{Introduction\label{sec:Introduction}}

Network dynamics is a central topic of contemporary physics, due to
its ability to describe a wide variety of biological, social and artificial
systems \citep{Dorogovtsev08}. Indeed, through different arrangements
of their connectivity, networks are able to manifest the most disparate
complex behaviors. A prime example are neural networks: local circuits
across different cortical areas specialize their connectivity to implement
very diverse computational functions \citep{Kandel13}, such as motor
control, object recognition or abstract reasoning.

Many observables of brain dynamics indeed appear to be fine-tuned
for specific computational needs. For example, in the primary visual
cortex (V1) of mice, the principal component (PC) spectrum of neuronal
activity is found to have an optimal slope for image encoding \citep{Stringer19_361}.
At the same time, dynamical observables are tuned differently and
flexibly across cortical areas, depending on the area's computational
needs. For example, neuronal activity is found to be either low \citep{Sadtler14,Gallego18_1,Semedo19_249}
or high dimensional \citep{Rigotti2013_585,Stringer19_361} in different
regions of cortex \citep{sorscher22}, with both high and low dimensionality
optimizing different computational functions.

A fundamental quest of neuroscience is to identify the connectivity
structures that allow for such flexible, yet precise tuning of network
dynamics. This is complicated by the fact that connectivity in cortical
microcircuits is highly heterogeneous, appearing to a large extent
as random, and thus strikingly similar across cortical areas \citep{Braitenberg91}.
Therefore, the challenge is often to find structure \emph{within}
randomness, that is structure in the connectivity statistics. In order
to provide focus and direction to this complex search, an important
role of theory is to answer this first-principles question: What are
the minimal and fundamental structures in a connectivity statistics,
which allow networks to fine-tune their dynamics?

One of such structures has been recently identified in motifs \citep{Hu13_P03012,Hu22,Dahmen22_365072v3}:
minimal connectivity patterns between few neurons, which occur with
higher chance than random \citep{Song05_0507}. Second order motifs
have been shown to have some control over neuronal activity, such
as its dimensionality \citep{Hu13_P03012,Hu22,Dahmen22_365072v3}.
Still, there remain many of the aforementioned network behaviors which
do not seem to be explainable by these structures alone. For example,
it was shown that second order motifs cannot fine-tune the power-law
decay of the PC spectrum of neuronal activity \citep{Hu22}. Also,
near criticality, where this and other power-laws ubiquitously observed
in brain dynamics emerge \citep{Beggs03_11167,Meshulam19,Fontenele19_208101,Stringer19_361},
neuronal activity is always low-dimensional, and cannot be tuned to
be high dimensional \citep{Dahmen22_365072v3}. This naturally raises
a question: Are there other fundamental connectivity structures that
we are missing, which can account for these unexplained behaviors?

Here we identify such structures by proposing a paradigm shift: If
motifs are \emph{local} structures, encoded in the connections between
few neurons, we argue that there is also a fundamentally different
type of structures, which are intrinsically \emph{collective}. These
are encoded in the space that naturally relates to collective network
dynamics - the space of the connectivity eigenmodes. Various works
have highlighted the importance of eigenmode structures. For example,
specific forms of eigenvalue distributions are shown to emerge after
training \citep{Hennequin14_1394}, and the angle between eigenvectors
often has important functional consequences, such as in low-rank connectivity
structures \citep{Mastrogiuseppe18_609}. However, thus far the field
is lacking a mathematical framework to directly impose eigenmode structures
on heterogeneous connectivity matrices, allowing to systematically
characterize their effect on network dynamics.

We thus develop a novel theory for large random connectivity matrices
with specifiable eigenmode statistics. In particular, as outlined
in \prettyref{sec:setting}, we can specify the degree of non-orthogonality
between eigenvectors, and the shape of the eigenvalue distribution.
This allows us to rigorously characterize network dynamics as a function
of these eigenmode structures. As shown in \prettyref{sec:Effect-on-Dynamics},
we find the shape of the eigenvalue distribution to be a particularly
effective structure, controlling dynamical observables such as the
autocorrelation, autoresponse, PC spectrum, and dimensionality of
neuronal activity. Specifically, the scaling exponents of these quantities
are directly controlled by the scaling exponent of the density of
nearly critical eigenvalues, in analogy to how the system's spatial
dimension affects scaling in classical critical phenomena (\prettyref{subsec:Autocorrelation-and-autoresponse}).
In particular, this mechanism can account for the the widely observed
fine-tuning of functionally relevant measures of network dynamics,
which so far remained unexplained by motif structures alone: varying
the density of nearly critical eigenvalues, the PC spectrum can be
fine-tuned into its optimal slope for stimulus encoding (\prettyref{subsec:Principal-components-spectrum}),
as observed in V1 of mice \citep{Stringer19_361}, and dimensionality
of neural activity can flexibly transition between both the high and
the low dimensional regimes observed throughout cortex \citep{Rigotti2013_585,Sadtler14,Gallego18_1,Semedo19_249,Stringer19_361,sorscher22},
while remaining at the critical point (\prettyref{subsec:Principal-components-spectrum}).

\begin{figure*}
\centering{}\includegraphics[width=0.8\textwidth]{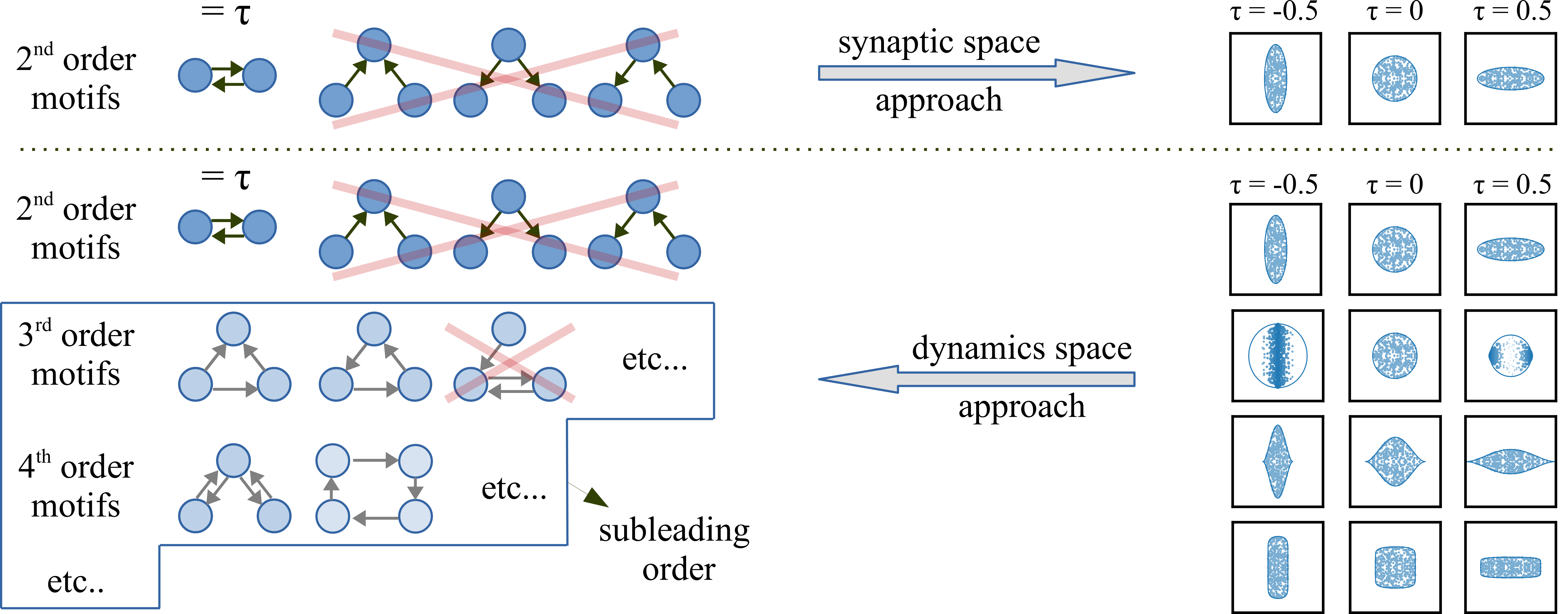}\caption{Schematic representation of the traditional synaptic space approach
vs our dynamics space approach. Top: synaptic space approach. Some
motif statistics are assumed (left) and the eigenmode statistics (right)
are derived. Specifically, here we show an archetypal approach which
assumes only the presence of second order reciprocal motifs (here
parameterized by $\tau$, \prettyref{eq:g_tau_def}). The corresponding
eigenvalue distribution is uniform on an ellipse. The eigenvalues
are relatively more spread along the real or the imaginary axis depending
on the value of $\tau$. Bottom: dynamics space approach. We specify
the eigenmode statistics, in particular allowing for any desired shape
of the eigenvalue distribution (right). The derived synaptic statistics
(left) to leading order only predict reciprocal motifs, as in the
archetypal approach. These are parameterized again by the relative
spread $\tau$ of the eigenvalues, here generalized to any distribution
shape. Higher order motif structures are present, but are subleading
in the number of neurons. Thus, the synaptic structures responsible
for different shapes of the eigenvalue distribution (thus different
dynamics) are hidden to a motifs analysis. \label{fig:Synaptic-vs-dynamics}}
\end{figure*}

From the space of eigenmodes, we also derive the connectivity statistics
in the more direct space of synaptic strengths. The result, presented
in \prettyref{sec:connectivity-statistics}, shows how the here considered
eigenmode stuctures are of a fundamentally different nature than motifs,
complementing them not only from a functional, but also from a structural
perspective. Indeed, these structures are intrinsically collective:
to leading order in the number of neurons, they are not reflected
by local motif configurations, and the connectivity always appears
as random Gaussian (see \prettyref{fig:Synaptic-vs-dynamics}). The
result provides an intriguing insight into the apparent paradox of
structure within randomness: there can in fact be much structure hidden
in the eigenmodes of an apparently random Gaussian connectivity, which
has profound effects on the network dynamics. This structure remains
hidden to a local motifs analysis. However, it can be revealed by
a more collective analysis of the connectivity eigenmodes.

\section{Setting\label{sec:setting}}

We consider the recurrent network of linear rate neurons

\begin{equation}
\tau\partial_{t}x_{i}\left(t\right)=-x_{i}\left(t\right)+\sum_{j}J_{ij}x_{j}\left(t\right)+\xi_{i}\left(t\right)\,,\label{eq:SCS_model}
\end{equation}
where $x_{i}\left(t\right)$ is the rate activity of neuron $i$ at
time $t$; $i=1,\ldots N$. Here $J_{ij}$ describes the connection
from neuron $j$ to $i$ and $\tau$ is the characteristic timescale
of neuronal response. The network is driven by Gaussian white noise
$\xi_{i}\left(t\right)$, with zero mean and variance $\text{\ensuremath{\left\langle \xi_{i}\left(t\right)\xi_{j}\left(t'\right)\right\rangle }=\ensuremath{D\delta_{ij}\delta\left(t-t'\right)}}$.
In the following, we consider \prettyref{eq:SCS_model} in dimensionless
units, setting $\tau=D=1$. In the linear regime, the second order
statistics of various network models, comprising integrate-fire and
inhomogenous Poisson neurons, is well captured by \prettyref{eq:SCS_model}
\citep{Lindner05_061919,Pernice11_e1002059,Pernice12_031916,Grytskyy13_131,Trousdale12_e1002408}.
For this reason, \prettyref{eq:SCS_model} is a minimal model for
characterizing network dynamics analytically, as a function of $J$
\citep{Grytskyy13_131,Dahmen19_13051,Hu22,Dahmen22_365072v3}.

Local cortical circuits present strong variability in their connectivity.
A common approach is to model such variability as disorder, choosing
$J$ as a random matrix. Most likely, brain networks are not completely
disordered: we expect them to have some function, thus some underlying
connectivity structure. In the random connectivity setting, information
about the network function or structure can be included in the model
through the choice of statistics for $J$.

To constrain the connectivity statistics, previous works have focused
on the observed abundance of local connectivity patterns involving
few neurons, such as the connectivity motifs \citep{Song05_0507}
depicted in \prettyref{fig:Synaptic-vs-dynamics}. In particular,
analytical studies characterizing the effect of motifs on network
dynamics are typically constrained to second order motifs \citep{Hu13_P03012,Hu22,Dahmen22_365072v3}.
The relative abundance of a motif with respect to a completely random
network can be modeled by a non-vanishing moment of the elements of
$J$ involved in the motif \citep{Hu22}. In an archetypal approach
\citep{Sompolinsky88_259,Sommers88}, for example, $J$ is assumed
Gaussian with statistics
\begin{equation}
g^{2}\equiv N\langle J_{ij}^{2}\rangle\qquad\tau\equiv\langle J_{ij}J_{ji}\rangle/\langle J_{ij}^{2}\rangle,\quad i\neq j\,.\label{eq:g_tau_def}
\end{equation}
Two parameters are present: the synaptic gain $g$ models the overall
strength of recurrent connections, while the degree of (anti)symmetry
$\tau$ controls correlations between reciprocal connections, which
can be associated with an abundance of reciprocal motifs. Throughout
the manuscript, we will refer to this connectivity choice as the archetypal
choice for $J$ and use it for comparison with our approach. Note
that any other second order motif (those shown in the top row of \ref{fig:Synaptic-vs-dynamics}
with a bar on top) can always be introduced later as a low rank perturbation
to this bulk connectivity, and their effect on network dynamics can
be characterized by a finite number of outliers perturbing the bulk
spectrum of the neuronal activity's covariance matrix \citep{Hu22}.
Therefore, here we focus on the bulk connectivity.

As discussed in \prettyref{sec:Introduction}, local motifs can only
partially account for the optimal fine-tuning of dynamics observed
in cortical networks. To fill this gap, here we introduce the study
of another fundamental type of collective, rather than local structures.
These are encoded in the space that naturally controls network dynamics
-- the space of eigenmodes. The connectivity $J$ can always be decomposed
into its eigenvalues $\lambda_{\alpha}$ and right eigenvectors $V_{\alpha}$
as

\begin{equation}
J_{ij}=\sum_{\alpha=1}^{N}V_{i\alpha}\lambda_{\alpha}V_{\alpha j}^{-1}\,,\label{eq:eigen_decomposition}
\end{equation}
were the left eigenvectors' matrix $V^{-1}$ is the inverse of the
right eigenvectors' matrix $V$. For typical random connectivity matrices,
this decomposition can be performed without loss of generality (see
Appendix \prettyref{subsec:Details-on-the-eigenmode-statistics}).
Eigenvalues and eigenvectors have a direct dynamical interpretation:
a linear combination of neurons $\sum_{i}V_{\alpha i}^{-1}x_{i}$
represents a collective mode of the linearized neuronal activity,
with dynamical response $\propto\exp\left(-\left(1-\lambda_{\alpha}\right)t\right)$.
Thus, calling an eigenvalue $\lambda=\lambda_{x}+\imath\lambda_{y}$,
the associated mode has decay constant $1-\lambda_{x}$ and oscillation
frequency $\lambda_{y}$. The distribution of eigenvalues therefore
characterizes the dynamical repertoire available to the network, while
the overlap between eigenvectors characterizes how much the network
input co-excites different eigenmodes \citep{Hennequin_12}. Both
the shape of the eigenvalue distribution and the eigenvectors' overlap
are therefore important structural properties of connectivity, which
are indeed shown to be associated with network function \citep{Hennequin14_1394,Mastrogiuseppe18_609}.

Through the archetypal choice of $J$, these eigenmode structures
are implicitly fixed: eigenvalues are uniformly distributed on an
ellipse centered in the complex plane \citep{Sommers88} (\prettyref{fig:Synaptic-vs-dynamics},
top), while eigenvectors are strongly non-orthogonal, as discussed
later in greater detail. Given their high relevance for network dynamics,
however, here we want to explicitly constrain the form of these connectivity
structures to systematically characterize their impact on neural activity.
Concretely, we approach connectivity modeling in a direction symmetric
to the archetypal approach (\prettyref{fig:Synaptic-vs-dynamics}):
In the archetypal \emph{synaptic space approach}, one specifies the
statistics of synaptic strengths (the entries of $J$) and later derives
the corresponding eigenmode statistics; in our \emph{dynamics space
approach}, we first specify the eigenmode statistics and later derive
the synaptic strength statistics. We define the eigenmode statistics
as

\begin{align}
\lambda & \sim p\left(\lambda\right)\label{eq:def_eigenvalue_statistics}\\
V & =O+\nu G\label{eq:def_eigenvectors_statistics}
\end{align}

\prettyref{eq:def_eigenvalue_statistics} states that $\lambda$ can
be drawn from a distribution $p$ of any arbitrary shape, provided
its moments are bounded. In particular, one is not constrained to
a uniform elliptical distribution as in the archetypal approach (see
\prettyref{fig:Synaptic-vs-dynamics}). In turn, we will show in \prettyref{sec:Effect-on-Dynamics}
how the shape of $p$ is direcly linked to the dynamics, in particular
to power-law exponents ubiquitously observed in cortical networks,
and thus can be fixed according to a certain observed or desired network
behavior. The eigenvalues are drawn independently, following the same
simplicity principle typically applied in the synaptic space approach:
introducing minimal information in an otherwise disordered connectivity.

Following the same principle, we want to be agnostic with regard to
the direction that a certain eigenmode takes in neuronal space. The
eigenvector matrix $V$ is therefore drawn independently from the
eigenvalues. \prettyref{eq:def_eigenvectors_statistics} defines it
as a combination of a unitary and a complex Gaussian random matrix
$O$ and $G$, respectively, with interpolation parameter $\nu\in\left[0,1\right)$
controlling the degree of non-normality of the network. A precise
definition of $O$ and $G$ can be found in \prettyref{subsec:Details-on-the-eigenmode-statistics}.
Intuitively, varying $\nu$ from $0$ to $1$ controls whether eigenmodes
are orthogonal to each other ($\nu=0)$, or can take on more and more
random, overlapping directions ($\nu\to1)$. The value $\nu=1$ is
an upper limit at which eigenvectors are too overlapping and the synaptic
gain $g$ diverges (see \prettyref{sec:connectivity-statistics} and
Appendix \ref{subsec:Insights-into-the}).

A difference with the archetypal case of Gaussian $J$ is that, there,
eigenvalues and eigenvectors are found to be tightly correlated, and
eigenvectors are in a strongly non-orthogonal regime, similar to choosing
$\nu\sim1$ in our ensemble. In \prettyref{sec:connectivity-statistics}
we show that, in this strongly non-normal regime, a fine-tuned correlation
structure is necessary to contrast the strong eigenvectors' overlap
and thus keep the synaptic gain $g$ of $\mathcal{O}\left(1\right)$.
We then devise a method to initialize connectivities with such fine-tuned
correlation structure, while still allowing for any shape of the eigenvalue
distribution. In this way, through numerical simulations, we are able
to characterize also the strongly non-normal regime. Including this
strongly non-normal ensemble, our study therefore encompasses the
archetypal case of Gaussian $J$.

\begin{figure}
\centering{}\includegraphics[width=1\columnwidth]{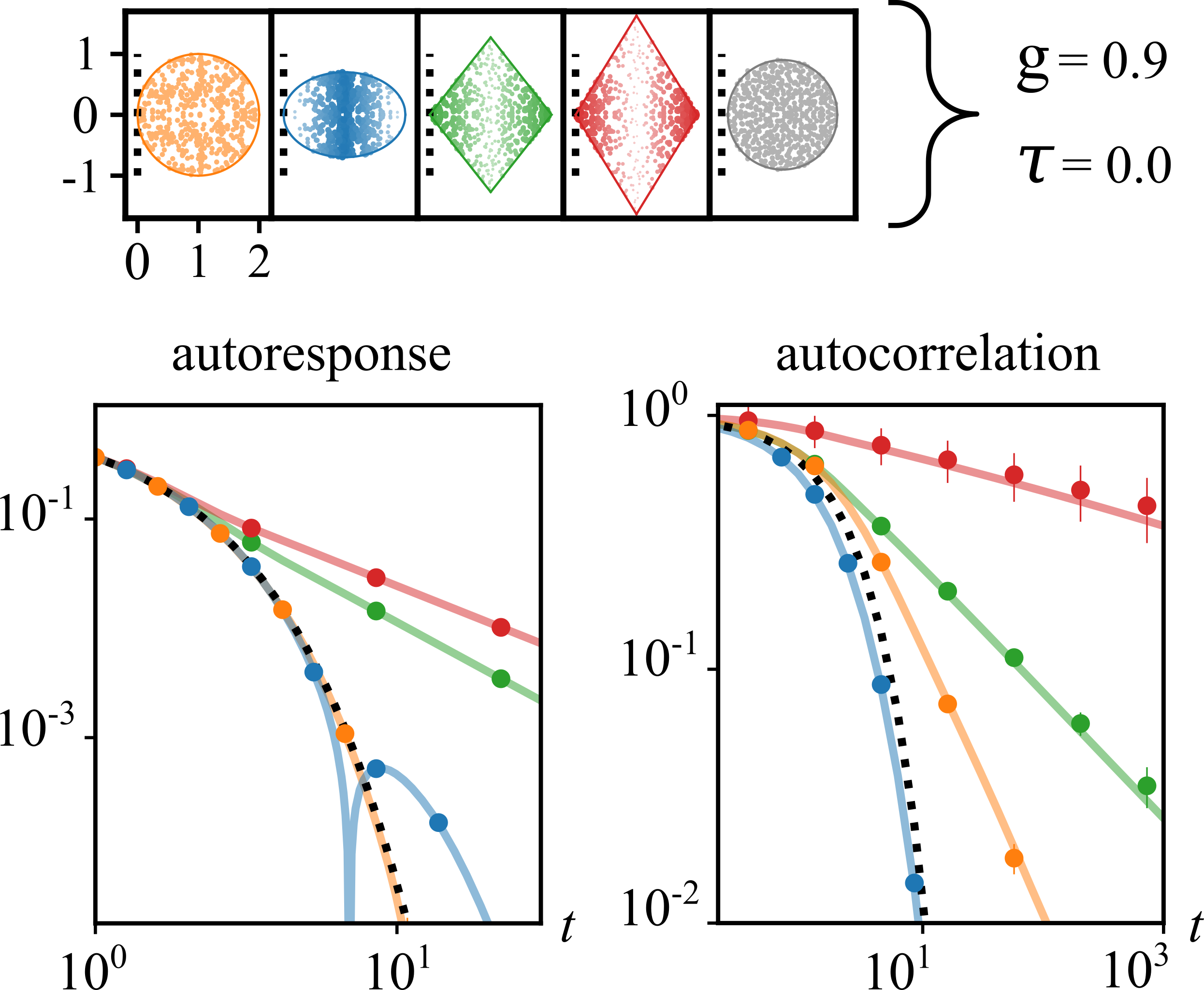}\caption{Same synaptic statistics correspond to different dynamics. Top: Various
eigenvalue distributions, corresponding to different connectivities,
which to leading order in the number of neurons share the same synaptic
statistics. Specifically, $g=0.9$ and $\tau=0$ (\prettyref{eq:g_tau_def}),
corresponding to independently distributed synaptic strengths following
Gaussian statistics. The eigenvalue distributions are plotted in the
complex plane $k=1-\lambda$ and are nearly critical, touching the
line of linear instability $k_{x}=0$. The archetypal case of an exactly
Gaussian connectivity with the same synaptic statistics is also shown
in black. Bottom: Autocorrelation and autoresponse functions decay
in time, colored accordingly. Despite sharing the same leading order
synaptic statistics, the different connectivities feature different
eigenvalue distributions, and thus correspond to very different network
dynamics: the autocorrelation and autoresponse can either decay exponentially
(blue, orange) or with power-laws with varying exponents (red, green).
Curves: theory; markers: simulation; dotted, black line: known theory
for the archetypal case of a Gaussian connectivity. Other parameters:
For a given eigenvalue distribution, $\nu$ is fixed according to
\prettyref{eq:g_relation} and \prettyref{eq:tau_relation} so to
have $g=0.9$ and $\tau=0$. Going from left to right in the top panel,
$\nu=0.49,0.7,0.52,0.37$.\label{fig:same_g_tau}}
\end{figure}

As we show in \prettyref{sec:Effect-on-Dynamics}, varying the shape
of the eigenvalue distribution causes a wide range of different dynamical
behaviors. Surprisingly, however, we find this is not reflected by
the leading order statistics of synaptic strengths. Just as for the
archetypal connectivity, we find the synaptic strength statistics
to be Gaussian, with only reciprocal motifs. In terms of the eigenmode
statistics, the leading order synaptic strength statistics are given
by

\begin{align}
g^{2} & =\frac{1+\nu^{2}}{1-\nu^{2}}\left(\left\langle \lambda_{x}^{2}\right\rangle _{\lambda}+\left\langle \lambda_{y}^{2}\right\rangle _{\lambda}\right)\label{eq:g_relation}\\
\tau & =\frac{1-\nu^{2}}{1+\nu^{2}}\frac{\left\langle \lambda_{x}^{2}\right\rangle _{\lambda}-\left\langle \lambda_{y}^{2}\right\rangle _{\lambda}}{\left\langle \lambda_{x}^{2}\right\rangle _{\lambda}+\left\langle \lambda_{y}^{2}\right\rangle _{\lambda}}\label{eq:tau_relation}
\end{align}
where $\left\langle \,\right\rangle _{\lambda}$ denotes statistical
averaging over the random variable $\lambda$. Other second order
motifs are absent. Higher order motifs are present, but only with
a probability that is vanishingly small in the number of neurons,
and are therefore very hard to detect. As a result, connectivities
with the same leading order motif structure can correspond to different
eigenmode structures, thus to very different network dynamics, as
exemplified in \prettyref{fig:same_g_tau}. In other words, there
are structures that have a strong impact in shaping the network dynamics,
while remaining hidden in the synaptic strength statistics of an apparently
random Gaussian connectivity. More precisely, these structures remain
hidden to a motif analysis, as they cannot be reduced to the generic
abundance of local connectivity patterns. However, they become apparent
to a more collective analysis of the connectivity's spectral properties,
in particular of its eigenvalue distribution. Our findings on the
synaptic strength statistics are discussed in more detail in \prettyref{sec:connectivity-statistics}.

\section{Effect on dynamics\label{sec:Effect-on-Dynamics}}

In this section, we characterize how the eigenvalue distribution affects
the network dynamics. We focus on networks at criticality. This regime
is of particular interest because of the wealth of characteristic
features of criticality, like power-laws, that are experimentally
observed in brain networks \citep{Meshulam19,Fontenele19_208101,Stringer19_361},
feeding the so-called critical brain hypothesis \citep{Chialvo10_744}.
For example, a power-law decay in the PC spectrum, whose observed
fine-tuning \citep{Stringer19_361} we want to describe in this work,
only arises when the model \eqref{eq:SCS_model} is close to criticality
\citep{Hu22}. Theoretical studies also suggest the edge of criticality
as an optimal computational regime \citep{Langton90_12,Bertschinger04_1413,Toyoizumi11_051908}.

It is natural to consider the shifted eigenvalues $k\equiv\lambda-1$,
that is the eigenvalues of the dynamics' Jacobian $J-\mathbb{I}$.
Thus, calling $k\equiv k_{x}+\imath k_{y}$, a mode has decay constant
$k_{x}$ and oscillation frequency $k_{y}$. The network is at the
critical point when the eigenvalue distribution touches the line of
linear instability $k_{x}=0$ from the right (see e.g. \prettyref{fig:same_g_tau},
top). Occasionally, we parameterize a small distance away from criticality
by adding a small leak term $-\delta\cdot x$ to the \emph{rhs} of
\prettyref{eq:SCS_model}. This effectively shifts the eigenvalues
of the Jacobian $k\to k+\delta$, so the longest living mode have
decay constant $\delta$, rather than $0$.

The main result of this section is that the density of nearly critical
eigenvalues controls various dynamical observables that are found
to be fine-tuned in cortical networks, including the dimensionality
and PC spectrum of neural activity. This happens thanks to a direct
link between the density of nearly critical eigenvalues and the power-law
scaling exponents characterizing these observables. Precisely, we
show that these scaling exponents are algebraic functions of the exponents
$d$ or $\bar{d}$, controlling the density of nearly critical eigenvalues
through $p\left(k_{x}\right)\overset{k\to0}{\sim}k_{x}^{d-1}$ or
$p\left(\rho\right)\overset{k\to0}{\sim}\rho^{\bar{d}-1}$, where
in the first case we consider the marginal distribution of the eigenvalues'
real part $k_{x}$, while the second case describes the marginal distribution
of the eigenvalues' radial component $\rho$, with $k\equiv\rho e^{\imath\phi}$.
We call the exponent $d$ the network's effective spatial dimension,
in analogy with classical critical phenomena, as exposed later.

Note that, while here we focus on nearly critical systems, our theory
works also away from criticality, showing qualitatively similar results.
The difference is that, at the critical point, the effect of the eigenvalue
distribution on the dynamics is more apparent and readily quantifiable
in terms of power-law scaling exponents.

\subsection{Autocorrelation and autoresponse\label{subsec:Autocorrelation-and-autoresponse}}

Before focusing on the dimensionality and PC component spectrum of
neural activity, it is instructive to consider two other quantities
commonly studied in disordered networks \citep{Sompolinsky88_259}:
the population averaged autocorrelation and autoresponse functions.
These help to illustrate the effect of the distribution of nearly
critical eigenvalues on the dynamics. We show that these functions
have a power-law decay in time, whose exponent is controlled by the
density of nearly critical modes. Furthermore, we show that these
power-laws only emerge for certain distributions of oscillation frequencies,
which control a transition from exponential to power-law decay.

We define the population averaged autocorrelation as $A\left(t\right)=\frac{1}{N}\sum_{i}\left\langle x_{i}\left(t\right)x_{i}\left(0\right)\right\rangle _{\xi}$,
and the population averaged autoresponse as $r\left(t\right)=\frac{1}{N}\sum_{i}\lim_{\epsilon\to0}\frac{1}{\epsilon}\left\langle x_{i}^{\epsilon}\left(t\right)-x_{i}\left(t\right)\right\rangle _{\xi}$,
where $x_{i}^{\epsilon}\left(t\right)$ is the neural activity if
\prettyref{eq:SCS_model} is perturbed by a term $\epsilon\delta\left(t\right)$
along direction $i$.

Standard linear response theory gives
\begin{equation}
r\left(t>0\right)=\frac{1}{N}\sum_{\alpha}\exp\left(-k_{\alpha}t\right)\stackrel{N\to\infty}{\to}\int\mathcal{D}k\exp\left(-kt\right)\,,\label{eq:response}
\end{equation}
where in the last step we have taken the limit of the sum of eigenvalues
to an integral over their probability density, with integration measure
$\mathcal{D}k\equiv p\left(k\right)dk$, valid for large $N$. In
this limit, using our random matrix theory (see Appendix \ref{sec:Derivation-of-dynamical}
and the Supplemental Material \citep{Supplement}), we find $A\left(t\right)$
has the expression
\begin{multline}
A\left(t\right)=\frac{1+\nu^{2}}{1-\nu^{2}}\int\frac{\mathcal{D}k}{2k_{x}}\exp\left(-k\left|t\right|\right)\\
-\frac{2\nu^{2}}{1-\nu^{2}}\int\frac{\mathcal{D}k_{1}\mathcal{D}k_{2}}{k_{1}+k_{2}}\exp\left(-k_{1}\left|t\right|\right)\,.\label{eq:correlation}
\end{multline}
The autoresponse $r\left(t\right)$ depends only on the eigenvalue
distribution, and not on the eigenvectors. The autocorrelation $A\left(t\right)$
instead also depends on the eigenvectors' distribution, as reflected
by the parameter $\nu$. The first term in \prettyref{eq:correlation}
is the only one present in the limit of a normal network $\nu\to0$,
while the second term reflects a non-vanishing overlap between eigenvectors
in the non-normal case (see Appendix \ref{sec:Derivation-of-dynamical}).
For reference, we also report the known expression for the archetypal
choice of a Gaussian connectivity, in the $\tau=0$ case: $r\left(t>0\right)\overset{\delta\to0}{\sim}\exp\left(-t\right)$
and $A\left(t\right)\overset{\delta\to0}{\sim}\frac{D}{4\delta}\exp(-\delta t$).
\begin{figure*}
\centering{}\includegraphics[width=1\textwidth]{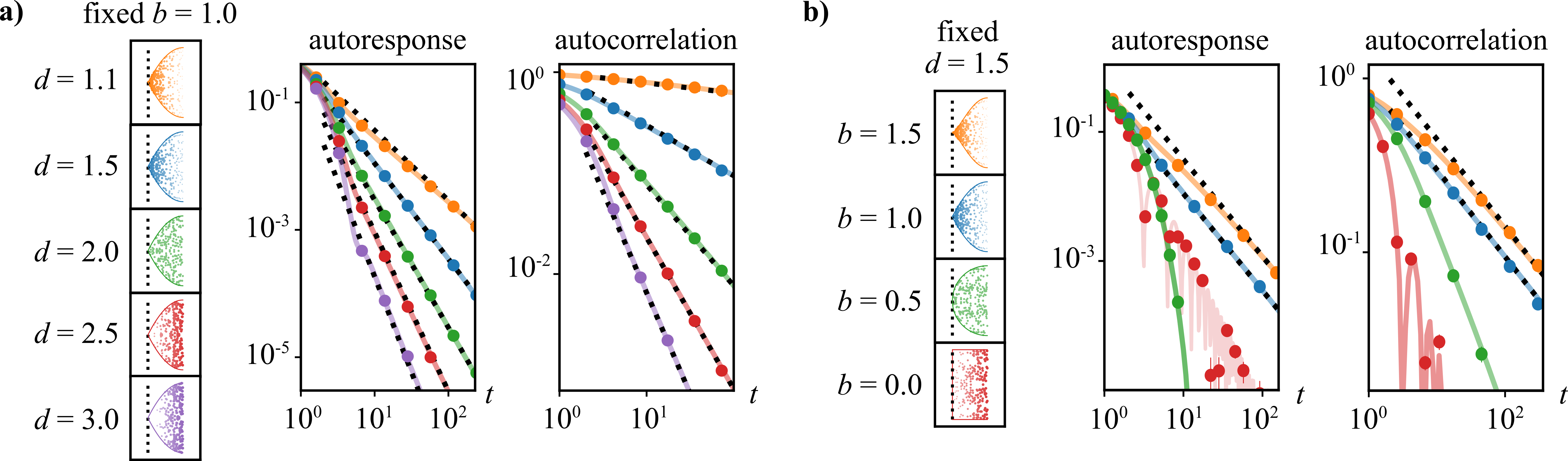}\caption{(a) Power-law decay, controlled by the density of nearly-critical
eigenvalues. On the left, different distributions of nearly critical
eigenvalues are shown in the $k_{x}$-$k_{y}$ plane for fixed $b=1$
and varying $d$. The dashed vertical line represents the critical
line of instability at $k=0$. Corresponding decay in time of the
autoresponse $r\left(t\right)$ and normalized autocorrelation $A\left(t\right)/A\left(0\right)$,
colored accordingly. Solid curves: theory; markers: simulations; dashed,
black lines: power-law decay with exponent $d$ for the autoresponse
and $d-1$ for the autocorreation. (b) Transition from power-law to
faster decay, controlled by the distribution of oscillation frequencies.
The autoresponse and normalized autocorrelation\textbf{ }are shown
for fixed $d=1.5$ and varying $b$. Solid curves: theory; markers:
simulation. For $b\protect\geq1$, we see a power law decay (dashed,
black lines) with exponent fixed by the value of $d$, regardless
of the value of $b$. For \textbf{$b<1$} the decay is faster. Other
parameters: $N=10^{2}$, $\nu=1/\sqrt{3}$. \label{fig:response_correation}}
\end{figure*}

Before looking at the time dependence of $r\left(t\right)$ and $A\left(t\right)$,
it is instructive to focus on the equal-time variance $A\left(0\right)$,
so as to introduce the concept of the network's effective spatial
dimension. As for the archetypal choice of $J$, the variance can
diverge as we approach criticality $\delta\to0$. It's divergent behavior
near criticality is described by the exponent $d$ characterizing
the density of nearly critical modes 
\begin{equation}
p\left(k_{x}\right)\overset{k_{x}\to0}{\sim}k_{x}^{d-1}.\label{eq:def_exponent_d}
\end{equation}
Indeed, we note that, when the variance diverges, the first term in
\prettyref{eq:correlation} dominates, behaving like
\begin{equation}
\sim\int\frac{p\left(k_{x}\right)dk_{x}}{k_{x}+\delta}\sim\int\frac{k_{x}^{d-1}dk_{x}}{k_{x}+\delta}\propto\begin{cases}
\delta^{d-1} & d<1\\
const & d>1
\end{cases},\label{eq:variance_scaling}
\end{equation}
where we introduced a small distance $\delta$ from criticality. We
can see that the variance remains finite for $d>d_{0}=1$, while it
diverges for smaller $d$.

Note that an analogous expression to \prettyref{eq:variance_scaling}
can be found for the variance of classical critical phenomena \citep{Wilson75_773,Hohenberg77,Taeuber14},
with $d$ playing the role of the system's spatial dimension. In fact,
for a very special choice of $J$, \prettyref{eq:SCS_model} becomes
the stochastic heat equation, a linear model of reference for out
of equilibrium critical phenomena \citep{Hohenberg77,Taeuber14}.
Specifically, one needs to choose $J$ such that $J-\mathbb{I}$ implements
a discretization of the Laplace operator (i.e. the kinetic term) on
a $d$-dimensional lattice. Then, the connectivity eigenmodes become
a very specific set of modes: the Fourier modes, with associated wave-vector
$\overrightarrow{k}$. The density of nearly critical Fourier modes
is fixed to be $p\left(\left\Vert \overrightarrow{k}\right\Vert \right)\sim\left\Vert \overrightarrow{k}\right\Vert ^{d-1}$,
where $d$ cannot take arbitrary values as in a generic network, but
corresponds to the system's spatial dimension. In analogy, we call
$d$ the network's effective spatial dimension. A difference between
networks and classical systems is that $d$ can take a continuous,
rather than discrete range of values, thanks to all possible shapes
of the connectivity's eigenvalue distribution. As in classical critical
phenomena, we will see how $d$ controls the scaling laws of dynamical
observables.

Having introduced the network's effective spatial dimension, we shall
now focus on the asymptotic long-time dependence of $r\left(t\right)$
and $A\left(t\right)$. As for $A\left(0\right)$, this is naturally
characterized by the marginal distribution $p\left(k_{x}\right)$
of decay constants of nearly critical modes, through the exponent
$d$. In addition, the conditional distribution of oscillation frequencies
$p\left(k_{y}|k_{x}\right)$ for a given decay constant is now important.
We find this quantity to control whether the decay in time is power-law
or exponential. Differently than for $A$$\left(0\right)$, a characterization
for a generic $p\left(k_{y}|k_{x}\right)$ is hard in this case. To
enlighten the role of oscillation frequencies, we consider the case
of a uniform distribution of oscillation frequencies

\begin{equation}
p\left(k_{y}|k_{x}\right)\propto\theta\left(B\left(k_{x}\right)-\left|k_{y}\right|\right),\quad B\left(k_{x}\right)\overset{k_{x}\to0}{\sim}k_{x}^{b}.\label{eq:def_exponent_b}
\end{equation}
As illustrated in \prettyref{fig:response_correation}, two exponents
parameterize the scaling properties of the distribution of nearly
critical modes. Exponent $d$ parameterizes the scaling of the density
of nearly critical modes \prettyref{eq:def_exponent_d}, while exponent
$b$ parameterizes the scaling of the maximum oscillation frequency
of nearly critical modes $\left|k_{y}\right|_{max}\propto k_{x}^{b}$,
i.e. the boundary of the eigenvalue distribution. We find that these
exponents also characterize the dynamics in a very direct way, as
illustrated in \prettyref{fig:response_correation}. Exponent $b$
controls how the autocorrelation and autoresponse decay in time (\prettyref{fig:response_correation}(b)).
In particular, it determines a transition from a power-law decay to
a faster decay: for slow enough oscillations of nearly critical modes
($b\geqslant1$) we have the power-law decay; when oscillations are
too fast ($b<1$), instead, the decay is slower, potentially exponential
and oscillating. Exponent $d$ controls the power-law exponent in
the case of exponential decay (\prettyref{fig:response_correation}(a)):
$r\left(t\right)\sim t^{-d}$ and $A\left(t\right)\sim t^{-(d-1)}$.

Considering $r\left(t\right)$ as an example, this can be easily seen
by integrating \prettyref{eq:response} over $k_{y}$ and passing
to the dimensionless variable $p\equiv k_{x}t$, obtaining
\begin{equation}
r\left(t\right)\overset{t\to\infty}{\sim}\frac{1}{t^{a+2}}\int_{0}^{\infty}dpp^{a}\exp\left(-p\right)\sin\left(p^{b}t^{\left(1-b\right)}\right).\label{eq:response_scaling}
\end{equation}
We can see that, if $b>1$, oscillations are unimportant in the long-time
limit, as we can expand the oscillating term $\sin\left(x\right)\sim x$
in the integrand, recovering the power-law scaling $r\left(t\right)\sim t^{-d}$.
If $b<1$, instead, oscillations interfere with the build up of nearly
critical modes into a pure power law. The reasoning is completely
analogous for $A\left(t\right)$, noting that the first term in \prettyref{eq:correlation}
dominates in the long-time limit. This means that, though the degree
$\nu$ of non-normality of the network affects quantitatively $A\left(t\right)$,
it does not affect its power-law decay exponent (see \prettyref{fig:Irrelevance-of-details}).

Note that in the archetypal case of a Gaussian connectivity, the decay
of both $r\left(t\right)$ and $A\left(t\right)$ is known to be exponential.
The emergence of power-laws is a novelty of certain connectivities
here considered, whose nearly critical modes have sufficiently slow
oscillation frequencies ($b>1$). This phenomenon may help accounting
for the emergence of power-laws ubiquitously observed in cortical
networks, both in correlation functions \citep{Meshulam19}, and in
the network's response, as in neural avalanches \citep{Beggs03_11167,Haldeman05,Fontenele19_208101}.
Of particular interest for neural avalanches is the here discovered
ability of connectivity to continuously vary the system's dynamical
scaling exponents, in line with the recently observed variability
in avalanche power-law exponents \citep{Fontenele19_208101}, and
in contrast with the classical notion of exponents being determined
by the universality class alone. While we leave investigating neural
avalanches to future work, we note that this variability may be explained
here by a fluctuation in the network's operational point, causing
a fluctuation in the network's effective connectivity, thus in the
distribution of nearly critical eigenvalues controlling power-law
exponents. 

We conclude remarking that the power-law scaling found in this section,
as well as in the following sections, only depends on a very generic
property of the eigenvalue distribution, namely the scaling \prettyref{eq:def_exponent_d}.
Other properties of the eigenvalue distribution can affect quantitatively
the shape of $A\left(t\right)$ and $r\text{\ensuremath{\left(t\right)}}$,
but are irrelevant with regards to the power-law scaling. For example,
\prettyref{fig:response_correation}(b) (cases $b=1.0,\,1.5$) illustrates
how varying the shape of the distribution boundary, here controlled
by $b$, does not affect the power-law exponents, which is fixed by
$d$. We further exemplify this in \prettyref{fig:Irrelevance-of-details}.
For example, we also show how the power-law scaling is not altered
by a stretching of the eigenvalue distribution along the imaginary
axis (cf. \prettyref{eq:precise_eigenval_d}). By \prettyref{eq:tau_relation},
this alters the degree of symmetry $\tau$ of the connectivity matrix.
Thus, in particular, also symmetry does not affect the power-law decay
exponent.

\subsection{Dimensionality\label{subsec:Dimensionality}}

In this subsection we characterize the effect of the density of nearly
critical eigenvalues on the dimensionality of neural activity. Recent
research has shown the high relevance of this measure for neural computation
\citep{Rigotti2013_585,Chung18_031003,Stringer19_361,sorscher22}.
Dimensionality indeed appears to be optimized into both a low \citep{Sadtler14,Gallego18_1,Semedo19_249}
and high \citep{Rigotti2013_585,Stringer19_361} dimensional regime,
depending on the cortical area and its specific computational tasks.
Here we show how the density of nearly critical eigenvalues can flexibly
tune dimensionality across both regimes, thanks to a novel transition
between high and low dimensional activity.

Dimensionality is typically defined through the principal components
spectrum of the neuronal activities' covariance matrix. Here we consider
both the case of the equal-time covariance $C_{ij}=\left\langle x_{i}\left(t\right)x_{j}\left(t\right)\right\rangle $
and the long time-window covariance $C_{ij}=\lim_{T\to\infty}\frac{1}{T}\left\langle \hat{x}_{i}\left(0\right)\hat{x}_{j}\left(0\right)\right\rangle $,
where we defined the Fourier transformed neural activity $\hat{x}_{i}\left(\omega\right)=\int_{-\frac{T}{2}}^{\frac{T}{2}}dte^{-i\omega t}x_{i}\left(t\right)$.
The so-called principal components of the covariance matrix are its
eigenmodes, defined through the decomposition
\begin{equation}
C=\sum_{\alpha}U_{\alpha}c_{\alpha}U_{\alpha}^{T}\,,\label{eq:def_principal_components}
\end{equation}
with the eigenvectors $U_{\alpha}$ identifying orthogonal directions
of neuronal variability, and the eigenvalues $c_{\alpha}\geq0$ its
intensity along that direction. Dimensionality is an estimate of how
many of the strongest principal components are required to explain
most of neuronal variability. A commonly adopted measure of dimensionality
is the participation ratio \citep{Hu22,Dahmen22_365072v3}
\begin{equation}
D\equiv\frac{\left(\sum_{\alpha}c_{\alpha}\right)^{2}}{\sum_{\alpha}c_{\alpha}^{2}}=\frac{\left(\mathrm{Tr}\left[C\right]\right)^{2}}{\mathrm{Tr}\left[C^{2}\right]},\label{eq:def_participation_ratio}
\end{equation}
which can conveniently be reduced to traces of powers of the covariance
matrix. The latter we are able to compute with our random matrix theory
(see Appendix \ref{sec:Derivation-of-dynamical} and the Supplemental
Material \citep{Supplement}). We now proceed to characterize the
behavior of dimensionality near criticality.

\subsubsection{Equal time covariance}

\begin{figure}
\centering{}\includegraphics[width=1\columnwidth]{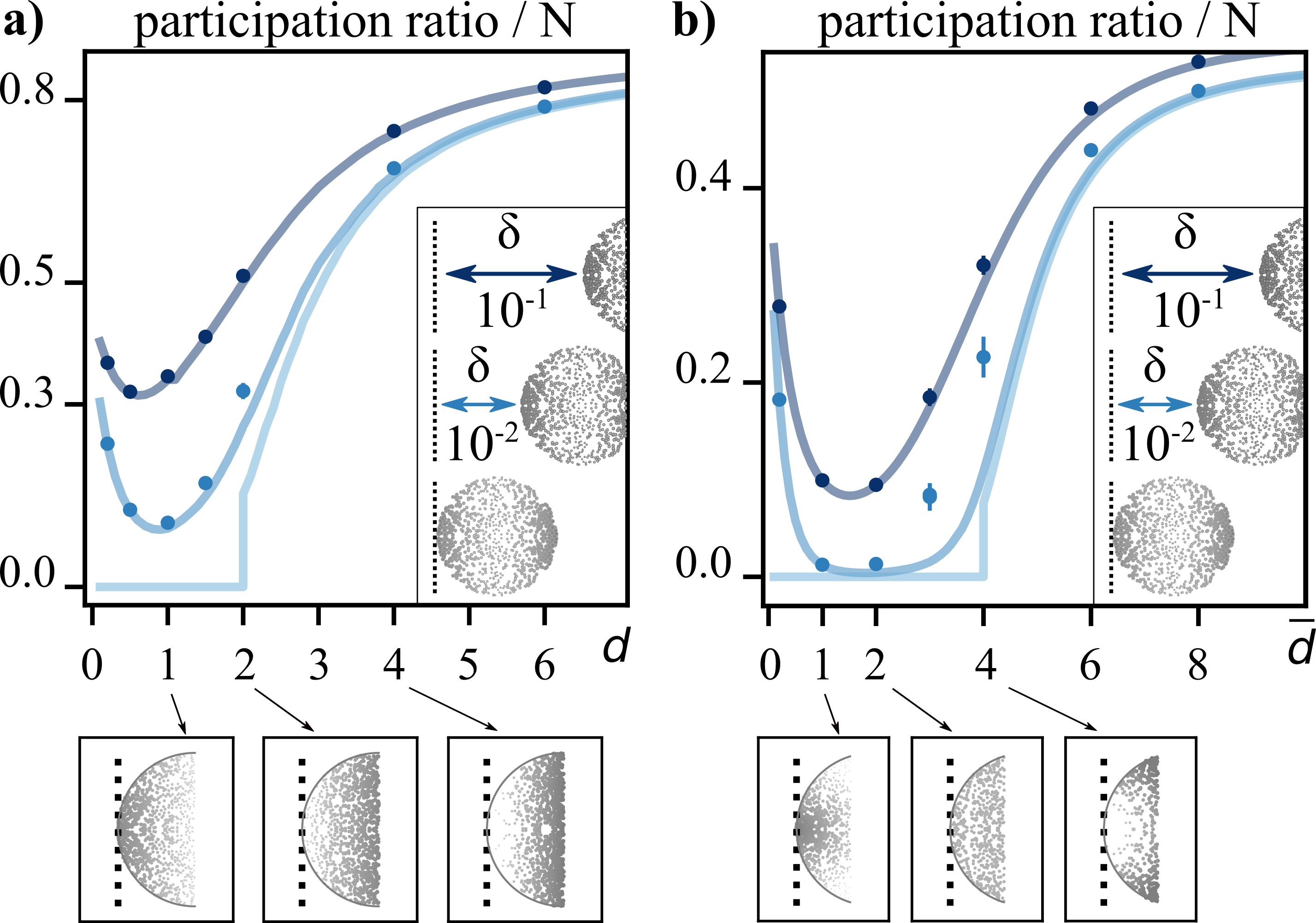}\caption{Transition from high to low dimensionality, controlled by the density
of nearly critical eigenvalues. (a) participation ratio for the equal-time
covariance, as a function of $d$.\textbf{ }Curves in lighter blue
correspond to decreasing values of $\delta$ (see inset). Solid curves:
theory; markers: simulation. Only theory is shown for $\delta=0$
(lightest blue), at which the transition is infinitely sharp. At the
bottom, distributions of nearly critical eigenvalues are shown for
some example values of $d$. (b) is a analogous to (a), showing the
participation ratio for the long time-window covariance as a function
of $\bar{d}.$ Note an eventual offset between theory and simulation
for the smaller $\delta$ is due to finite size effects (see \prettyref{fig:finite_size_dimensionality}
in the Supplemental Material \citep{Supplement}). Other parameters:
(a) $N=10^{3}$, $\nu=1/\sqrt{3}$, $b=0.5$; (b) $N=8\cdot10^{3}$.
\label{fig:dimensionality}}
\end{figure}

Let us start considering the equal-time covariance. Once again, we
find that dimensionality is controlled by the density of nearly critical
modes \prettyref{eq:def_exponent_d} through the exponent $d$. The
full expression for the dimensionality is made intricate by its dependence
on the degree $\nu$ of non-orthogonality, and is given in the Supplemental
Material \citep{Supplement}. However, we can get a readily interpretable
picture of its dependence on $d$ by looking at the asymptotic behavior
at criticality ($\delta\to0$). Note the numerator in \prettyref{eq:def_participation_ratio}
corresponds to $A\left(t=0\right)^{2}$ given in \prettyref{eq:correlation}.
We already discussed its diverging behavior near criticality in \prettyref{eq:variance_scaling}.
The full expression for the denominator $\mathrm{Tr}\left[C^{2}\right]$
is lengthy and is reported in the Supplemental Material \citep{Supplement}.
The expression is analogous to \prettyref{eq:correlation}, containing
a first term surviving for $\nu\to0$, which is the one dominating
the divergent behavior near criticality, in addition to terms due
to the network non-normality, analogous to the second term in \prettyref{eq:correlation}.
The dominating term behaves as
\begin{equation}
\sim\int\frac{p\left(k_{x}\right)dk_{x}}{\left(k_{x}+\delta\right)^{2}}\sim\int\frac{k_{x}^{d-1}dk_{x}}{\left(k_{x}+\delta\right)^{2}}\propto\begin{cases}
\delta^{d-2} & d<2\\
const & d>2
\end{cases}.\label{eq:spread_scaling}
\end{equation}
Combining \prettyref{eq:variance_scaling} and \prettyref{eq:spread_scaling}
into \prettyref{eq:def_participation_ratio}, we therefore distinguish
the behavior of dimensionality into three regions (\prettyref{fig:dimensionality}(a)):
for $d>2$, we are in a high-dimensional regime, in which dimensionality
is finite even at criticality; for $1<d<2$ we have low-dimensional
activity, decaying as $\delta^{2-d}$; for $d<1$ again we have low-dimensional
activity, but decaying as $\delta^{d}$.\textbf{ }We therefore have
a transition from high to low dimensional activity at $d=2$. The
transition becomes infinitely sharp at criticality. Otherwise, dimensionality
forms a bell shape in the low-dimensional regime $d<2$, with an optimum
for connectivities with dimension $d\sim1$, at which the dimensionality
of the input noise is reduced the most. This last observation may
also be interesting for initializing the connectivity of artificial
networks at an optimal starting point for dimensionality reduction,
from which to begin the network training. Note how the asymptotic
description we have just given well describes the full analytical
expression for the dimensionality even further away from criticality,
e.g. for $\delta=0.1$.

Once again, we show in \prettyref{fig:Irrelevance-of-details} that
both the degree of non-normality and other details in the shape of
the eigenvalue distribution do not affect qualitatively our predictions.

\subsubsection{Long time-window covariance}

For the long-time window covariance the reasoning is completely analogous.
Integrals of the same kind as \prettyref{eq:variance_scaling} appear
(see Appendix \ref{sec:Derivation-of-dynamical}). The main difference
is that, while in \prettyref{eq:variance_scaling} the relevant direction
of integration approaching the critical point $k=0$ was the one along
the real axis $k_{x}$, for the long-time window covariance, instead,
the relevant direction is the radial one $\rho$, where we defined
$k\equiv\rho e^{i\phi}$. We therefore introduce a radial dimension
$\bar{d}$ controlling the density of nearly critical modes along
the radial direction
\begin{equation}
p\left(\rho\right)\overset{\rho\to0}{\sim}\rho^{\bar{d}-1}.\label{eq:def_d_bar}
\end{equation}
The results are then completely analogous to those for the equal-time
covariance, and are shown in\textbf{ }\prettyref{fig:dimensionality}(b).
In this case, the transition between high and low dimensionality occurs
at $\text{\ensuremath{\bar{d}=4}}.$ In the low-dimensional regime,
dimensionality is $\sim\delta^{4-\bar{d}}$ for $2<\bar{d}<4$, and
$\sim\delta^{\bar{d}}$ for $\bar{d}<2$. The optimum is at $\bar{d}=2$. 

Note that, in the archetypal case of Gaussian $J$, we are in the
low-dimensional regime and dimensionality decays as $\delta^{2}$
\citep{Dahmen19_13051,Hu22,Dahmen22_365072v3}. This is reproduced
also for our matrix with a matching distribution of eigenvalues (i.e.
uniform on the circle), which has radial dimension $\bar{d}=2$. Going
beyond, here we show that with different shapes of the eigenvalue
distribution, any other scaling behavior can be obtained. In practice
this implies that, through connectivity, dimensionality can be flexibly
tuned within a wide range of values (cf. \prettyref{fig:dimensionality}).
Most importantly, also a high-dimensional regime can be accessed,
in which dimensionality remains finite even at the critical point.
This is of relevance because other connectivity structures like second
order motifs can only control dimensionality within the low-dimensional
regime, when at criticality \citep{Dahmen22_365072v3,Hu22}. In the
density of nearly critical eigenvalues, instead, we have identified
a connectivity structure that can flexibly tune dimensionality across
the entire range of low and high dimensional activity which is observed
throughout cortex \citep{Rigotti2013_585,Sadtler14,Gallego18_1,Semedo19_249,Stringer19_361}.
The ability to control dimensionality while staying at the critical
point is important in the light of the ubiquitous evidence of criticality
found in brain dynamics \citep{Chialvo10_744}, in particular the
characteristic presence of power-laws \citep{Beggs03_11167,Meshulam19,Fontenele19_208101,Stringer19_361},
such as, for example, that in PC spectrum \citep{Stringer19_361}
discussed below.

\subsection{Principal components spectrum\label{subsec:Principal-components-spectrum}}

\begin{figure*}
\centering{}\includegraphics[width=1\textwidth]{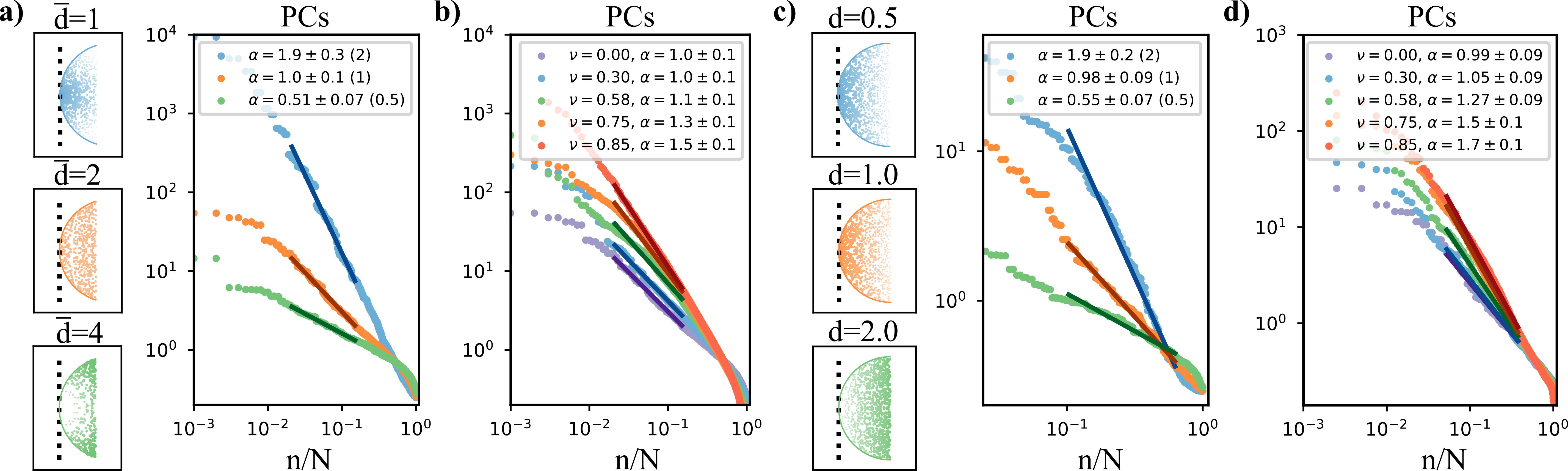}\caption{The density of nearly critical eigenvalues controls the slope of the
PC spectrum. PC spectrum plotted against the fractional rank $n/N$.
(a-b) Case of the long time-window covariance. (a) Case $\nu=0$.
Shown for varying densities of nearly critical eigenvalues, parameterized
by $\bar{d}$ (shown on the left and colored accordingly). Markers:
simulation; lines: fitting power-law; legend: mean and standard deviation
of the power law exponent $\alpha$, found for $48$ independent realizations
of the connectivity. Theoretical prediction in parenthesis. (b) Same
as (a), but for fixed $\bar{d}=2$ and varying $\nu$ (see legend).
(c-d) Same as (a-b), but for the case of the equal-time covariance:
In (c) we fix $\nu=0$ and vary $d$; In (d) we fix $d=1$ and vary
$\nu$. Other parameters: (a-b) $N=10^{3}$, $\delta=0.01$, $b=0.5$;
(c-d) $N=4\cdot10^{2}$, $\delta=0.01$. \label{fig:PCs}}
\end{figure*}
Another dynamical observable that has recently received considerable
interest is the PC spectrum of neural activity, i.e. the full set
of PC eigenvalues $c$ appearing in \prettyref{eq:def_principal_components},
plotted from largest to smallest. For example, a recent experimental
study has considered the PC spectrum of neural activity in V1 of mice,
during the encoding of image stimuli \citep{Stringer19_361}. A power-law
decay with exponent $\alpha=1$ is found, which is theoretically argued
to be optimal for encoding visual stimuli: larger $\alpha$ would
correspond to fewer details encoded, while a smaller $\alpha$ would
be too sensitive to details and correspond to a fractal representation.
This optimal exponent has also been found in artificial recurrent
networks trained to perform image classification \citep{Morales21_702}.
Recent theoretical work has derived a similar power-law decay for
a network governed by \prettyref{eq:SCS_model}, for the archetypal
choice of $J$ \citep{Hu22}. There, the power-law exponent is $\alpha=1.5$,
different from the one experimentally observed. Furthermore, it is
shown that no second order motif structure can alter this exponent.
Here we show how the density of nearly critical eigenvalues can instead
control this exponent, thus identifying a connectivity structure that
can account for the experimental observation of an optimized PC spectrum.

Deriving an expression for $p\left(c\right)$ for any value of the
degree of non-normality $\nu$ is a hard task that we leave for future
work. However, we can easily derive $p\left(c\right)$ for the normal
case $\nu=0$, and study numerically what happens for $\nu\neq0$.

Let us first consider the principal components of the time integrated
covariance. For $\nu=0$, the principal components eigenvalues $c$
are simply related to the connectivity eigenvalues $k=\rho e^{\imath\phi}$
by $c=\rho^{-2}$ (see Appendix \ref{sec:Derivation-of-dynamical}).
Therefore, $p\left(c\right)=p\left(\rho\left(c\right)\right)c^{-\frac{3}{2}}$.
Once again, we see that a power-law emerges, that is controlled by
the density of nearly critical modes \prettyref{eq:def_d_bar}: $p\left(c\right)\overset{c\to\infty}{\sim}c^{-\frac{\bar{d}+2}{2}}$.
It follows that the power-law decay of $c$ with its rank $n$ is
$c\sim n^{-\alpha}$, with $\alpha=2/\bar{d}$. Therefore, as illustrated
in \prettyref{fig:PCs}(a), varying the connectivity's density of
nearly critical modes tunes the power-law decay of the principal components
spectrum. Numerically, we find that also the degree of non-normality
$\nu$ can partially control the decay exponent. As illustrated in
\prettyref{fig:PCs}(b), for a fixed $\bar{d}$, increasing $\nu$
increases the exponent $\alpha$. This effect is weak for small or
intermediate values of $\nu$, and starts becoming apparent for large
values of $\nu$. Finally, let us note that a uniform elliptical eigenvalue
distribution corresponds to $\bar{d}=2$ and therefore to $\alpha=1$
in our ensemble. This is different from the value $\alpha=1.5$ found
for the archetypal $J$, which has the same eigenvalue distribution.
The difference is caused by the different statistics for the eigenvectors,
which in the archetypal case are strongly non-orthogonal and correlated
to the eigenvalues. Interestingly, the exponent $\alpha=1.5$ is approached
in our case by increasing $\nu$ to higher degrees of non-normality
$\nu\sim0.85$. We characterize the strongly non-normal regime in
\prettyref{subsec:Strongly-nonnormal-regime}, finding a similar qualitative
dependence of $\alpha$ on $\bar{d}$. In particular, the mechanism
by which the density of nearly critical eigenvalues can fine-tune
the slope of the PC spectrum is preserved also in this regime.

As illustrated in \prettyref{fig:PCs}(c-d), completely analogous
results follow for the equal-time covariance. The only difference
is that here $c=k_{x}^{-1}$ (see Appendix \ref{sec:Derivation-of-dynamical}),
thus the power-law exponent $\alpha$ is still controlled by the density
of nearly critical modes \prettyref{eq:def_exponent_d}, but the one
obtained approaching the critical point along the real axis. Specifically,
we have a power-law decay with rank $c\sim n^{-\alpha}$, with $\alpha=1/d$. 

\section{Connectivity statistics\label{sec:connectivity-statistics}}

In this section, we present in more detail our results deriving the
synaptic strength statistics of the connectivity matrices in our ensemble.

\subsection{Leading order synaptic statistics\label{subsec:Leading-order-synaptic}}

First, we comment on the leading order synaptic statistics. To leading
order, the connectivity is Gaussian and presents only reciprocal motifs,
just as in the archetypal case. The synaptic strength statistics are
given by \prettyref{eq:g_relation} and \prettyref{eq:tau_relation}.
These expressions depend on both the degree of non-normality $\nu$
and the second moments of the eigenvalue distribution. 

Let us first fix $\nu$ and discuss the effect of the eigenvalue distribution.
Notice the synaptic gain $g$ is controlled by the overall spread
of eigenvalues on the complex plane. Unsurprisingly, increasing the
spectral radius of the eigenvalue distribution corresponds to proportionally
increasing the synaptic gain. 

\begin{figure}
\centering{}\includegraphics[width=1\columnwidth]{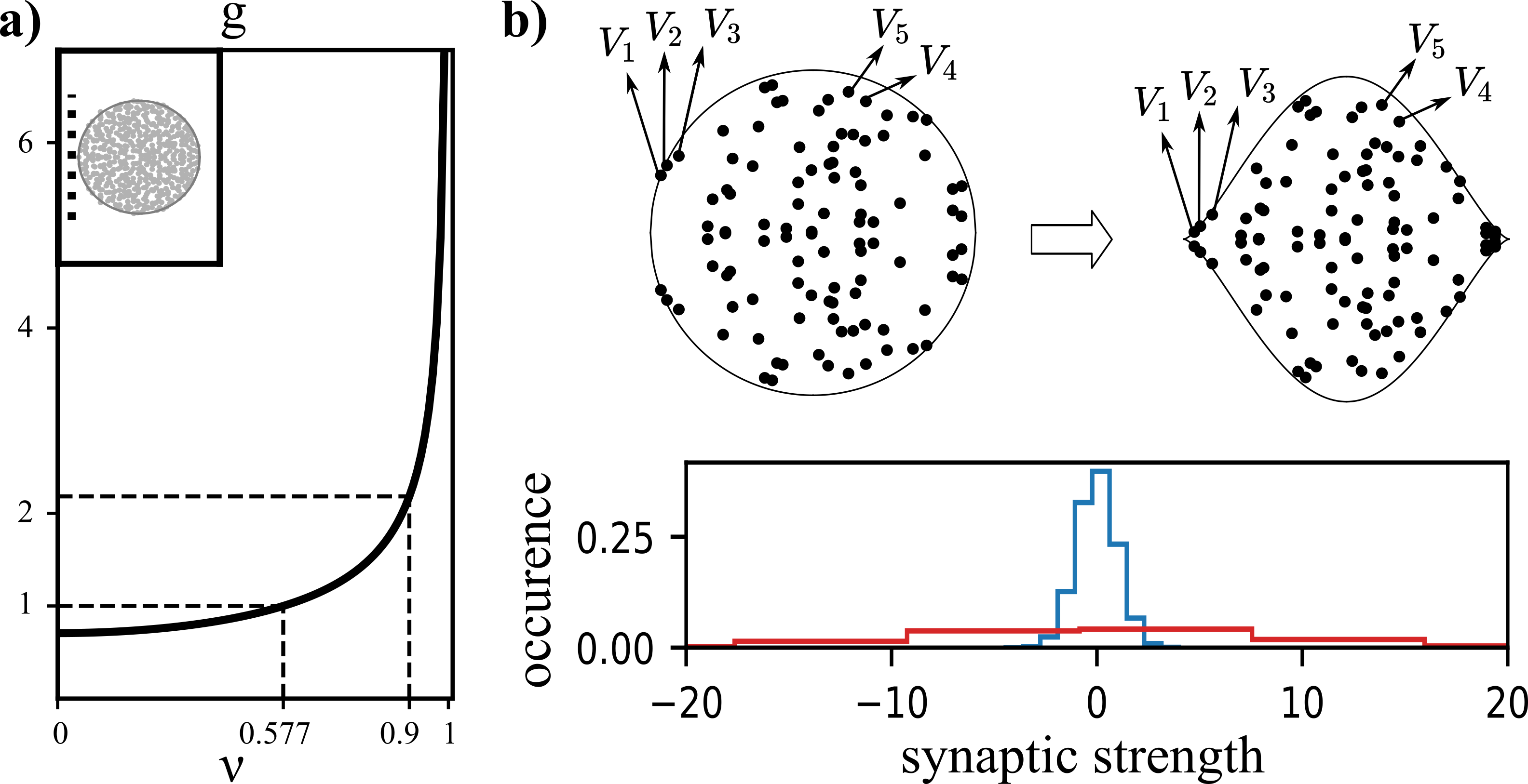}\caption{(a) Synaptic gain $g$ as a function of the degree of non-normality
$\nu$, for the case of a uniform circular distribution (inset), as
that found in the archetypal connectivity for $g=1$ and $\tau=0$.
The dotted lines highlight the special value $\nu=1/\sqrt{3}\sim0.577$,
for which $g=1$ is recollected. Also the value $\nu=0.9$ is highlighted,
for which $g$ is still of $\mathcal{O}\left(1\right)$. (b) Method
for initializing connectivities with arbitrary eigenvalue distribution
in the strongly non-normal regime. Top: starting from the archetypal
Gaussian connectivity, we continuously shift its eigenvalues (left)
into a new distribution (right). Each eigenvector remains associated
to the original eigenvalue, so as to preserve the tight correlation
structure that ensures $g=\mathcal{O}\left(1\right)$. This can be
checked in the blue histogram of scaled synaptic strengths $\sqrt{N}J_{ij}$
at the bottom, which has an $\mathcal{O}\left(1\right)$ variance.
If instead the strongly non-normal eigenvectors where randomly assigned
to the eigenvalues of the new distribution, we would have a divergent
synaptic gain (red histogram).\label{fig:eigenvectors_stats}}
\end{figure}

The degree of symmetry $\tau$ is instead controlled by the relative
spread of the distribution along the real and imaginary axes. Notice
that \prettyref{eq:tau_relation} generalizes to any eigenvalue distribution
in our ensemble a fact that could already be observed for the elliptical
distribution corresponding to the archetypal $J$: there, varying
$\tau$ corresponds to stretching the ellipse along the imaginary
or real axis (see e.g. \prettyref{fig:Synaptic-vs-dynamics}). In
fact, provided we choose $\nu^{2}=\frac{1-\tau^{2}}{3+\tau^{2}}$,
our ensemble reproduces the values of $g$ and $\tau$ of the archetypal
connectivity, when it is initialized with the same elliptical eigenvalue
distribution. For the special case $\tau=0$, we have to choose $\nu=1/\sqrt{3}\sim0.577$
(see \prettyref{fig:eigenvectors_stats}(a)). Note, however, that
the connectivities in the two ensembles are still different: On the
one hand, synaptic strength statistics of our ensemble have a more
complicated subleading order structure, which accounts for the different
possible shapes of the eigenvalue distribution. On the other hand,
the archetypal connectivity has strongly non-orthogonal eigenvectors
and a fine-tuned correlation structure between eigenvalues and eigenvectors,
which we introduce in our analysis later in \prettyref{subsec:Strongly-nonnormal-regime}.

For a fixed eigenvalue distribution, notice $g$ and $\tau$ can still
be partially tuned by the degree of non-normality $\nu$. For example,
stronger normality of the network (i.e. smaller $\nu$) corresponds
to stronger (anti)symmetry. Note that, as should be expected, perfect
(anti)symmetry is only achieved in the special limit of a normal network
$\nu=0$ and an eigenvalue distribution collapsed on the (imaginary)
real axis. 

Regarding the effect of $\nu$ on the synaptic gain $g$, note that
this remains of $\mathcal{O}\left(1\right)$ for considerably strong
$\nu\sim0.9$ (\prettyref{fig:eigenvectors_stats}(a)). However, a
too strong non-normality $\nu\to1$ makes $g$ diverge. More extreme
degrees of non-normality require correlations between eigenvalues
and eigenvectors, in order to keep $g$ of order unity. We discuss
this case in \prettyref{subsec:Strongly-nonnormal-regime}. Finally,
note that this divergence of the synaptic gain is a non-trivial effect
of the eigenvectors' non-orthogonality. It occurs for any fixed spectral
radius of the eigenvalue distribution. In particular, it cannot be
compensated by a trivial rescaling of the synaptic strengths, as this
would also shrink the spectral radius. 

\subsection{Hidden synaptic structures\label{subsec:Subleading-order-synaptic}}

Despite the connectivity being Gaussian to leading order in $N$,
it still contains all sorts of higher order statistics. These are
expected, as they must account for all eigenvalue distributions that
are possible within our ensemble. The surprising result that we show
here, however, is that these higher order statistics appear only at
subleading order in $N$.

Writing the connectivity in dimensionless units $\bar{J}\coloneqq J\sqrt{\frac{N}{g}}$,
we prove (see Appendix \prettyref{subsec:Derivation-of-the-synaptic-statistics}
and the Supplemental Material \citep{Supplement}) that higher order
cumulants $\langle\langle\bar{J}_{i_{1}i_{2}}\ldots\bar{J}_{i_{2n-1}i_{2n}}\rangle\rangle$
are of order $O$$\left(\frac{1}{\sqrt{N}}\right)$ for $n=3$ and
of order $\mathcal{O}$$\left(\frac{1}{N}\right)$ for any other $n>3$.
We also prove that the only nonvanishing cumulants are those for which
all indices $i_{1},\ldots,i_{2n}$ are matched in pairs. For second
order cumulants, this means only reciprocal motifs, associated with
the cumulant $\langle\langle J_{ij}J_{ji}\rangle\rangle$, are present.
For example, instead, cumulants like $\langle\langle J_{ij}J_{kj}\rangle\rangle$,
$\langle\langle J_{ji}J_{jk}\rangle\rangle$ or $\langle\langle J_{ij}J_{jk}\rangle\rangle$,
associated respectively with divergent, convergent, or chain motifs
are null (see \prettyref{fig:Synaptic-vs-dynamics} for a schematic
representation). For a numerical validation of our predictions on
the synaptic strength statistics, see \prettyref{fig:motifs_statistics_skewed}
and \prettyref{fig:motifs_statistics_variable_d_b} in the Supplemental
Material \citep{Supplement}.

The above results show that the here studied eigenmode structures
remain hidden to an analysis of local connectivity motifs. Indeed,
different shapes of the eigenvalue distribution and different degrees
of the eigenvectors' non-orthogonality are in large part not reflected
by motif structures. In particular, only the second order statistics
of the eigenvalues distribution affect the synaptic strength statistics,
through Equations \eqref{eq:g_relation} and \eqref{eq:tau_relation}.
All other details about the shape of the eigenvalue distribution remain
hidden in correlations that are vanishingly small in the number of
neurons. In particular, details like the density of nearly critical
eigenvalues, which we have shown in \prettyref{sec:Effect-on-Dynamics}
to significantly control network dynamics, remain hidden (cf. \prettyref{fig:same_g_tau}).
Note that this does not imply that these eigenmode structures are
undetectable -- in fact, they are clearly revealed by an eigenmode
analysis of connectivity, for example through different shapes of
the eigenvalue distribution. What this result shows is that these
structures are intrinsically collective: they cannot be reduced to
any local connectivity pattern, but rather involve all connections
concertedly.

These observations also provide a new insight into the paradox of
structure within randomness in highly heterogeneous circuits. We have
indeed shown that there can be much structure encoded in the eigenmode
statistics of an apparently random connectivity, which remains hidden
within the synaptic strength statistics. This opens new possibilities
for modeling neural networks: these eigenmode structures correspond
to additional degrees of freedom which can be appropriately specified,
without affecting other experimentally observed parameters, which
are traditionally fixed in connectivity models. For example, as illustrated
in \prettyref{fig:same_g_tau}, the density of nearly critical eigenvalues
can be specified without affecting the variance of synaptic strengths
or the abundance of reciprocal motifs. Other second other motifs are
also unaffected, as these are represented by additive low rank perturbations
to the here studied bulk connectivity \citep{Hu22}. From the opposite
angle, fixing these traditional parameters does not fix the density
of nearly critical eigenvalues. Rather, for the typical choice of
a Gaussian connectivity, it selects the specific case of a uniform
elliptical eigenvalue distribution. Given its impact on the dynamics,
however, the density of nearly-critical eigenvalues could instead
be fixed using our theory, so to match experimentally observed power-law
exponents, for example those found in the PC spectrum of neural activity,
in the case of V1.

\subsection{Strongly non-normal regime\label{subsec:Strongly-nonnormal-regime}}

\begin{figure*}
\centering{}\includegraphics[width=1\textwidth]{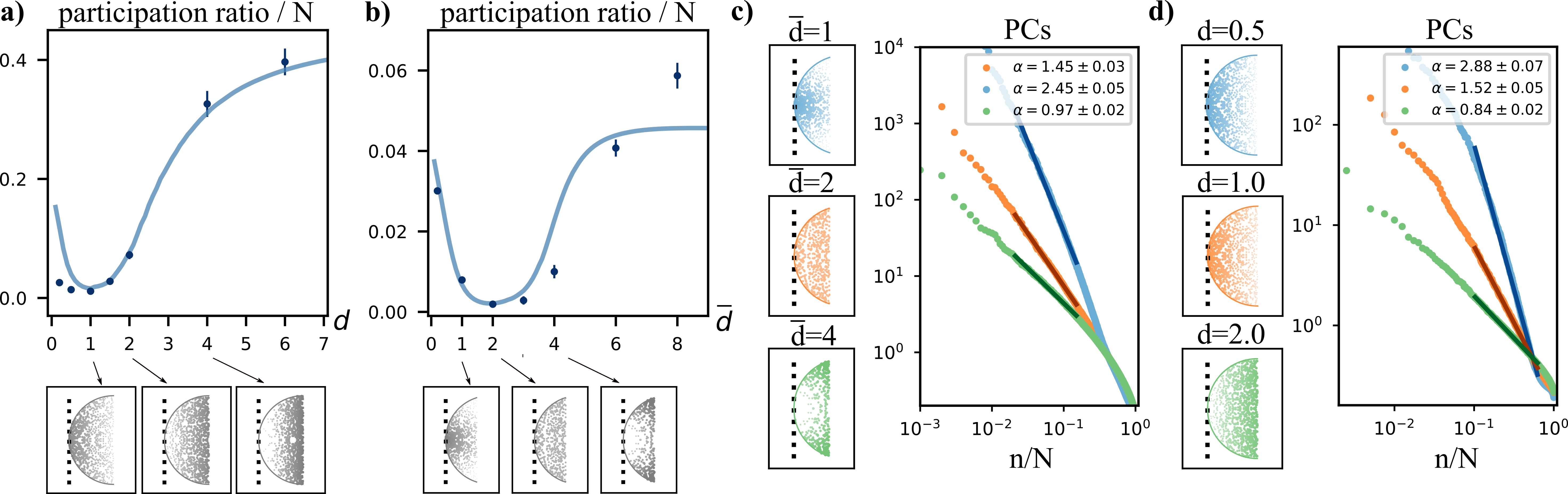}\caption{Strongly non-normal ensemble. (a-b) are analogous to \prettyref{fig:dimensionality}(a-b),
while (c-d) are analogous to \prettyref{fig:PCs}(a) and \prettyref{fig:PCs}(c).
The same eigenvalue distributions are considered, but with strongly
non-orthogonal eigenvectors. (a-b) participation ratio of the equal-time
(a) and long time-window (b) covariance. $\delta=0.01$. Markers:
simulation. Solid line: theory for the original ensemble at $\nu=0.8$
(a) and $\nu=0.9$ (b), shown for comparison. (c-d) Principal components
spectrum for the long time-window (c) and equal-time (d) covariance.
\label{fig:strongly-nonormal-regime}}
\end{figure*}
In this subsection we numerically extend our analysis to the case
of strongly non-orthogonal eigenvectors. With this extension, our
analysis encompasses the case of the archetypal choice of Gaussian
connectivity. We will see that most results obtained in \prettyref{sec:Effect-on-Dynamics}
qualitatively apply also to this regime.

In \prettyref{subsec:Leading-order-synaptic} we noticed that the
synaptic gain $g$ diverges in the strongly non-normal regime $\nu\to1$.
In Appendix \prettyref{subsec:Insights-into-the} we give an intuition
for this divergence in terms of the eigenvalues of the eigenvector
matrix. Here we rather focus on giving an intuition on how $g$ can
be kept of $\mathcal{O}\left(1\right)$ even in this strongly non-normal
regime, and provide a method to explore this regime numerically.

It is instructive to inspect the archetypal Gaussian connectivity.
This has $g=\mathcal{O}\left(1\right)$, while at the same time presenting
strongly non-orthogonal eigenvectors (see \prettyref{fig:overlaps-1}
(a) in the Supplemental Material \citep{Supplement}). This is made
possible by a fine-tuned correlation structure between eigenvalues
and eigenvectors, which we do not have in our original ensemble. We
can see this by randomly shuffling the association between eigenvectors
and eigenvalues and then reconstruct the connectivity through \prettyref{eq:eigen_decomposition}:
the synaptic gain then diverges with $N$, just like it does in our
ensemble, in which no correlation structure is assumed (see \prettyref{fig:overlaps-1}
(b-c) in the Supplemental Material \citep{Supplement}). This shows
that the original association was fine-tuned to tightly balance the
summation on the r.h.s. of \prettyref{eq:eigen_decomposition} to
result in finite synaptic strengths.

Leveraging on these insights, we devise a method to initialize connectivities
also in this strongly non-normal regime, while still allowing for
any shape of the eigenvalue distribution. The idea is to preserve
the necessary correlation structure found in the archetypal connectivity,
even when choosing an eigenvalue distribution different than uniform
elliptical. The method's details are given in Appendix \prettyref{subsec:appendix-Strongly-nonnormal-regime}.
The idea can be summarized as follows (\prettyref{fig:eigenvectors_stats}(b)):
We start from the archetypal $J$ for $\tau=0$, which naturally implements
strongly non-normal eigenvectors and the necessary correlations with
eigenvalues. It's eigenvalue distribution will be uniform on the circle.
We obtain other eigenvalue distributions by continuously shifting
the eigenvalues into some new position in the complex plane, such
that they are distributed according to the new desired distribution.
The intuition is that the continuous shift preserves to some extent
the tight correlations between eigenvalues and eigenvectors, such
that the connectivity remains well defined even if the eigenvectors
are strongly non-orthogonal. We verify numerically that indeed this
method ensures a well defined $g=\mathcal{O}\left(1\right)$ (see
\prettyref{fig:eigenvectors_stats}(b) and \prettyref{fig:overlaps-1}
in the Supplemental Material \citep{Supplement}). We call the ensemble
of connectivites obtained with this method the \emph{strongly non-normal
ensemble}, and refer to the ensemble discussed in the rest of the
manuscript, for which we derived analytical predictions, as the \emph{original
ensemble}.

Our numerical simulations for the strongly non-normal ensemble show
that our theory developed for the original ensemble still qualitatively
captures most of the phenomena described in \prettyref{sec:Effect-on-Dynamics}.
As shown in \prettyref{fig:strongly-nonormal-regime}(a-b), dimensionality
as a function of the density of nearly critical modes, as parameterized
by $d$ or $\bar{d}$, presents a similar shape as that shown in \prettyref{fig:dimensionality}.
As could be expected, it is for strong degrees of non-normality that
our theory for the original ensemble is most similar to the here presented
simulations. \prettyref{fig:supp_dimensionality_shaped} in the Supplemental
Material \citep{Supplement} shows additional simulations further
away from criticality.

As can be seen in \prettyref{fig:strongly-nonormal-regime}(c-d),
$d$ or $\bar{d}$ also control the power-law decay exponent of the
principal components spectrum. As we already noted in \prettyref{subsec:Principal-components-spectrum},
the decay is steeper than in our original ensemble at $\nu=0$, but
is in fact very similar to our original ensemble at large $\nu$.

For the autoresponse function, the results are exactly the same as
those for the original ensemble, as this function does not depend
on the eigenvector statistics. The only different behavior is shown
by the autocorrelation function (see \prettyref{fig:supp_autocorr_shaped}
in the Supplemental Material \citep{Supplement}). There, we find
the autocorrelation to diverge in both its amplitude $A\left(0\right)$
and exponential decay time as $\delta\to0$, even for $d>d_{0}$ (cf.
\prettyref{eq:variance_scaling}). Moreover, regardless of the value
of $b$, the decay is always exponential, rather than power-law. This
suggests that, in the case of the autocorrelation, the interplay between
strong non-orthogonality and correlations between eigenvalues and
eigenvectors has a role that is not qualitatively captured by our
original ensemble.

Not only the dynamics, but also the motifs statistics of these strongly
non-normal connectivities appear to be qualitatively captured by our
theory for the original ensemble (see \prettyref{fig:motifs_statistics_skewed}
and \prettyref{fig:motifs_statistics_variable_d_b} in the Supplemental
Material \citep{Supplement}). In particular our simulations suggest
that, also in this case, hidden synaptic structures control the shape
of the different eigenvalue distributions and hence the different
dynamics.

\section{Discussion\label{sec:Discussion}}

A fundamental quest of neuroscience is to identify the connectivity
structures that underlie the observed fine tuning of neural activity
into computationally optimal states. For cortical microcircuits, whose
connectivity is highly heterogeneous, the challenge is often to identify
these structures within an apparent randomness, that is in the connectivity
statistics. Within this setting, the literature has mainly focused
on the statistical abundance of local structures, such as connectivity
motifs \citep{Song05_0507,Hu13_P03012,Hu22,Dahmen22_365072v3}. However,
despite their impact on neural activity, for example its dimensionality
\citep{Hu13_P03012,Dahmen22_365072v3}, there still remain many computationally
relevant observables of network dynamics which cannot be controlled
by these structures alone.

Here we discover another type of connectivity structures which can
fine-tune such dynamical observables. These structures are of a complementary
and fundamentally different nature than motifs: they are \emph{collective},
rather than \emph{local}, and are encoded in the connectivity's eigenmode
statistics. We develop a novel random matrix theory that allows imposing
these eigenmode structures on the connectivity, such as the shape
of the eigenvalue distribution and the degree of the eigenvectors'
non-orthogonality, enabling us to systematically explore their effect
on network dynamics. In particular, we find the density of nearly-critical
eigenvalues to control dynamical observables that are found to be
fine-tuned in brain networks, and whose optimization remained so far
inaccessible to motif structures.

\subsection{Structures controlling dynamics}

Specifically, we show in \prettyref{subsec:Principal-components-spectrum}
that the slope of the PC spectrum of neural activity can be controlled
in a continuous fashion. In particular, it can be fine-tuned into
an optimal value for stimulus encoding, as observed in V1 of mice
\citep{Stringer19_361}. This functionally relevant feature of the
PC spectrum is not controllable by any type of second order motifs
\citep{Hu22}. Furthermore, we show in \prettyref{subsec:Dimensionality}
that dimensionality can be flexibly tuned across the entire range
of low and high dimensional activity, even when the network is kept
at criticality. This can happen thanks to a novel transition between
low and high dimensional activity discovered in this manuscript, which
is controlled by the density of nearly critical eigenvalues. In contrast,
second order motifs can only control dimensionality within the low
dimensional regime, when the network is critical \citep{Dahmen22_365072v3}.
Identifying a connectivity structure with such a broad and flexible
control of dimensionality is important to account for the observation
of both high \citep{Rigotti2013_585,Stringer19_361} and low \citep{Sadtler14,Gallego18_1,Semedo19_249}
dimensional activity in different cortical areas \citep{sorscher22},
which are thought to select either of the two regimes to optimize
their area-specific functions. Controlling dimensionality while staying
at the critical point is important in the light of the ubiquitous
evidence of criticality found in brain dynamics \citep{Chialvo10_744},
in particular the characteristic presence of power-laws \citep{Beggs03_11167,Meshulam19,Fontenele19_208101,Stringer19_361},
such as, for example, that in PC spectrum \citep{Stringer19_361}
discussed above.

The novel mechanisms described above are here formulated in terms
of a solid mathematical theory linking power-law exponents, ubiquitously
observed in brain dynamics \citep{Beggs03_11167,Meshulam19,Fontenele19_208101,Stringer19_361},
and connectivity. As we show in \prettyref{subsec:Autocorrelation-and-autoresponse},
the power-law scaling exponents of dynamical observables are directly
related to the scaling exponent parameterizing the density of nearly
critical eigenvalues, in analogy to how the system's spatial dimension
controls scaling in classical critical phenomena. Similar analogies
had been previously restricted to symmetric networks \citep{Bradde2010,Tuncer2015,brinkman2023},
thus limiting their application to neuroscience, which needs to consider
brain networks' asymmetric connections. The latter case presents new
technical challenges, such as non-orthogonal eigenvectors and complex
eigenvalues, that are here tackled by a novel kind of random matrix
theory.

Regarding complex eigenvalues, we find them to give rise to new phenomena
specific to the asymmetric case: As we show in \prettyref{subsec:Autocorrelation-and-autoresponse},
the distribution of oscillation frequencies (i.e. the eigenvalues'
imaginary part) controls a transition from exponential to power-law
decay of the autocorrelation and autoresponse functions. The emergence
of power-laws in the asymmetric case is in particular a novelty of
certain connectivities here considered, which may be relevant for
describing neural avalanches in disordered networks (cf. \prettyref{subsec:Autocorrelation-and-autoresponse}). 

Regarding the eigenvectors' non-orthogonality, we find it to only
weakly affect scaling exponents -- a result that further consolidates
the here found link between connectivity eigenvalues and dynamics.
This link holds qualitatively even in the strongly non-orthogonal
regime (see \prettyref{subsec:Strongly-nonnormal-regime}), especially
for those observables of most interest to this work, such as the PC
spectrum and dimensionality.

\subsection{Fundamental, yet hidden structures}

The eigenmode structures here discovered are of a new fundamental
kind, which is different and largely complements traditional motifs.
Their fundamental nature is exposed by our results in \prettyref{sec:Effect-on-Dynamics}
and \prettyref{sec:connectivity-statistics}, respectively from a
functional and structural perspective. Functionally, we have just
revised how these structures largely complement motifs, controlling
dynamical features inaccessible to the latter. But also structurally,
these eigenmode structures are fundamentally different: their nature
is \emph{collective}, involving all connections concertedly, rather
than \emph{local}, involving connections between a few neurons. Indeed,
we have shown in \prettyref{sec:connectivity-statistics} that, to
leading order in the number of neurons, different shapes of the eigenvalue
distribution and different degrees of eigenvectors' non-orthogonality
are in large part not reflected by local motifs.

Importantly, this last result exposes the existence of connectivity
structures which, despite their high relevance for network dynamics,
remain hidden to a traditional motif analysis. This poses solid theoretical
grounds to look for these structures in connectivity data. These are
revealed by an eigenmode analysis of the whole connectivity, which
is accessible to experimental methods acquiring snapshots of a whole
cortical microcircuit \citep{Berning15,yin2020}. A good surrogate
of experimental data are digital reconstructions of cortical circuits
\citep{Markram2015_456}, which are already being used to characterize
local motif structures, like simplices \citep{ecker23}.

\subsection{Structure within randomness}

The high heterogeneity of cortical microcircuits makes them appear,
to a first approximation, as randomly connected and thus strikingly
similar across cortical areas \citep{Braitenberg91}. Analyses and
models of their connectivity are thus faced with the paradox of identifying
structure within this randomness. The results in \prettyref{sec:connectivity-statistics}
provide a new insight into this paradox, showing that much structure
can be encoded in the eigenmode statistics of an apparently random
connectivity, while remaining hidden in the synaptic statistics.

Our random matrix theory enables direct control of these eigenmode
structures, thus opening new possibilities for modeling cortical microcircuits.
As shown in \prettyref{subsec:Subleading-order-synaptic}, these structures
represent new hidden features of connectivity, which can be fixed
in addition to more traditional parameters, like the mean and variance
of synaptic strengths, or the abundance of specific motifs. Note that
fixing only these traditional parameters corresponds to implicitly
making a particular choice for the eigenmode statistics, in particular
selecting a uniform elliptical eigenvalue distribution. However, given
the here discovered impact of the density of nearly critical eigenvalues
on functionally relevant observables, it is sensible to explicitly
fix this property of connectivity using our theory linking it to measurable
power-law exponents. For example, in the case of V1, this density
can be fixed to match the observed fine-tuning in the slope of the
PC spectrum \citep{Stringer19_361}.

We have shown how these eigenmode structures have a profound impact
on the dynamics, controlling previously inaccessible observables that
are also found to be fine-tuned in brain networks. It would thus be
interesting to further explore the effect of these structures on network
dynamics, in particular their interplay with other features of neural
networks, such as non-linear interactions, spiking neurons, or additional
low rank structures, whether of an organized \citep{Mastrogiuseppe18_609}
or statistical nature like non-reciprocal second order motifs \citep{Hu22}.
All of these features are readily added to the here studied system
\prettyref{eq:SCS_model}. In fact, we note that our method of modeling
connectivity is general and can be implemented in many network models.
In particular, it can also be adopted in models implementing inhibitory
and excitatory populations, as well as sparse connections. For example,
the connectivity can be decomposed as $S\odot\left(J+M\right)$ \citep{harris2022},
with $S$ a Boolean random matrix indicating the presence of a synapse,
implementing sparse connectivity, and $M$ a low rank matrix encoding
the inhibitory and excitatory mean synaptic strengths \citep{kadmon15,Hu22}.
In this case, the here studied random connectivity $J$ models fluctuations
around the mean synaptic strengths. This and similar extensions and
generalizations will reveal the impact of the here found hidden structures
in shaping the dynamics and function of other state-of-the-art neural
network models.
\begin{acknowledgments}
This work was partially supported by the Helmholtz Association, the
Initiative and Networking Fund under project number SO-092 (Advanced
Computing Architectures, ACA), and the German Federal Ministry for
Education and Research (BMBF Grant 01IS19077A). Open access publication
funded by the Deutsche Forschungsgemeinschaft (DFG, German Research
Foundation) -- 491111487.
\end{acknowledgments}

\appendix

\section{Connectivity statistics}

\renewcommand\thefigure{A\arabic{figure}}\setcounter{figure}{0}

\subsection{Details on the definition of the eigenmode statistics\label{subsec:Details-on-the-eigenmode-statistics}}

Let us give the remaining details on the definition of the eigenmode
statistics, presented in \prettyref{sec:setting}. For simplicity,
let us consider $N$ even. Generalizing to $N$ odd is straightforward.
The realness of the elements of $J$ imposes the condition that for
any complex $\lambda_{\alpha}$ and $V_{\alpha}$, there is an eigenmode
index $\alpha^{*}$ such that $\lambda_{\alpha^{*}}=\lambda_{\alpha}^{*}$
and $V_{\alpha^{*}}=V_{\alpha}^{*}$. Note that since eigenmodes are
randomly drawn in our ensemble, the probability that an eigenmode
is real has null measure. Therefore we can assume all eigenmodes to
be complex. By the constraint of real-valued $J$ we only have $N/2$
eigenmodes that are independent. We assign these eigenmodes to the
first $\alpha=1,\ldots,N/2$ of the eigenmode indices. The corresponding
complex conjugate eigenmodes are assigned the index $\alpha^{*}=\alpha+N/2$.
The first $N/2$ eigenvalues are drawn independently according to
the desired distribution $p\left(\lambda\right)$. The remaining half
is obtained by complex conjugation.

The matrix of eigenvectors $V$ is a linear combination of $O$ and
$G$. The scheme by which these two matrices are defined is the same.
Let us consider, for example, $G$. Only the first $\alpha=1,\ldots,N/2$
eigenvectors are independent, and each has a real and an imaginary
component. So we need to draw $\frac{N}{2}2N=N^{2}$ random variables.
We draw a real matrix $\gamma$ whose entries $\gamma_{i\alpha}\sim\mathcal{N}\left(0,1/N\right)$
are independent, normally distributed variables. The first $\alpha=1,\ldots,N/2$
vectors $G_{.,\alpha}$ are then defined as
\[
G_{\cdot,\alpha}=\frac{1}{\sqrt{2}}\gamma_{\cdot,\alpha}+\frac{\imath}{\sqrt{2}}\gamma_{\cdot,\alpha+N/2}
\]
In words, the first $\alpha=1,\ldots,N/2$ columns of $\gamma$ constitute
the real part of the vectors $G_{\cdot,\alpha}$, the remaining columns
$\alpha+N/2$ constitute the imaginary part. The remaining vectors
$G_{\cdot,\alpha^{*}}$ are obtained through complex conjugation.

The matrix $O$ is defined in the same way, substituting $G\to O$
and $\gamma\to o$. The only difference is that $o$ is a real orthonormal
matrix, drawn from the Haar distribution (i.e. the uniform distribution
over orthogonal matrices). One can easily check that $O$ satisfies
the complex orthonormality relation $OO^{\dagger}=\mathbb{I}$.

Let us note that here we considered eigenvalue distributions for which
the mean $\left\langle \lambda\right\rangle =0$. A non-vanishing
mean is trivial, as it amounts to sending $J\to J+\left\langle \lambda\right\rangle \mathbb{I}$,
which can be reabsorbed in the definition of the leak term $-x\to-\left(1-\left\langle \lambda\right\rangle \right)x$
in \prettyref{eq:SCS_model}.

We finally comment on $J$ being diagonalizable, and on the left eigenvectors'
matrix being the inverse of the right eigenvectors' matrix (cf. \prettyref{eq:eigen_decomposition}).
The subset of non-diagonalizable matrices has null measure. Therefore
it has null probability, unless one assumes some very specific distribution
that is singular on the subset, which is neither the case for our
ensemble nor for the archetypal Gaussian ensemble. The same reasoning
applies to the subset of matrices that have degenerate or null eigenvalues,
or both. Therefore, left and right eigenvectors are inverse of each
other if properly normalized. Note that here we are modeling the random
part of a network's connectivity. In this case, we have shown that
the aforementioned properties can be reasonably assumed. This is not
necessarily the case, instead, if one is modeling specific structures
that are added on top of the random connectivity, such as, for example,
low rank perturbations \citep{Mastrogiuseppe18_609}.

\subsection{Insights into the non-normality parameter $\nu$\label{subsec:Insights-into-the}}

Here we give further intuition regarding the non-normality parameter
$\nu$ and the specific definition \prettyref{eq:def_eigenvectors_statistics}
of the eigenvectors. This also allows us to get better insights into
why the synaptic gain $g$ diverges in the strongly non-normal regime
$\nu\to1$.

\begin{figure}
\centering{}\includegraphics[width=1\columnwidth]{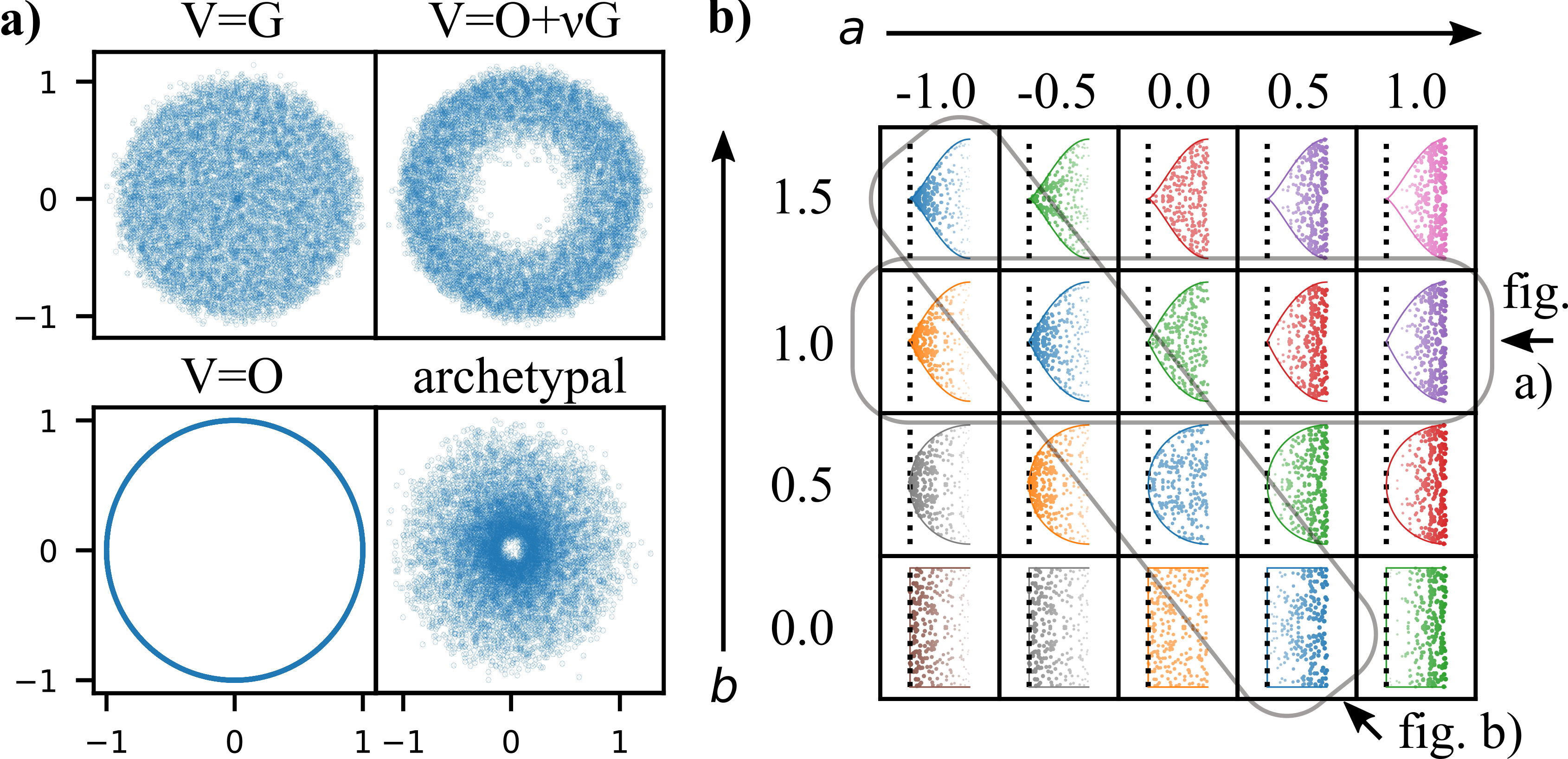}\caption{(a) Eigenvalues of the normalized eigenvector matrix $V$, for different
choices of $V$. Parameters: $N=10^{2}$. We show eigenvalues from
$10^{2}$ independent realizations of the eigenvectors. For the case
$V=O+\nu G$, we use a value of $\nu=0.8$. (b) Grid showing different
distributions of nearly critical eigenvalues for varying $a$ and
$b$. Distributions with the same $d=a+b+1$ share the same color
(diagonals). We highlight the eigenvalue distributions used for the
simulations in \prettyref{fig:response_correation}(a) (fixed $b$
and varying $a$, i.e. varying $d$) and in \prettyref{fig:response_correation}(b)
(fixed $d$ and varying $b$). \label{fig:vectors_eig_and_grid}}
\end{figure}

It is instructive to look at a naive approach in defining the eigenvector
statistics. As we stated in \prettyref{sec:setting}, we want to be
agnostic regarding the direction taken by the random eigenvectors
in neuronal space. The simplest - but too naive - way of implementing
this would be to initialize $V=G$ as random Gaussians, rather than
$V=O+\nu G$ as in \prettyref{eq:def_eigenvectors_statistics}. The
problem is that eigenvectors defined in this way take on too random
directions, having too strong overlaps. This causes the synaptic gain
$g$ to take on arbitrarily large values, when it should instead be
of order unity to have physical meaning. One helpful way to visualize
this is by looking at the eigenvalues of $V$ (\prettyref{fig:vectors_eig_and_grid}(a)).
For $V=G$, these are uniformly distributed in a circle, and can get
arbitrarily close to zero (\prettyref{fig:vectors_eig_and_grid}(a),
top left). This means $V^{-1}$, which also appears in the definition
of the connectivity $J,$\prettyref{eq:eigen_decomposition}, can
have arbitrarily large eigenvalues, and is thus unbounded, causing
also $g$ to be unbounded as a consequence. On the other hand, notice
that in the trivial case of a normal network, $V=O$, the eigenvalues
are exactly constrained to the unit circle, and the problem does not
occur (\prettyref{fig:vectors_eig_and_grid}(a), bottom left). To
have a well defined connectivity with $g=\mathcal{O}\left(1\right)$,
we introduce non-orthogonality in the eigenvectors gradually, shifting
away from the orthogonal case by increasing $\nu$, through the choice
$V=O+\nu G$. Notice that in this case the eigenvalues of $V$ cannot
get arbitrarily close to zero, but are constrained to be outside of
an inner circle, which will shrink back to zero as $\nu\to1$ (\prettyref{fig:vectors_eig_and_grid}(a),
top right). We observe the same mechanism occurring in the eigenvectors
of the archetypal connectivity (\prettyref{fig:vectors_eig_and_grid}(a),
bottom right). Notice that in this case the inner circle has a very
small radius, which in our approach would correspond to values of
$\nu$ very close to $1$ and a nonphysical synaptic gain $g=\mathcal{O}\left(\sqrt{N}\right)$,
diverging with the system size. In fact, as discussed in \prettyref{subsec:Strongly-nonnormal-regime},
such strong degree of non-normality in the archetypal $J$ is only
achieved through a fine-tuned correlation between eigenvalues and
eigenvectors, which ensures that all diverging contributions in \prettyref{eq:eigen_decomposition}
are tightly balanced.

\subsection{Eigenvalue distributions used in simulations\label{subsec:Eigenvalue-distributions-used}}

Here we report the precise eigenvalue distributions used for the simulation
of the dynamical quantities considered in \prettyref{sec:Effect-on-Dynamics}.
Note that only the shape of the distribution near criticality is important
in determining the scaling properties of dynamical quantities. Such
shape has been already reported in the main text and figures. The
choice of the distribution for eigenvalues further away from criticality
is arbitrary and irrelevant. We report it below for completeness.

For all quantities controlled by the exponent $d$, that is the density
of nearly critical modes along the real axis, we used the following
eigenvalue distribution
\begin{equation}
p\left(\boldsymbol{k}\right)=k_{x}^{a}\theta\left(S\left(1-\left(1-k_{x}\right)^{2}\right)^{b}-\left|k_{y}\right|\right)\quad\forall k_{x}\leq1\,,\label{eq:precise_eigenval_d}
\end{equation}
with $\theta$ being the Heaviside function. Note that \ref{eq:precise_eigenval_d}
obeys the asymptotic scaling \ref{eq:def_exponent_d} and \ref{eq:def_exponent_b}
for nearly critical eigenvalues, with $d=a+b+1$. The parameter $S$
allows to stretch the distribution along the imaginary axis. It is
fixed to $1$ in all simulations, unless stated otherwise. In \prettyref{fig:vectors_eig_and_grid}(b)
we show the distribution for the first half of the eigenvalues ($0<k_{x}<1$),
as a function of the parameters $a$ and $b$. The remaining half
of the eigenvalues, with $1<k_{x}<2$ and thus further away from criticality,
are drawn in a symmetric manner: we send $k\to2-k$, draw the new
variable according to \prettyref{eq:precise_eigenval_d}, and transform
back to the original variable (see for example the distributions shown
in \prettyref{fig:same_g_tau}). As stated above, the precise shape
of the eigenvalue distribution in this second half of the complex
plane is irrelevant. Thus, the specific choice of drawing eigenvalues
in a symmetric manner is only taken for technical convenience, because
then $\left\langle \lambda\right\rangle =0$ and there is no need
to rescale the leak term in \prettyref{eq:SCS_model} (see \prettyref{subsec:Details-on-the-eigenmode-statistics}).
In fact, even the precise shape of the distribution of the first half
of the eigenvalues, \prettyref{eq:precise_eigenval_d}, is irrelevant:
Only its limiting behavior for $k_{x}\to0$ is important. The specific
shape of the distribution has been chosen because for $a=0$ and $b=0.5$
it corresponds to a uniform elliptical distribution, as that of the
archetypal $J$.

For all quantities controlled by the exponent $\bar{d}$, that is
the density of nearly critical modes along the radial direction, we
used the following eigenvalue distribution
\begin{equation}
p\left(\rho,\phi\right)=\rho^{\bar{d}-1}\theta\left(\arccos\left(\frac{\rho}{2}\right)-\left|\phi\right|\right)\quad\forall\rho\leq1\label{eq:precise_eigenval_radial}
\end{equation}
where $k\equiv\rho e^{\imath\phi}$. Note the expression in the Heaviside
function constrains eigenvalues to lie within a circle. We fixed this
shape for convenience, because for $\bar{d}=2$ it reduces to the
uniform distribution on the circle, as that of the archetypal $J$.
Again, the exact shape of the distribution away from criticality is
irrelevant, but we report it here for completeness. The remaining
eigenvalues in the first half of the circle, i.e. with $1<\rho<\sqrt{2}$,
are drawn according to \prettyref{eq:precise_eigenval_radial}, but
are rejected and redrawn if they fall out of the semicircle, that
is if $k_{x}>1$. The remaining half of the eigenvalues are drawn
symmetrically, as discussed in the paragraph above.

\subsection{Strongly non-normal regime\label{subsec:appendix-Strongly-nonnormal-regime}}

Here we report details on the method presented in \prettyref{subsec:Strongly-nonnormal-regime},
to numerically implement connectivities in the strongly non-normal
regime that have a desired shape of the eigenvalue distribution. The
method produces the distributions presented in \prettyref{subsec:Eigenvalue-distributions-used}.

We start by initializing the archetypal Gaussian $J$, with $g=1$
and $\tau=0$. Its eigenvalues $\lambda_{\alpha}$ and associated
right and left eigenvectors $V_{\alpha}$ and $V_{\alpha}^{-1}$ are
derived numerically. The eigenvalues $\lambda$ are uniformly distributed
within a circle of unit radius. We start from this distribution and
continuously shift the eigenvalues into some new position, so that
they follow a new desired distribution. The continuous shift of the
eigenvalues preserves to some extent the correlation structure between
eigenvalues and eigenvectors, which in the strongly non-normal regime
is necessary to have a synaptic gain $g=\mathcal{O}\left(1\right)$.

We first focus on producing the eigenvalue distribution \prettyref{eq:precise_eigenval_d}.
This distribution is characterized by the parameters $S$, $b$ and
$d$ (or equivalently $a$). The starting uniform circular distribution
has parameters $S_{0}=1$, $b_{0}=0.5$ and $d_{0}=1.5$. Consider
$k=1-\lambda$. To produce the eigenvalue distribution \prettyref{eq:precise_eigenval_d}
we shift the eigenvalues $k\to\bar{k}$ through the transformation
\begin{gather}
\bar{k}_{x}=k_{x}^{\frac{d_{0}}{d}}\label{eq:kx_shift}\\
\bar{k}_{y}=k_{y}\frac{B\left(\bar{k}_{x};S,b\right)}{B\left(k_{x};S_{0},b_{0}\right)}\label{eq:ky_shift}
\end{gather}
where we defined the distribution's boundary function
\[
B\left(k_{x};S,b\right)\equiv S\left(1-\left(1-k_{x}\right)^{2}\right)^{b}\,.
\]
This transformation is applied to all eigenvalues with $k_{x}\leq1$.
The remaining half of the eigenvalues, with $1<k_{x}<2$, are transformed
in the symmetric manner described in \prettyref{subsec:Eigenvalue-distributions-used}.
Intuitively, \prettyref{eq:kx_shift} readjusts the eigenvalues closer
or further to the critical point according to the new $d$. \prettyref{eq:ky_shift}
rescales the imaginary part so that it fits the new boundary of the
distribution. 

To obtain the distribution \prettyref{eq:precise_eigenval_radial}
we use an analogous method, only that the transformation is performed
on the polar coordinates of $k=\rho e^{\imath\phi}$. The target distribution
is characterized by the parameter $\bar{d}$, and the original distribution
has a parameter $\bar{d}_{0}=2$. We apply the transformation
\begin{gather*}
\bar{\rho}=\rho^{\frac{\bar{d}_{0}}{\bar{d}}}\\
\bar{\phi}=\phi\frac{B\left(\bar{\rho}\right)}{B\left(\rho\right)}
\end{gather*}
with boundary function 
\[
B\left(\rho\right)\equiv\arccos\left(\frac{\rho}{2}\right)
\]
This transformation is applied to all eigenvalues with $\rho\leq1$.
As stated in \prettyref{subsec:Eigenvalue-distributions-used}, the
shape of the distribution for eigenvalues further from criticality
is arbitrary and irrelevant. We give it here for completeness. The
remaining eigenvalues with $1<\rho<\sqrt{2}$ which also are in the
first semicircle $k_{x}\leq1$ are left untouched. The remaining eigenvalues
in the second semicircle are transformed in the symmetric manner described
in \prettyref{subsec:Eigenvalue-distributions-used}.

\subsection{Derivation of the synaptic statistics\label{subsec:Derivation-of-the-synaptic-statistics}}

We develop a method to compute moments (or cumulants) of the matrix
elements of $J$. The details are reported in \prettyref{sec:Synaptic-strength-statistics}
of the Supplemental Material \citep{Supplement}. Here we summarize
the main ideas behind the method. 

Looking at the definition \prettyref{eq:eigen_decomposition}, we
can see that this involves being able to compute moments of the elements
of the eigenvector matrix $V=O+\nu G$ and its inverse $V^{-1}$ (commonly
called the matrices of the right and left eigenvectors, respectively).
For example, computing the second moment of $J$ corresponds to 
\[
\left\langle J_{ij}J_{hk}\right\rangle =\sum_{\alpha\beta}\left\langle \lambda_{\alpha}\lambda_{\beta}\right\rangle _{\lambda}\left\langle V_{i\alpha}V_{\alpha j}^{-1}V_{h\beta}V_{\beta k}^{-1}\right\rangle _{O,G}
\]
To proceed, we note that the inverse can be written as the infinite
series
\[
V^{-1}=\sum_{n=0}^{\infty}\left(-\nu\right)^{n}\left(O^{\dagger}G\right)^{n}O^{\dagger}\,,
\]
where we used that $O$ is unitary. Computing a certain moment of
$V$ and $V^{-1}$ thus corresponds to computing an infinite number
of moments of $O$ and $G$. Being Gaussian, the moments of $G$ can
be computed using Wick calculus \citep{Helias20_970}, that is using
the known result that moments of $G$ factorize into the expectation
of pairs of $G$, summing over all possible ways of pairing the $G$s.
Similarly, moments of $O$ can be computed using Weingarten calculus
\citep{weingarten78,collins06,matsumoto11}, the analogous of Wick
calculus for orthogonal matrices. Weingarten calculus is more complicated,
but in the limit of large $N$ it reduces to leading order to Wick
calculus \citep{weingarten78}.

There is still an infinite number of moments to compute and, at each
order of $\nu$, a large number of terms arising from the combinatorics
involved in Wick calculus. At each order of $\nu$, however, only
a few terms are of leading order in $N$. Using a Feynman-diagram
representation \citep{Helias20_970}, we are able to keep track of
these leading order terms, which can be identified based on the topology
of the associated diagrams. Once the terms of leading order in $N$
are computed for any given order of $\nu$, we are able to resum all
orders of $\nu$ exactly. Note that therefore our results are exact
in $\nu$ and perturbative in $N$, which is naturally large.

With this method, we compute the second moments of $J,$\prettyref{eq:g_relation}
and \prettyref{eq:tau_relation}. In \prettyref{subsec:Third-order-motifs}
of the Supplemental Material \citep{Supplement}, we also compute
the third cumulants of $J$. We do not compute explicitly higher order
moments of $J$. Computing these would involve considering subleading
order deviations of Weingarten calculus from Wick calculus. However,
we are able to use the properties of the full Weingarten calculus
to prove the results presented in \prettyref{subsec:Subleading-order-synaptic},
i.e. identifying which higher order moments do not vanish, and proving
that these non-vanishing moments are still of subleading order in
$N$.

\section{Derivation of dynamical quantities\label{sec:Derivation-of-dynamical}}

Here we give details on the derivation of the dynamical quantities
considered in \prettyref{sec:Effect-on-Dynamics}.

\subsection{Autocorrelation and autoresponse\label{subsec:Autocorrelation-and-autoresponse-1}}

Let us start by considering the system's linear response matrix $R\left(t\right)$,
which is the Green function of \prettyref{eq:SCS_model}. In frequency
domain this is defined as the solution to 
\begin{equation}
(\imath\omega\mathbb{I}+\mathbb{I}-J)R\left(\omega\right)=\mathbb{I}\,,\label{eq:greens_function}
\end{equation}
which is 
\begin{equation}
R\left(\omega\right)=(\imath\omega\mathbb{I}+\mathbb{I}-J)^{-1}\,.\label{eq:response_matrix_frequency}
\end{equation}
Using the eigenmode decomposition of $J$, \prettyref{eq:eigen_decomposition},
we can rewrite \prettyref{eq:response_matrix_frequency} as 
\begin{equation}
R_{ij}\left(\omega\right)=\sum_{\alpha}\frac{1}{\imath\omega+k_{\alpha}}V_{i\alpha}V_{\alpha j}^{-1}\label{eq:response_matrix_frequncy_eigen}
\end{equation}
which in time domain reads
\begin{equation}
R_{ij}\left(t\right)=\sum_{\alpha}\exp\left(-k_{\alpha}t\right)V_{i\alpha}V_{\alpha j}^{-1}\label{eq:response_matrix_time_eigen}
\end{equation}
The expression for the population averaged autoresponse \prettyref{eq:response},
considered in \prettyref{subsec:Autocorrelation-and-autoresponse},
directly follows from computing $r\left(t\right)=\frac{1}{N}\sum_{i}R_{ii}\left(t\right)$,
noticing that $\sum_{i}V_{i\alpha}V_{\alpha i}^{-1}=\delta_{\alpha,\alpha}=1$.
The latter identity makes $r\left(t\right)$ independent of the eigenvectors. 

Let us now consider the time-lagged covariance matrix $C_{ij}\left(t\right)\equiv\left\langle x_{i}\left(t\right)x_{j}\left(0\right)\right\rangle _{\xi}$.
This can be derived by plugging into its definition the formal solution
$x_{i}\left(t\right)=\int_{t'}\sum_{j}R_{ij}$$\left(t-t'\right)\xi_{j}\left(t'\right)$
and averaging over the noise. In frequency domain, the result is 
\begin{equation}
C\left(\omega\right)=(\imath\omega\mathbb{I}+\mathbb{I}-J)^{-1}(\imath\omega\mathbb{I}+\mathbb{I}-J)^{-\dagger}\label{eq:lagged_covariance_frequency}
\end{equation}
Using the eigenmode decomposition of $J$, \prettyref{eq:eigen_decomposition},
we can rewrite \prettyref{eq:lagged_covariance_frequency} as 
\begin{gather}
C_{ij}\left(\omega\right)=\sum_{\alpha\beta}F_{\alpha\beta}\left(\omega\right)V_{i\alpha}\left(\sum_{h}V_{\alpha h}^{-1}V_{\beta h}^{-1}\right)V_{j\beta}\,,\nonumber \\
F_{\alpha\beta}\left(\omega\right)\equiv\frac{1}{\left(\omega-\imath k_{\alpha}\right)\left(\omega+\imath k_{\beta}\right)}\label{eq:lagged_corr_frequency_eigen}
\end{gather}
which in time domain reads
\begin{gather}
C_{ij}\left(t\right)=\sum_{\alpha\beta}F_{\alpha\beta}\left(t\right)V_{i\alpha}\left(\sum_{h}V_{\alpha h}^{-1}V_{\beta h}^{-1}\right)V_{j\beta}\,,\nonumber \\
F_{\alpha\beta}\left(t\right)\equiv\frac{\theta\left(t\right)\exp\left(-k_{\alpha}t\right)+\theta\left(-t\right)\exp\left(k_{\beta}t\right)}{k_{\alpha}+k_{\beta}}\,.\label{eq:lagged_corr_time_eigen}
\end{gather}
The population averaged autocorrelation considered in \prettyref{subsec:Autocorrelation-and-autoresponse}
is given by $A\left(t\right)=\frac{1}{N}\sum_{i}C_{ii}\left(t\right)$,
which reads
\begin{equation}
A\left(t\right)=\frac{1}{N}\sum_{\alpha\beta}F_{\alpha\beta}\left(t\right)L_{\alpha\beta}\,,\label{eq:autocorr_not_self_averaged}
\end{equation}
where we defined the so-called overlap matrix 
\begin{equation}
L_{\alpha\beta}=\sum_{i}V_{i\alpha}V_{i\beta}\sum_{h}V_{\alpha h}^{-1}V_{\beta h}^{-1}\label{eq:overlap_def}
\end{equation}
which is a measure of how much different modes overlap in neuronal
space. It is diagonal in the case $\nu=0$ of orthonormal $V=O$.
For large number of neurons $N$, $A$ is self-averaging, meaning
$A\sim\left\langle A\right\rangle _{O,G}$ apart from fluctuations
of subleading order in $N$. In the expression for $A$, \prettyref{eq:autocorr_not_self_averaged},
we can therefore substitute $L$ with $\left\langle L\right\rangle _{O,G}$.
The latter we can compute using the same methods summarized in \prettyref{subsec:Derivation-of-the-synaptic-statistics}
(see \prettyref{subsec:Autocorrelation} of the Supplemental Material
\citep{Supplement} for the derivation). The result is, to leading
order in $N,$
\begin{equation}
\left\langle L_{\alpha\beta}\right\rangle =\frac{1+\nu^{2}}{1-\nu^{2}}\delta_{\beta,\alpha^{*}}-\frac{2}{N}\frac{\nu^{2}}{1-\nu^{2}}\,.\label{eq:overlap_result}
\end{equation}
As commented in \prettyref{subsec:Autocorrelation-and-autoresponse},
the first term is the only one present in the limit of orthonormal
eigenvectors $\nu\to0$, while the second term reflects a non-vanishing
overlap between eigenvectors for any other $\nu\neq0$. \prettyref{eq:correlation}
in \prettyref{subsec:Autocorrelation-and-autoresponse} is obtained
by plugging \prettyref{eq:overlap_result} into \prettyref{eq:autocorr_not_self_averaged}
and taking the limit of the sum over eigenvalues to an integral over
their probability density, with integration measure $\mathcal{D}k\equiv p\left(k\right)dk$.

\subsection{Dimensionality\label{subsec:Dimensionality-1}}

Let us recall the definition of the participation ratio for a generic
covariance matrix $C$
\begin{equation}
D\equiv\frac{\left(\mathrm{Tr}\left[C\right]\right)^{2}}{\mathrm{Tr}\left[C^{2}\right]}\,.\label{eq:appendix_def_participation_ratio}
\end{equation}
For the equal-time covariance, $C$ corresponds to $C\left(t=0\right)$
given by \prettyref{eq:lagged_corr_time_eigen}, while for the long
time-window covariance $C$ corresponds to $C\left(\omega=0\right)$
given by \prettyref{eq:lagged_corr_frequency_eigen}. 

\paragraph{Equal-time covariance}

As noted in \prettyref{subsec:Dimensionality}, for the equal-time
covariance, the numerator $\left(\mathrm{Tr}\left[C\right]\right)^{2}$
corresponds to $A\left(t=0\right)^{2}$ given in \prettyref{eq:correlation},
whose diverging behavior near criticality is discussed in \prettyref{eq:variance_scaling}.
Using \prettyref{eq:lagged_corr_time_eigen} the expression for the
denominator reads
\begin{equation}
\mathrm{Tr}\left[C^{2}\right]=\frac{1}{N^{2}}\sum_{\alpha\beta\gamma\delta}F_{\alpha\beta}\left(t=0\right)F_{\gamma\delta}\left(t=0\right)L_{\alpha\beta\gamma\delta}^{\left(2\right)}\,,\label{eq:spread_non_self_avg}
\end{equation}
where we defined the overlap tensor 
\begin{equation}
L_{\alpha\beta\gamma\delta}^{\left(2\right)}=\sum_{i}V_{i\alpha}V_{i\gamma}\sum_{j}V_{j\beta}V_{j\delta}\sum_{h}V_{\alpha h}^{-1}V_{\beta h}^{-1}\sum_{k}V_{\gamma k}^{-1}V_{\delta k}^{-1}\,.\label{eq:tensor_overlap}
\end{equation}
Also $\mathrm{Tr}\left[C^{2}\right]$ is self-averaging, so we can
substitute $L^{\left(2\right)}$ with $\left\langle L^{\left(2\right)}\right\rangle _{O,G}$
and compute it with the same methods used for $\left\langle L\right\rangle _{O,G}$.
The full result and its derivation are lengthy and are reported in
\prettyref{subsec:Crosscorrelations-spread} of the Supplemental Material
\citep{Supplement}. Here we only report the term that dominates in
the diverging behavior of $\mathrm{Tr}\left[C^{2}\right]$ near criticality
\[
\left\langle L_{\alpha\beta\gamma\delta}^{\left(2\right)}\right\rangle _{O,G}\sim\left(\frac{1+\nu^{2}}{1-\nu^{2}}\right)^{2}\delta_{\beta,\alpha^{*}}\delta_{\gamma,\alpha^{*}}\delta_{\delta,\alpha}
\]
which plugged into \prettyref{eq:spread_non_self_avg} gives \prettyref{eq:spread_scaling}
in \prettyref{subsec:Dimensionality}.

\paragraph{Long time-window covariance}

The reasoning for the long time-window covariance is completely analogous.
By comparing \prettyref{eq:lagged_corr_time_eigen} with \prettyref{eq:lagged_corr_frequency_eigen}
we notice that one simply needs to replace $F_{\alpha\beta}\left(t=0\right)\to F_{\alpha\beta}\left(\omega=0\right)$.
This leads to the results given in \prettyref{subsec:Dimensionality}.
Notice that, while $F_{\alpha\alpha^{*}}\left(t=0\right)=\frac{1}{\mathrm{Re}k_{\alpha}}$,
instead $F_{\alpha\alpha^{*}}\left(\omega=0\right)=\frac{1}{\left|k_{\alpha}\right|^{2}}$,
so the relevant direction along which to approach criticality is the
radial one, instead of the real axis. 

\subsection{Principal components spectrum\label{subsec:Principal-components-spectrum-1}}

From \prettyref{eq:lagged_corr_time_eigen} and \prettyref{eq:lagged_corr_frequency_eigen}
we immediately see that for $\nu=0$ the eigenvalues $c_{\alpha}$
of the equal time or long time-window covariance are, respectively,
$F_{\alpha\alpha^{*}}\left(t=0\right)=\frac{1}{\mathrm{Re}k_{\alpha}}$
and $F_{\alpha\alpha^{*}}\left(\omega=0\right)=\frac{1}{\left|k_{\alpha}\right|^{2}}$.
Indeed, for $\nu=0$ the eigenvectors are orthonormal and so the term
$\sum_{h}V_{\alpha h}^{-1}V_{\beta h}^{-1}=\delta_{\beta,\alpha^{*}}$.
From this observation follow the results discussed in \prettyref{subsec:Principal-components-spectrum}.

\section{Additional figures}

Here we provide some supplementary figures. Further supplementary
figures are provided in the Supplemental Material \citep{Supplement}.

\prettyref{fig:Irrelevance-of-details} shows the qualitative irrelevance
of details in the eigenmode statistics beyond the density of nearly
critical eigenmodes in controlling the dynamical quantities studied
in \prettyref{sec:Effect-on-Dynamics}. In particular, \prettyref{fig:Irrelevance-of-details}(a)
shows the autoresponse and autocorrelation functions for different
connectivities, which share the same density of nearly critical eigenvalues,
but differ in other properties: we consider different relative spreads
of the eigenvalues along the imaginary and real axis, controlling
the level of symmetry in the connectivity, along with different levels
of non-normality. Even if these parameters vary, the power-law decay
of both functions is only controlled by $d$ and remains the same.
Analogously, \prettyref{fig:Irrelevance-of-details}(b) shows that
the slope in the PC spectrum does not change for different relative
spreads or different shapes of the boundary of the eigenvalue distribution.
Figures \prettyref{fig:Irrelevance-of-details}(c-d) show dimensionality
as all of the aforementioned parameters are varied. Again, while we
have quantitative differences, the qualitative behavior of dimensionality
is not altered. The transition from high to low dimensionality still
occurs at $d=2$, and a clear minimum is present at $d\sim1$.

\prettyref{fig:motifs_statistics_skewed} shows a numerical validation
of our analytical predictions for the motifs statistics. A connectivity
with an asymmetric eigenvalue distribution is chosen, so that third
order motifs do not vanish. The numerics for the strongly non-normal
regime are also reported and compared with the predictions for our
ensemble. Notice that simulations agree well with theory. Also in
the case of the strongly non-normal connectivity, we have a qualitative
agreement for the non-vanishing motifs, and even a quantitative agreement
for the motifs that are expected to vanish exactly (see the red markers
partially hidden below the blue markers). Note that the values of
the third and fourth order cumulants are shown rescaled by a factor
$\sqrt{N}$ and $N$, respectively. The fact that they are of order
unity in the plot, therefore, validates our prediction that these
higher order motifs are subleading in the number of neurons.

\renewcommand\thefigure{C\arabic{figure}}\setcounter{figure}{0}
\begin{figure*}
\centering{}\includegraphics[width=1\textwidth]{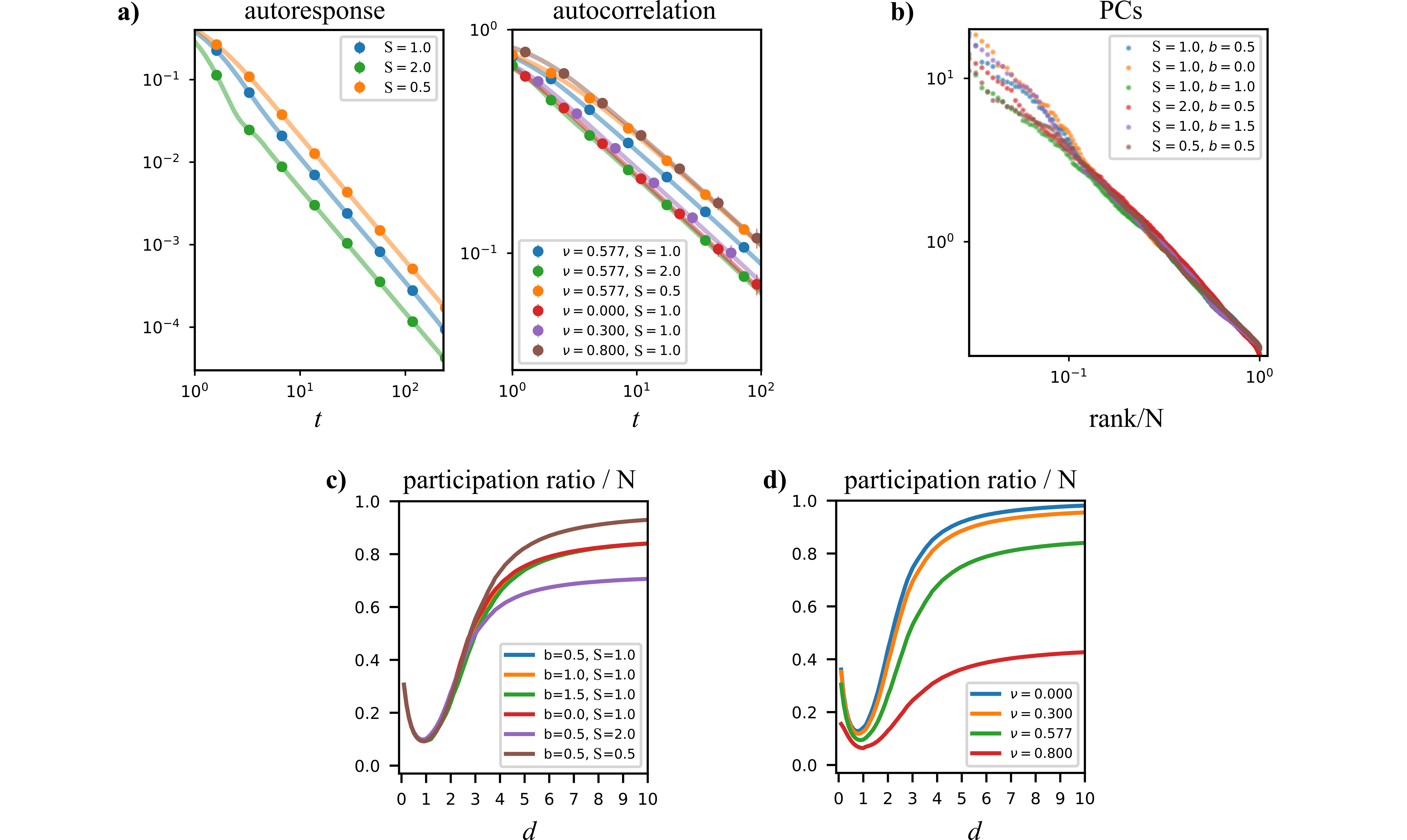}\caption{Irrelevance of details in the shape of the eigenvalue distribution
and the degree of non-normality. (a) Autoresponse and autocorrelation,
for $d=1.5$. Shown for varying stretching factor $S$ of the eigenvalue
distribution along the imaginary axis (cfr. \prettyref{eq:precise_eigenval_d})
and for varying degree of non-normality $\nu$ (autocorrelation only).
Markers: simulation; full curves: theory. Other parameters: $b=1.0$,
$N=10^{2}$. (b) Principal component spectrum of the equal-time covariance,
for $d=1.0$. Shown for varying $S$ and varying $b$, controlling
the boundary of the eigenvalue distribution (cfr. \prettyref{eq:def_exponent_b}).
Other parameters: $\nu=1/\sqrt{3}$, $N=4\cdot10^{2}$, $\delta=0.01$.
(c) Dimensionality of the equal-time covariance, for $\delta=0.01$.
Shown for varying $S$ and $b$. Full curves: theory. Other parameters:
$\nu=1/\sqrt{3}$. (d) Dimensionality of the equal-time covariance,
for $\delta=0.01$. Shown for varying $\nu$. Full curves: theory.
Other parameters: $S=1$, $b=0.5$.\label{fig:Irrelevance-of-details}}
\end{figure*}
 
\begin{figure*}
\centering{}\includegraphics[width=1\textwidth]{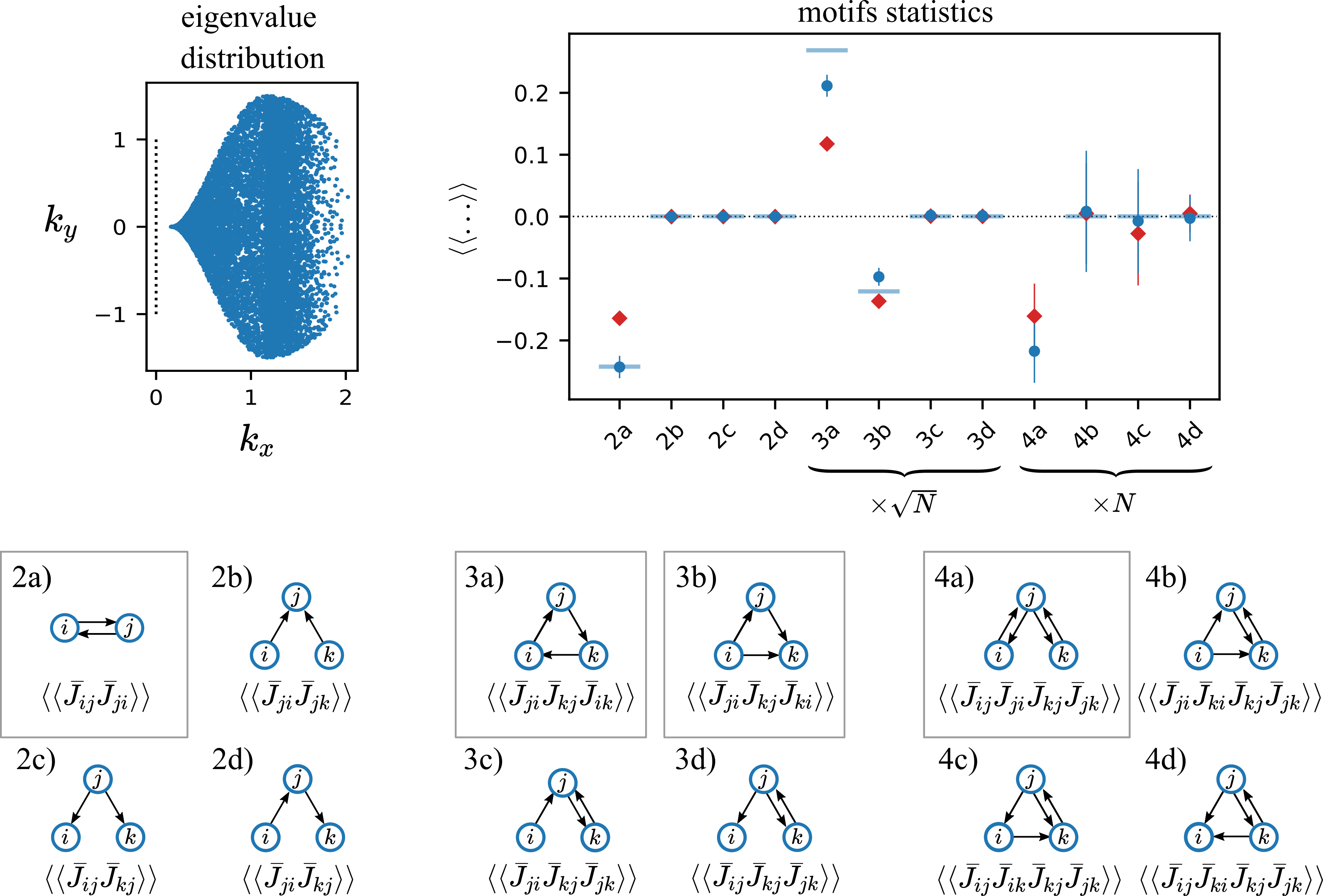}\caption{Motifs statistics. Bottom: catalog of all motifs up to fourth order
involving up to three different neurons, alongside with the associated
cumulant of the normalized synaptic strengths matrix $\bar{J}$ (cf.
\prettyref{subsec:Subleading-order-synaptic}). Top: empirical measurement
of these cumulants (i.e. averaging over synapses within a single connectivity
realization), for a connectivity with eigenvalue distribution shown
on the left. Blue dots: case of a connectivity in our ensemble, for
$\nu=0.577$; red dots: case of a conenctivity in the strongly non-normal
ensemble. Note that third and fourth order cumulants are plotted,
rescaled by a factor $\sqrt{N}$ and $N$, respectively. Blue dash:
theoretical prediction for the original ensemble. All motifs that
are not highlighted with boxes in the bottom catalog are predicted
by our theory to vanish exactly. For the non-vanishing motifs up to
third order, the analytical prediction is given by \prettyref{eq:tau_relation}
for the second order motifs and equations (\ref{eq:third_motifs_1}-\ref{eq:third_motifs_2})
in the Supplemental Material \citep{Supplement} for the third order
motifs. \label{fig:motifs_statistics_skewed}}
\end{figure*}

\clearpage\bibliographystyle{apsrev4-1_prx}
\bibliography{brain,supplement}

\begin{thebibliography}{51}%
\makeatletter
\providecommand \@ifxundefined [1]{%
 \@ifx{#1\undefined}
}%
\providecommand \@ifnum [1]{%
 \ifnum #1\expandafter \@firstoftwo
 \else \expandafter \@secondoftwo
 \fi
}%
\providecommand \@ifx [1]{%
 \ifx #1\expandafter \@firstoftwo
 \else \expandafter \@secondoftwo
 \fi
}%
\providecommand \natexlab [1]{#1}%
\providecommand \enquote  [1]{``#1''}%
\providecommand \bibnamefont  [1]{#1}%
\providecommand \bibfnamefont [1]{#1}%
\providecommand \citenamefont [1]{#1}%
\providecommand \href@noop [0]{\@secondoftwo}%
\providecommand \href [0]{\begingroup \@sanitize@url \@href}%
\providecommand \@href[1]{\@@startlink{#1}\@@href}%
\providecommand \@@href[1]{\endgroup#1\@@endlink}%
\providecommand \@sanitize@url [0]{\catcode `\\12\catcode `\$12\catcode
  `\&12\catcode `\#12\catcode `\^12\catcode `\_12\catcode `\%12\relax}%
\providecommand \@@startlink[1]{}%
\providecommand \@@endlink[0]{}%
\providecommand \url  [0]{\begingroup\@sanitize@url \@url }%
\providecommand \@url [1]{\endgroup\@href {#1}{\urlprefix }}%
\providecommand \urlprefix  [0]{URL }%
\providecommand \Eprint [0]{\href }%
\providecommand \doibase [0]{http://dx.doi.org/}%
\providecommand \selectlanguage [0]{\@gobble}%
\providecommand \bibinfo  [0]{\@secondoftwo}%
\providecommand \bibfield  [0]{\@secondoftwo}%
\providecommand \translation [1]{[#1]}%
\providecommand \BibitemOpen [0]{}%
\providecommand \bibitemStop [0]{}%
\providecommand \bibitemNoStop [0]{.\EOS\space}%
\providecommand \EOS [0]{\spacefactor3000\relax}%
\providecommand \BibitemShut  [1]{\csname bibitem#1\endcsname}%
\let\auto@bib@innerbib\@empty
\bibitem [{\citenamefont {Dorogovtsev}\ \emph {et~al.}(2008)\citenamefont
  {Dorogovtsev}, \citenamefont {Goltsev},\ and\ \citenamefont
  {Mendes}}]{Dorogovtsev08}%
  \BibitemOpen
  \bibfield  {author} {\bibinfo {author} {\bibfnamefont {S.~N.}\ \bibnamefont
  {Dorogovtsev}}, \bibinfo {author} {\bibfnamefont {A.~V.}\ \bibnamefont
  {Goltsev}}, \ and\ \bibinfo {author} {\bibfnamefont {J.~F.~F.}\ \bibnamefont
  {Mendes}},\ }\bibfield  {title} {\bibinfo {title} {Critical phenomena in
  complex networks},\ }\href {\doibase 10.1103/RevModPhys.80.1275} {\bibfield
  {journal} {\bibinfo  {journal} {Rev. Mod. Phys.}\ }\textbf {\bibinfo {volume}
  {80}},\ \bibinfo {pages} {1275} (\bibinfo {year} {2008})}\BibitemShut
  {NoStop}%
\bibitem [{\citenamefont {Kandel}\ \emph {et~al.}(2013)\citenamefont {Kandel},
  \citenamefont {Schwartz}, \citenamefont {Jessell}, \citenamefont
  {Siegelbaum}, \citenamefont {Hudspeth},\ and\ \citenamefont
  {Mack}}]{Kandel13}%
  \BibitemOpen
  \bibfield  {author} {\bibinfo {author} {\bibfnamefont {E.~R.}\ \bibnamefont
  {Kandel}}, \bibinfo {author} {\bibfnamefont {J.~H.}\ \bibnamefont
  {Schwartz}}, \bibinfo {author} {\bibfnamefont {T.~M.}\ \bibnamefont
  {Jessell}}, \bibinfo {author} {\bibfnamefont {S.~A.}\ \bibnamefont
  {Siegelbaum}}, \bibinfo {author} {\bibfnamefont {A.}~\bibnamefont
  {Hudspeth}}, \ and\ \bibinfo {author} {\bibfnamefont {S.}~\bibnamefont
  {Mack}},\ }\href@noop {} {\emph {\bibinfo {title} {Principles of Neural
  Science}}},\ \bibinfo {edition} {5th}\ ed.\ (\bibinfo  {publisher}
  {McGraw-Hill},\ \bibinfo {address} {New York},\ \bibinfo {year}
  {2013})\BibitemShut {NoStop}%
\bibitem [{\citenamefont {Stringer}\ \emph {et~al.}(2019)\citenamefont
  {Stringer}, \citenamefont {Pachitariu}, \citenamefont {Steinmetz},
  \citenamefont {Carandini},\ and\ \citenamefont {Harris}}]{Stringer19_361}%
  \BibitemOpen
  \bibfield  {author} {\bibinfo {author} {\bibfnamefont {C.}~\bibnamefont
  {Stringer}}, \bibinfo {author} {\bibfnamefont {M.}~\bibnamefont
  {Pachitariu}}, \bibinfo {author} {\bibfnamefont {N.}~\bibnamefont
  {Steinmetz}}, \bibinfo {author} {\bibfnamefont {M.}~\bibnamefont
  {Carandini}}, \ and\ \bibinfo {author} {\bibfnamefont {K.~D.}\ \bibnamefont
  {Harris}},\ }\bibfield  {title} {\bibinfo {title} {High-dimensional geometry
  of population responses in visual cortex},\ }\href@noop {} {\bibfield
  {journal} {\bibinfo  {journal} {Nature}\ }\textbf {\bibinfo {volume} {571}},\
  \bibinfo {pages} {361} (\bibinfo {year} {2019})}\BibitemShut {NoStop}%
\bibitem [{\citenamefont {Sadtler}\ \emph {et~al.}(2014)\citenamefont
  {Sadtler}, \citenamefont {Quick}, \citenamefont {Golub}, \citenamefont
  {Chase}, \citenamefont {Ryu}, \citenamefont {Tyler-Kabara}, \citenamefont
  {Yu},\ and\ \citenamefont {Batista}}]{Sadtler14}%
  \BibitemOpen
  \bibfield  {author} {\bibinfo {author} {\bibfnamefont {P.~T.}\ \bibnamefont
  {Sadtler}}, \bibinfo {author} {\bibfnamefont {K.~M.}\ \bibnamefont {Quick}},
  \bibinfo {author} {\bibfnamefont {M.~D.}\ \bibnamefont {Golub}}, \bibinfo
  {author} {\bibfnamefont {S.~M.}\ \bibnamefont {Chase}}, \bibinfo {author}
  {\bibfnamefont {S.~I.}\ \bibnamefont {Ryu}}, \bibinfo {author} {\bibfnamefont
  {E.~C.}\ \bibnamefont {Tyler-Kabara}}, \bibinfo {author} {\bibfnamefont
  {B.~M.}\ \bibnamefont {Yu}}, \ and\ \bibinfo {author} {\bibfnamefont {A.~P.}\
  \bibnamefont {Batista}},\ }\bibfield  {title} {{\selectlanguage
  {English}\bibinfo {title} {Neural constraints on learning},\ }}\href
  {\doibase 10.1038/nature13665} {\bibfield  {journal} {\bibinfo  {journal}
  {Nature}\ }\textbf {\bibinfo {volume} {512}},\ \bibinfo {pages} {423}
  (\bibinfo {year} {2014})},\ \bibinfo {note} {number: 7515 Publisher: Nature
  Publishing Group}\BibitemShut {NoStop}%
\bibitem [{\citenamefont {Gallego}\ \emph {et~al.}(2018)\citenamefont
  {Gallego}, \citenamefont {Perich}, \citenamefont {Naufel}, \citenamefont
  {Ethier}, \citenamefont {Solla},\ and\ \citenamefont {Miller}}]{Gallego18_1}%
  \BibitemOpen
  \bibfield  {author} {\bibinfo {author} {\bibfnamefont {J.~A.}\ \bibnamefont
  {Gallego}}, \bibinfo {author} {\bibfnamefont {M.~G.}\ \bibnamefont {Perich}},
  \bibinfo {author} {\bibfnamefont {S.~N.}\ \bibnamefont {Naufel}}, \bibinfo
  {author} {\bibfnamefont {C.}~\bibnamefont {Ethier}}, \bibinfo {author}
  {\bibfnamefont {S.~A.}\ \bibnamefont {Solla}}, \ and\ \bibinfo {author}
  {\bibfnamefont {L.~E.}\ \bibnamefont {Miller}},\ }\bibfield  {title}
  {\bibinfo {title} {Cortical population activity within a preserved neural
  manifold underlies multiple motor behaviors},\ }\href@noop {} {\bibfield
  {journal} {\bibinfo  {journal} {Nat. Commun.}\ }\textbf {\bibinfo {volume}
  {9}},\ \bibinfo {pages} {1} (\bibinfo {year} {2018})}\BibitemShut {NoStop}%
\bibitem [{\citenamefont {Semedo}\ \emph {et~al.}(2019)\citenamefont {Semedo},
  \citenamefont {Zandvakili}, \citenamefont {Machens}, \citenamefont {Byron},\
  and\ \citenamefont {Kohn}}]{Semedo19_249}%
  \BibitemOpen
  \bibfield  {author} {\bibinfo {author} {\bibfnamefont {J.~D.}\ \bibnamefont
  {Semedo}}, \bibinfo {author} {\bibfnamefont {A.}~\bibnamefont {Zandvakili}},
  \bibinfo {author} {\bibfnamefont {C.~K.}\ \bibnamefont {Machens}}, \bibinfo
  {author} {\bibfnamefont {M.~Y.}\ \bibnamefont {Byron}}, \ and\ \bibinfo
  {author} {\bibfnamefont {A.}~\bibnamefont {Kohn}},\ }\bibfield  {title}
  {\bibinfo {title} {Cortical areas interact through a communication
  subspace},\ }\href@noop {} {\bibfield  {journal} {\bibinfo  {journal}
  {Neuron}\ }\textbf {\bibinfo {volume} {102}},\ \bibinfo {pages} {249}
  (\bibinfo {year} {2019})}\BibitemShut {NoStop}%
\bibitem [{\citenamefont {Rigotti}\ \emph {et~al.}(2013)\citenamefont
  {Rigotti}, \citenamefont {Barak}, \citenamefont {Warden}, \citenamefont
  {Wang}, \citenamefont {Daw}, \citenamefont {Miller},\ and\ \citenamefont
  {Fusi}}]{Rigotti2013_585}%
  \BibitemOpen
  \bibfield  {author} {\bibinfo {author} {\bibfnamefont {M.}~\bibnamefont
  {Rigotti}}, \bibinfo {author} {\bibfnamefont {O.}~\bibnamefont {Barak}},
  \bibinfo {author} {\bibfnamefont {M.~R.}\ \bibnamefont {Warden}}, \bibinfo
  {author} {\bibfnamefont {X.-J.}\ \bibnamefont {Wang}}, \bibinfo {author}
  {\bibfnamefont {N.~D.}\ \bibnamefont {Daw}}, \bibinfo {author} {\bibfnamefont
  {E.~K.}\ \bibnamefont {Miller}}, \ and\ \bibinfo {author} {\bibfnamefont
  {S.}~\bibnamefont {Fusi}},\ }\bibfield  {title} {\bibinfo {title} {{The
  importance of mixed selectivity in complex cognitive tasks}},\ }\href
  {\doibase 10.1038/nature12160} {\bibfield  {journal} {\bibinfo  {journal}
  {Nature}\ }\textbf {\bibinfo {volume} {497}},\ \bibinfo {pages} {585}
  (\bibinfo {year} {2013})}\BibitemShut {NoStop}%
\bibitem [{\citenamefont {Sorscher}\ \emph {et~al.}(2022)\citenamefont
  {Sorscher}, \citenamefont {Ganguli},\ and\ \citenamefont
  {Sompolinsky}}]{sorscher22}%
  \BibitemOpen
  \bibfield  {author} {\bibinfo {author} {\bibfnamefont {B.}~\bibnamefont
  {Sorscher}}, \bibinfo {author} {\bibfnamefont {S.}~\bibnamefont {Ganguli}}, \
  and\ \bibinfo {author} {\bibfnamefont {H.}~\bibnamefont {Sompolinsky}},\
  }\bibfield  {title} {\bibinfo {title} {Neural representational geometry
  underlies few-shot concept learning},\ }\href@noop {} {\bibfield  {journal}
  {\bibinfo  {journal} {Proceedings of the National Academy of Sciences}\
  }\textbf {\bibinfo {volume} {119}},\ \bibinfo {pages} {e2200800119} (\bibinfo
  {year} {2022})}\BibitemShut {NoStop}%
\bibitem [{\citenamefont {Braitenberg}\ and\ \citenamefont
  {Sch\"{u}z}(1991)}]{Braitenberg91}%
  \BibitemOpen
  \bibfield  {author} {\bibinfo {author} {\bibfnamefont {V.}~\bibnamefont
  {Braitenberg}}\ and\ \bibinfo {author} {\bibfnamefont {A.}~\bibnamefont
  {Sch\"{u}z}},\ }\href@noop {} {\emph {\bibinfo {title} {Anatomy of the
  Cortex: Statistics and Geometry}}}\ (\bibinfo  {publisher}
  {Springer-Verlag},\ \bibinfo {address} {Berlin, Heidelberg, New York},\
  \bibinfo {year} {1991})\BibitemShut {NoStop}%
\bibitem [{\citenamefont {Hu}\ \emph {et~al.}(2013)\citenamefont {Hu},
  \citenamefont {Trousdale}, \citenamefont {Josi{\'c}},\ and\ \citenamefont
  {Shea-Brown}}]{Hu13_P03012}%
  \BibitemOpen
  \bibfield  {author} {\bibinfo {author} {\bibfnamefont {Y.}~\bibnamefont
  {Hu}}, \bibinfo {author} {\bibfnamefont {J.}~\bibnamefont {Trousdale}},
  \bibinfo {author} {\bibfnamefont {K.}~\bibnamefont {Josi{\'c}}}, \ and\
  \bibinfo {author} {\bibfnamefont {E.}~\bibnamefont {Shea-Brown}},\ }\bibfield
   {title} {\bibinfo {title} {Motif statistics and spike correlations in
  neuronal networks},\ }\href@noop {} {\bibfield  {journal} {\bibinfo
  {journal} {J. Stat. Mech. Theory Exp.}\ }\textbf {\bibinfo {volume} {2013}},\
  \bibinfo {pages} {P03012} (\bibinfo {year} {2013})}\BibitemShut {NoStop}%
\bibitem [{\citenamefont {Hu}\ and\ \citenamefont {Sompolinsky}(2022)}]{Hu22}%
  \BibitemOpen
  \bibfield  {author} {\bibinfo {author} {\bibfnamefont {Y.}~\bibnamefont
  {Hu}}\ and\ \bibinfo {author} {\bibfnamefont {H.}~\bibnamefont
  {Sompolinsky}},\ }\bibfield  {title} {\bibinfo {title} {The spectrum of
  covariance matrices of randomly connected recurrent neuronal networks with
  linear dynamics},\ }\href {\doibase 10.1371/journal.pcbi.1010327} {\bibfield
  {journal} {\bibinfo  {journal} {PLOS Computational Biology}\ }\textbf
  {\bibinfo {volume} {18}},\ \bibinfo {pages} {1} (\bibinfo {year}
  {2022})}\BibitemShut {NoStop}%
\bibitem [{\citenamefont {Dahmen}\ \emph {et~al.}(2022)\citenamefont {Dahmen},
  \citenamefont {Recanatesi}, \citenamefont {Jia}, \citenamefont {Ocker},
  \citenamefont {Campagnola}, \citenamefont {Jarsky}, \citenamefont {Seeman},
  \citenamefont {Helias},\ and\ \citenamefont
  {Shea-Brown}}]{Dahmen22_365072v3}%
  \BibitemOpen
  \bibfield  {author} {\bibinfo {author} {\bibfnamefont {D.}~\bibnamefont
  {Dahmen}}, \bibinfo {author} {\bibfnamefont {S.}~\bibnamefont {Recanatesi}},
  \bibinfo {author} {\bibfnamefont {X.}~\bibnamefont {Jia}}, \bibinfo {author}
  {\bibfnamefont {G.~K.}\ \bibnamefont {Ocker}}, \bibinfo {author}
  {\bibfnamefont {L.}~\bibnamefont {Campagnola}}, \bibinfo {author}
  {\bibfnamefont {T.}~\bibnamefont {Jarsky}}, \bibinfo {author} {\bibfnamefont
  {S.}~\bibnamefont {Seeman}}, \bibinfo {author} {\bibfnamefont
  {M.}~\bibnamefont {Helias}}, \ and\ \bibinfo {author} {\bibfnamefont
  {E.}~\bibnamefont {Shea-Brown}},\ }\bibfield  {title} {\bibinfo {title}
  {Strong and localized recurrence controls dimensionality of neural activity
  across brain areas},\ }\href@noop {} {\bibfield  {journal} {\bibinfo
  {journal} {BioRxiv}\ } (\bibinfo {year} {2022})}\BibitemShut {NoStop}%
\bibitem [{\citenamefont {Song}\ \emph {et~al.}(2005)\citenamefont {Song},
  \citenamefont {Sj\"{o}str\"{o}m}, \citenamefont {Reigl}, \citenamefont
  {Nelson},\ and\ \citenamefont {Chklovskii}}]{Song05_0507}%
  \BibitemOpen
  \bibfield  {author} {\bibinfo {author} {\bibfnamefont {S.}~\bibnamefont
  {Song}}, \bibinfo {author} {\bibfnamefont {P.}~\bibnamefont
  {Sj\"{o}str\"{o}m}}, \bibinfo {author} {\bibfnamefont {M.}~\bibnamefont
  {Reigl}}, \bibinfo {author} {\bibfnamefont {S.}~\bibnamefont {Nelson}}, \
  and\ \bibinfo {author} {\bibfnamefont {D.}~\bibnamefont {Chklovskii}},\
  }\bibfield  {title} {\bibinfo {title} {Highly nonrandom features of synaptic
  connectivity in local cortical circuits},\ }\href {\doibase
  10.1371/journal.pbio.0030068} {\bibfield  {journal} {\bibinfo  {journal}
  {PLOS Biol.}\ }\textbf {\bibinfo {volume} {3}},\ \bibinfo {pages} {e68}
  (\bibinfo {year} {2005})}\BibitemShut {NoStop}%
\bibitem [{\citenamefont {Beggs}\ and\ \citenamefont
  {Plenz}(2003)}]{Beggs03_11167}%
  \BibitemOpen
  \bibfield  {author} {\bibinfo {author} {\bibfnamefont {J.~M.}\ \bibnamefont
  {Beggs}}\ and\ \bibinfo {author} {\bibfnamefont {D.}~\bibnamefont {Plenz}},\
  }\bibfield  {title} {\bibinfo {title} {Neuronal avalanches in neocortical
  circuits},\ }\href {\doibase 10.1523/JNEUROSCI.23-35-11167.2003} {\bibfield
  {journal} {\bibinfo  {journal} {J. Neurosci.}\ }\textbf {\bibinfo {volume}
  {23}},\ \bibinfo {pages} {11167} (\bibinfo {year} {2003})}\BibitemShut
  {NoStop}%
\bibitem [{\citenamefont {Meshulam}\ \emph {et~al.}(2019)\citenamefont
  {Meshulam}, \citenamefont {Gauthier}, \citenamefont {Brody}, \citenamefont
  {Tank},\ and\ \citenamefont {Bialek}}]{Meshulam19}%
  \BibitemOpen
  \bibfield  {author} {\bibinfo {author} {\bibfnamefont {L.}~\bibnamefont
  {Meshulam}}, \bibinfo {author} {\bibfnamefont {J.~L.}\ \bibnamefont
  {Gauthier}}, \bibinfo {author} {\bibfnamefont {C.~D.}\ \bibnamefont {Brody}},
  \bibinfo {author} {\bibfnamefont {D.~W.}\ \bibnamefont {Tank}}, \ and\
  \bibinfo {author} {\bibfnamefont {W.}~\bibnamefont {Bialek}},\ }\bibfield
  {title} {\bibinfo {title} {Coarse graining, fixed points, and scaling in a
  large population of neurons},\ }\href {\doibase
  10.1103/PhysRevLett.123.178103} {\bibfield  {journal} {\bibinfo  {journal}
  {Phys. Rev. Lett.}\ }\textbf {\bibinfo {volume} {123}},\ \bibinfo {pages}
  {178103} (\bibinfo {year} {2019})}\BibitemShut {NoStop}%
\bibitem [{\citenamefont {Fontenele}\ \emph {et~al.}(2019)\citenamefont
  {Fontenele}, \citenamefont {de~Vasconcelos}, \citenamefont {Feliciano},
  \citenamefont {Aguiar}, \citenamefont {Soares-Cunha}, \citenamefont
  {Coimbra}, \citenamefont {Dalla~Porta}, \citenamefont {Ribeiro},
  \citenamefont {Rodrigues}, \citenamefont {Sousa}, \citenamefont {Carelli},\
  and\ \citenamefont {Copelli}}]{Fontenele19_208101}%
  \BibitemOpen
  \bibfield  {author} {\bibinfo {author} {\bibfnamefont {A.~J.}\ \bibnamefont
  {Fontenele}}, \bibinfo {author} {\bibfnamefont {N.~A.~P.}\ \bibnamefont
  {de~Vasconcelos}}, \bibinfo {author} {\bibfnamefont {T.}~\bibnamefont
  {Feliciano}}, \bibinfo {author} {\bibfnamefont {L.~A.~A.}\ \bibnamefont
  {Aguiar}}, \bibinfo {author} {\bibfnamefont {C.}~\bibnamefont
  {Soares-Cunha}}, \bibinfo {author} {\bibfnamefont {B.}~\bibnamefont
  {Coimbra}}, \bibinfo {author} {\bibfnamefont {L.}~\bibnamefont
  {Dalla~Porta}}, \bibinfo {author} {\bibfnamefont {S.}~\bibnamefont
  {Ribeiro}}, \bibinfo {author} {\bibfnamefont {A.~J.}\ \bibnamefont
  {Rodrigues}}, \bibinfo {author} {\bibfnamefont {N.}~\bibnamefont {Sousa}},
  \bibinfo {author} {\bibfnamefont {P.~V.}\ \bibnamefont {Carelli}}, \ and\
  \bibinfo {author} {\bibfnamefont {M.}~\bibnamefont {Copelli}},\ }\bibfield
  {title} {\bibinfo {title} {Criticality between cortical states},\ }\href
  {\doibase 10.1103/PhysRevLett.122.208101} {\bibfield  {journal} {\bibinfo
  {journal} {Phys. Rev. Lett.}\ }\textbf {\bibinfo {volume} {122}},\ \bibinfo
  {pages} {208101} (\bibinfo {year} {2019})}\BibitemShut {NoStop}%
\bibitem [{\citenamefont {Hennequin}\ \emph {et~al.}(2014)\citenamefont
  {Hennequin}, \citenamefont {Vogels},\ and\ \citenamefont
  {Gerstner}}]{Hennequin14_1394}%
  \BibitemOpen
  \bibfield  {author} {\bibinfo {author} {\bibfnamefont {G.}~\bibnamefont
  {Hennequin}}, \bibinfo {author} {\bibfnamefont {T.}~\bibnamefont {Vogels}}, \
  and\ \bibinfo {author} {\bibfnamefont {W.}~\bibnamefont {Gerstner}},\
  }\bibfield  {title} {\bibinfo {title} {Optimal control of transient dynamics
  in balanced networks supports generation of complex movements},\ }\href@noop
  {} {\bibfield  {journal} {\bibinfo  {journal} {Neuron}\ }\textbf {\bibinfo
  {volume} {82}},\ \bibinfo {pages} {1394} (\bibinfo {year}
  {2014})}\BibitemShut {NoStop}%
\bibitem [{\citenamefont {Mastrogiuseppe}\ and\ \citenamefont
  {Ostojic}(2018)}]{Mastrogiuseppe18_609}%
  \BibitemOpen
  \bibfield  {author} {\bibinfo {author} {\bibfnamefont {F.}~\bibnamefont
  {Mastrogiuseppe}}\ and\ \bibinfo {author} {\bibfnamefont {S.}~\bibnamefont
  {Ostojic}},\ }\bibfield  {title} {\bibinfo {title} {Linking connectivity,
  dynamics, and computations in low-rank recurrent neural networks},\
  }\href@noop {} {\bibfield  {journal} {\bibinfo  {journal} {Neuron}\ }\textbf
  {\bibinfo {volume} {99}},\ \bibinfo {pages} {609} (\bibinfo {year}
  {2018})}\BibitemShut {NoStop}%
\bibitem [{\citenamefont {Lindner}\ \emph {et~al.}(2005)\citenamefont
  {Lindner}, \citenamefont {Doiron},\ and\ \citenamefont
  {Longtin}}]{Lindner05_061919}%
  \BibitemOpen
  \bibfield  {author} {\bibinfo {author} {\bibfnamefont {B.}~\bibnamefont
  {Lindner}}, \bibinfo {author} {\bibfnamefont {B.}~\bibnamefont {Doiron}}, \
  and\ \bibinfo {author} {\bibfnamefont {A.}~\bibnamefont {Longtin}},\
  }\bibfield  {title} {\bibinfo {title} {Theory of oscillatory firing induced
  by spatially correlated noise and delayed inhibitory feedback},\ }\href
  {\doibase 10.1103/physreve.72.061919} {\bibfield  {journal} {\bibinfo
  {journal} {Phys. Rev. E}\ }\textbf {\bibinfo {volume} {72}},\ \bibinfo
  {pages} {061919} (\bibinfo {year} {2005})}\BibitemShut {NoStop}%
\bibitem [{\citenamefont {Pernice}\ \emph {et~al.}(2011)\citenamefont
  {Pernice}, \citenamefont {Staude}, \citenamefont {Cardanobile},\ and\
  \citenamefont {Rotter}}]{Pernice11_e1002059}%
  \BibitemOpen
  \bibfield  {author} {\bibinfo {author} {\bibfnamefont {V.}~\bibnamefont
  {Pernice}}, \bibinfo {author} {\bibfnamefont {B.}~\bibnamefont {Staude}},
  \bibinfo {author} {\bibfnamefont {S.}~\bibnamefont {Cardanobile}}, \ and\
  \bibinfo {author} {\bibfnamefont {S.}~\bibnamefont {Rotter}},\ }\bibfield
  {title} {\bibinfo {title} {How structure determines correlations in neuronal
  networks},\ }\href@noop {} {\bibfield  {journal} {\bibinfo  {journal} {PLOS
  Comput. Biol.}\ }\textbf {\bibinfo {volume} {7}},\ \bibinfo {pages}
  {e1002059} (\bibinfo {year} {2011})}\BibitemShut {NoStop}%
\bibitem [{\citenamefont {Pernice}\ \emph {et~al.}(2012)\citenamefont
  {Pernice}, \citenamefont {Staude}, \citenamefont {Cardanobile},\ and\
  \citenamefont {Rotter}}]{Pernice12_031916}%
  \BibitemOpen
  \bibfield  {author} {\bibinfo {author} {\bibfnamefont {V.}~\bibnamefont
  {Pernice}}, \bibinfo {author} {\bibfnamefont {B.}~\bibnamefont {Staude}},
  \bibinfo {author} {\bibfnamefont {S.}~\bibnamefont {Cardanobile}}, \ and\
  \bibinfo {author} {\bibfnamefont {S.}~\bibnamefont {Rotter}},\ }\bibfield
  {title} {\bibinfo {title} {Recurrent interactions in spiking networks with
  arbitrary topology},\ }\href@noop {} {\bibfield  {journal} {\bibinfo
  {journal} {Phys. Rev. E}\ }\textbf {\bibinfo {volume} {85}},\ \bibinfo
  {pages} {031916} (\bibinfo {year} {2012})}\BibitemShut {NoStop}%
\bibitem [{\citenamefont {Grytskyy}\ \emph {et~al.}(2013)\citenamefont
  {Grytskyy}, \citenamefont {Tetzlaff}, \citenamefont {Diesmann},\ and\
  \citenamefont {Helias}}]{Grytskyy13_131}%
  \BibitemOpen
  \bibfield  {author} {\bibinfo {author} {\bibfnamefont {D.}~\bibnamefont
  {Grytskyy}}, \bibinfo {author} {\bibfnamefont {T.}~\bibnamefont {Tetzlaff}},
  \bibinfo {author} {\bibfnamefont {M.}~\bibnamefont {Diesmann}}, \ and\
  \bibinfo {author} {\bibfnamefont {M.}~\bibnamefont {Helias}},\ }\bibfield
  {title} {\bibinfo {title} {A unified view on weakly correlated recurrent
  networks},\ }\href {\doibase 10.3389/fncom.2013.00131} {\bibfield  {journal}
  {\bibinfo  {journal} {Front. Comput. Neurosci.}\ }\textbf {\bibinfo {volume}
  {7}},\ \bibinfo {pages} {131} (\bibinfo {year} {2013})}\BibitemShut {NoStop}%
\bibitem [{\citenamefont {Trousdale}\ \emph {et~al.}(2012)\citenamefont
  {Trousdale}, \citenamefont {Hu}, \citenamefont {Shea-Brown},\ and\
  \citenamefont {Josic}}]{Trousdale12_e1002408}%
  \BibitemOpen
  \bibfield  {author} {\bibinfo {author} {\bibfnamefont {J.}~\bibnamefont
  {Trousdale}}, \bibinfo {author} {\bibfnamefont {Y.}~\bibnamefont {Hu}},
  \bibinfo {author} {\bibfnamefont {E.}~\bibnamefont {Shea-Brown}}, \ and\
  \bibinfo {author} {\bibfnamefont {K.}~\bibnamefont {Josic}},\ }\bibfield
  {title} {\bibinfo {title} {Impact of network structure and cellular response
  on spike time correlations.}\ }\href@noop {} {\bibfield  {journal} {\bibinfo
  {journal} {PLOS Comput. Biol.}\ }\textbf {\bibinfo {volume} {8}},\ \bibinfo
  {pages} {e1002408} (\bibinfo {year} {2012})}\BibitemShut {NoStop}%
\bibitem [{\citenamefont {Dahmen}\ \emph {et~al.}(2019)\citenamefont {Dahmen},
  \citenamefont {Gr\"{u}n}, \citenamefont {Diesmann},\ and\ \citenamefont
  {Helias}}]{Dahmen19_13051}%
  \BibitemOpen
  \bibfield  {author} {\bibinfo {author} {\bibfnamefont {D.}~\bibnamefont
  {Dahmen}}, \bibinfo {author} {\bibfnamefont {S.}~\bibnamefont {Gr\"{u}n}},
  \bibinfo {author} {\bibfnamefont {M.}~\bibnamefont {Diesmann}}, \ and\
  \bibinfo {author} {\bibfnamefont {M.}~\bibnamefont {Helias}},\ }\bibfield
  {title} {\bibinfo {title} {Second type of criticality in the brain uncovers
  rich multiple-neuron dynamics},\ }\href {\doibase 10.1073/pnas.1818972116}
  {\bibfield  {journal} {\bibinfo  {journal} {Proc. Natl. Acad. Sci. USA}\
  }\textbf {\bibinfo {volume} {116}},\ \bibinfo {pages} {13051} (\bibinfo
  {year} {2019})}\BibitemShut {NoStop}%
\bibitem [{\citenamefont {Sompolinsky}\ \emph {et~al.}(1988)\citenamefont
  {Sompolinsky}, \citenamefont {Crisanti},\ and\ \citenamefont
  {Sommers}}]{Sompolinsky88_259}%
  \BibitemOpen
  \bibfield  {author} {\bibinfo {author} {\bibfnamefont {H.}~\bibnamefont
  {Sompolinsky}}, \bibinfo {author} {\bibfnamefont {A.}~\bibnamefont
  {Crisanti}}, \ and\ \bibinfo {author} {\bibfnamefont {H.~J.}\ \bibnamefont
  {Sommers}},\ }\bibfield  {title} {\bibinfo {title} {Chaos in random neural
  networks},\ }\href {\doibase 10.1103/PhysRevLett.61.259} {\bibfield
  {journal} {\bibinfo  {journal} {Phys. Rev. Lett.}\ }\textbf {\bibinfo
  {volume} {61}},\ \bibinfo {pages} {259} (\bibinfo {year} {1988})}\BibitemShut
  {NoStop}%
\bibitem [{\citenamefont {Sommers}\ \emph {et~al.}(1988)\citenamefont
  {Sommers}, \citenamefont {Crisanti}, \citenamefont {Sompolinsky},\ and\
  \citenamefont {Stein}}]{Sommers88}%
  \BibitemOpen
  \bibfield  {author} {\bibinfo {author} {\bibfnamefont {H.}~\bibnamefont
  {Sommers}}, \bibinfo {author} {\bibfnamefont {A.}~\bibnamefont {Crisanti}},
  \bibinfo {author} {\bibfnamefont {H.}~\bibnamefont {Sompolinsky}}, \ and\
  \bibinfo {author} {\bibfnamefont {Y.}~\bibnamefont {Stein}},\ }\bibfield
  {title} {\bibinfo {title} {Spectrum of large random asymmetric matrices},\
  }\href@noop {} {\bibfield  {journal} {\bibinfo  {journal} {Phys. Rev. Lett.}\
  }\textbf {\bibinfo {volume} {60}},\ \bibinfo {pages} {1895} (\bibinfo {year}
  {1988})}\BibitemShut {NoStop}%
\bibitem [{\citenamefont {Hennequin}\ \emph {et~al.}(2012)\citenamefont
  {Hennequin}, \citenamefont {Vogels},\ and\ \citenamefont
  {Gerstner}}]{Hennequin_12}%
  \BibitemOpen
  \bibfield  {author} {\bibinfo {author} {\bibfnamefont {G.}~\bibnamefont
  {Hennequin}}, \bibinfo {author} {\bibfnamefont {T.}~\bibnamefont {Vogels}}, \
  and\ \bibinfo {author} {\bibfnamefont {W.}~\bibnamefont {Gerstner}},\
  }\bibfield  {title} {\bibinfo {title} {Non-normal amplification in random
  balanced neuronal networks},\ }\href@noop {} {\bibfield  {journal} {\bibinfo
  {journal} {Phys. Rev. E}\ }\textbf {\bibinfo {volume} {86}},\ \bibinfo
  {pages} {011909} (\bibinfo {year} {2012})}\BibitemShut {NoStop}%
\bibitem [{\citenamefont {Chialvo}(2010)}]{Chialvo10_744}%
  \BibitemOpen
  \bibfield  {author} {\bibinfo {author} {\bibfnamefont {D.~R.}\ \bibnamefont
  {Chialvo}},\ }\bibfield  {title} {{\selectlanguage {English}\bibinfo {title}
  {Emergent complex neural dynamics},\ }}\href {\doibase 10.1038/nphys1803}
  {\bibfield  {journal} {\bibinfo  {journal} {Nat. Phys.}\ }\textbf {\bibinfo
  {volume} {6}},\ \bibinfo {pages} {744} (\bibinfo {year} {2010})},\ \bibinfo
  {note} {number: 10 Publisher: Nature Publishing Group}\BibitemShut {NoStop}%
\bibitem [{\citenamefont {Langton}(1990)}]{Langton90_12}%
  \BibitemOpen
  \bibfield  {author} {\bibinfo {author} {\bibfnamefont {C.~G.}\ \bibnamefont
  {Langton}},\ }\bibfield  {title} {\bibinfo {title} {Computation at the edge
  of chaos: phase transitions and emergent computation},\ }\href@noop {}
  {\bibfield  {journal} {\bibinfo  {journal} {Physica D}\ }\textbf {\bibinfo
  {volume} {42}},\ \bibinfo {pages} {12} (\bibinfo {year} {1990})}\BibitemShut
  {NoStop}%
\bibitem [{\citenamefont {Bertschinger}\ and\ \citenamefont
  {Natschl{\"a}ger}(2004)}]{Bertschinger04_1413}%
  \BibitemOpen
  \bibfield  {author} {\bibinfo {author} {\bibfnamefont {N.}~\bibnamefont
  {Bertschinger}}\ and\ \bibinfo {author} {\bibfnamefont {T.}~\bibnamefont
  {Natschl{\"a}ger}},\ }\bibfield  {title} {\bibinfo {title} {Real-time
  computation at the edge of chaos in recurrent neural networks},\ }\href@noop
  {} {\bibfield  {journal} {\bibinfo  {journal} {Neural Comput.}\ }\textbf
  {\bibinfo {volume} {16}},\ \bibinfo {pages} {1413} (\bibinfo {year}
  {2004})}\BibitemShut {NoStop}%
\bibitem [{\citenamefont {Toyoizumi}\ and\ \citenamefont
  {Abbott}(2011)}]{Toyoizumi11_051908}%
  \BibitemOpen
  \bibfield  {author} {\bibinfo {author} {\bibfnamefont {T.}~\bibnamefont
  {Toyoizumi}}\ and\ \bibinfo {author} {\bibfnamefont {L.~F.}\ \bibnamefont
  {Abbott}},\ }\bibfield  {title} {\bibinfo {title} {Beyond the edge of chaos:
  Amplification and temporal integration by recurrent networks in the chaotic
  regime},\ }\href@noop {} {\bibfield  {journal} {\bibinfo  {journal} {Phys.
  Rev. E}\ }\textbf {\bibinfo {volume} {84}},\ \bibinfo {pages} {051908}
  (\bibinfo {year} {2011})}\BibitemShut {NoStop}%
\bibitem [{Sup()}]{Supplement}%
  \BibitemOpen
  \href@noop {} {}\bibinfo {note} {See Supplemental Material for supplementary
  figures and further details on the analytical derivations}\BibitemShut
  {NoStop}%
\bibitem [{\citenamefont {Wilson}(1975)}]{Wilson75_773}%
  \BibitemOpen
  \bibfield  {author} {\bibinfo {author} {\bibfnamefont {K.~G.}\ \bibnamefont
  {Wilson}},\ }\bibfield  {title} {\bibinfo {title} {The renormalization group:
  Critical phenomena and the kondo problem},\ }\href {\doibase
  10.1103/RevModPhys.47.773} {\bibfield  {journal} {\bibinfo  {journal} {Rev.
  Mod. Phys.}\ }\textbf {\bibinfo {volume} {47}},\ \bibinfo {pages} {773}
  (\bibinfo {year} {1975})}\BibitemShut {NoStop}%
\bibitem [{\citenamefont {Hohenberg}\ and\ \citenamefont
  {Halperin}(1977)}]{Hohenberg77}%
  \BibitemOpen
  \bibfield  {author} {\bibinfo {author} {\bibfnamefont {P.~C.}\ \bibnamefont
  {Hohenberg}}\ and\ \bibinfo {author} {\bibfnamefont {B.~I.}\ \bibnamefont
  {Halperin}},\ }\bibfield  {title} {\bibinfo {title} {Theory of dynamic
  critical phenomena},\ }\href {\doibase 10.1103/RevModPhys.49.435} {\bibfield
  {journal} {\bibinfo  {journal} {Rev. Mod. Phys.}\ }\textbf {\bibinfo {volume}
  {49}},\ \bibinfo {pages} {435} (\bibinfo {year} {1977})}\BibitemShut
  {NoStop}%
\bibitem [{\citenamefont {Taeuber}(2014)}]{Taeuber14}%
  \BibitemOpen
  \bibfield  {author} {\bibinfo {author} {\bibfnamefont {U.~C.}\ \bibnamefont
  {Taeuber}},\ }\href@noop {} {\emph {\bibinfo {title} {Critical dynamics: a
  field theory approach to equilibrium and non-equilibrium scaling behavior}}}\
  (\bibinfo  {publisher} {Cambridge University Press},\ \bibinfo {year}
  {2014})\BibitemShut {NoStop}%
\bibitem [{\citenamefont {Haldeman}\ and\ \citenamefont
  {Beggs}(2005)}]{Haldeman05}%
  \BibitemOpen
  \bibfield  {author} {\bibinfo {author} {\bibfnamefont {C.}~\bibnamefont
  {Haldeman}}\ and\ \bibinfo {author} {\bibfnamefont {J.~M.}\ \bibnamefont
  {Beggs}},\ }\bibfield  {title} {\bibinfo {title} {Critical branching captures
  activity in living neural networks and maximizes the number of metastable
  states},\ }\href {\doibase 10.1103/PhysRevLett.94.058101} {\bibfield
  {journal} {\bibinfo  {journal} {Phys. Rev. Lett.}\ }\textbf {\bibinfo
  {volume} {94}},\ \bibinfo {pages} {058101} (\bibinfo {year}
  {2005})}\BibitemShut {NoStop}%
\bibitem [{\citenamefont {Chung}\ \emph {et~al.}(2018)\citenamefont {Chung},
  \citenamefont {Lee},\ and\ \citenamefont {Sompolinsky}}]{Chung18_031003}%
  \BibitemOpen
  \bibfield  {author} {\bibinfo {author} {\bibfnamefont {S.}~\bibnamefont
  {Chung}}, \bibinfo {author} {\bibfnamefont {D.~D.}\ \bibnamefont {Lee}}, \
  and\ \bibinfo {author} {\bibfnamefont {H.}~\bibnamefont {Sompolinsky}},\
  }\bibfield  {title} {\bibinfo {title} {Classification and geometry of general
  perceptual manifolds},\ }\href {\doibase 10.1103/physrevx.8.031003}
  {\bibfield  {journal} {\bibinfo  {journal} {Phys. Rev. X}\ }\textbf {\bibinfo
  {volume} {8}} (\bibinfo {year} {2018}),\
  10.1103/physrevx.8.031003}\BibitemShut {NoStop}%
\bibitem [{\citenamefont {Morales}\ and\ \citenamefont
  {Mu\~{n}oz}(2021)}]{Morales21_702}%
  \BibitemOpen
  \bibfield  {author} {\bibinfo {author} {\bibfnamefont {G.~B.}\ \bibnamefont
  {Morales}}\ and\ \bibinfo {author} {\bibfnamefont {M.~A.}\ \bibnamefont
  {Mu\~{n}oz}},\ }\bibfield  {title} {\bibinfo {title} {Optimal input
  representation in neural systems at the edge of chaos},\ }\href {\doibase
  10.3390/biology10080702} {\bibfield  {journal} {\bibinfo  {journal}
  {Biology}\ }\textbf {\bibinfo {volume} {10}} (\bibinfo {year} {2021}),\
  10.3390/biology10080702}\BibitemShut {NoStop}%
\bibitem [{\citenamefont {Bradde}\ \emph {et~al.}(2010)\citenamefont {Bradde},
  \citenamefont {Caccioli}, \citenamefont {Dall'Asta},\ and\ \citenamefont
  {Bianconi}}]{Bradde2010}%
  \BibitemOpen
  \bibfield  {author} {\bibinfo {author} {\bibfnamefont {S.}~\bibnamefont
  {Bradde}}, \bibinfo {author} {\bibfnamefont {F.}~\bibnamefont {Caccioli}},
  \bibinfo {author} {\bibfnamefont {L.}~\bibnamefont {Dall'Asta}}, \ and\
  \bibinfo {author} {\bibfnamefont {G.}~\bibnamefont {Bianconi}},\ }\bibfield
  {title} {\bibinfo {title} {Critical fluctuations in spatial complex
  networks},\ }\href {\doibase 10.1103/PhysRevLett.104.218701} {\bibfield
  {journal} {\bibinfo  {journal} {Phys. Rev. Lett.}\ }\textbf {\bibinfo
  {volume} {104}},\ \bibinfo {pages} {218701} (\bibinfo {year}
  {2010})}\BibitemShut {NoStop}%
\bibitem [{\citenamefont {Tuncer}\ and\ \citenamefont
  {Erzan}(2015)}]{Tuncer2015}%
  \BibitemOpen
  \bibfield  {author} {\bibinfo {author} {\bibfnamefont {A.~i. e. i.~f.}\
  \bibnamefont {Tuncer}}\ and\ \bibinfo {author} {\bibfnamefont {A.~m.~c.}\
  \bibnamefont {Erzan}},\ }\bibfield  {title} {\bibinfo {title} {Spectral
  renormalization group for the gaussian model and ${\ensuremath{\psi}}^{4}$
  theory on nonspatial networks},\ }\href {\doibase 10.1103/PhysRevE.92.022106}
  {\bibfield  {journal} {\bibinfo  {journal} {Phys. Rev. E}\ }\textbf {\bibinfo
  {volume} {92}},\ \bibinfo {pages} {022106} (\bibinfo {year}
  {2015})}\BibitemShut {NoStop}%
\bibitem [{\citenamefont {Brinkman}(2023)}]{brinkman2023}%
  \BibitemOpen
  \bibfield  {author} {\bibinfo {author} {\bibfnamefont {B.~A.}\ \bibnamefont
  {Brinkman}},\ }\bibfield  {title} {\bibinfo {title} {Non-perturbative
  renormalization group analysis of nonlinear spiking networks},\ }\href@noop
  {} {\bibfield  {journal} {\bibinfo  {journal} {arXiv preprint
  arXiv:2301.09600}\ } (\bibinfo {year} {2023})}\BibitemShut {NoStop}%
\bibitem [{\citenamefont {Berning}\ \emph {et~al.}(2015)\citenamefont
  {Berning}, \citenamefont {Boergens},\ and\ \citenamefont
  {Helmstaedter}}]{Berning15}%
  \BibitemOpen
  \bibfield  {author} {\bibinfo {author} {\bibfnamefont {M.}~\bibnamefont
  {Berning}}, \bibinfo {author} {\bibfnamefont {K.~M.}\ \bibnamefont
  {Boergens}}, \ and\ \bibinfo {author} {\bibfnamefont {M.}~\bibnamefont
  {Helmstaedter}},\ }\bibfield  {title} {\bibinfo {title} {{SegEM}: Efficient
  image analysis for high-resolution connectomics},\ }\href {\doibase
  10.1016/j.neuron.2015.09.003} {\bibfield  {journal} {\bibinfo  {journal}
  {Neuron}\ }\textbf {\bibinfo {volume} {87}},\ \bibinfo {pages} {1193}
  (\bibinfo {year} {2015})}\BibitemShut {NoStop}%
\bibitem [{\citenamefont {Yin}\ \emph {et~al.}(2020)\citenamefont {Yin},
  \citenamefont {Brittain}, \citenamefont {Borseth}, \citenamefont {Scott},
  \citenamefont {Williams}, \citenamefont {Perkins}, \citenamefont {Own},
  \citenamefont {Murfitt}, \citenamefont {Torres}, \citenamefont {Kapner} \emph
  {et~al.}}]{yin2020}%
  \BibitemOpen
  \bibfield  {author} {\bibinfo {author} {\bibfnamefont {W.}~\bibnamefont
  {Yin}}, \bibinfo {author} {\bibfnamefont {D.}~\bibnamefont {Brittain}},
  \bibinfo {author} {\bibfnamefont {J.}~\bibnamefont {Borseth}}, \bibinfo
  {author} {\bibfnamefont {M.~E.}\ \bibnamefont {Scott}}, \bibinfo {author}
  {\bibfnamefont {D.}~\bibnamefont {Williams}}, \bibinfo {author}
  {\bibfnamefont {J.}~\bibnamefont {Perkins}}, \bibinfo {author} {\bibfnamefont
  {C.~S.}\ \bibnamefont {Own}}, \bibinfo {author} {\bibfnamefont
  {M.}~\bibnamefont {Murfitt}}, \bibinfo {author} {\bibfnamefont {R.~M.}\
  \bibnamefont {Torres}}, \bibinfo {author} {\bibfnamefont {D.}~\bibnamefont
  {Kapner}},  \emph {et~al.},\ }\bibfield  {title} {\bibinfo {title} {A
  petascale automated imaging pipeline for mapping neuronal circuits with
  high-throughput transmission electron microscopy},\ }\href@noop {} {\bibfield
   {journal} {\bibinfo  {journal} {Nature communications}\ }\textbf {\bibinfo
  {volume} {11}},\ \bibinfo {pages} {4949} (\bibinfo {year}
  {2020})}\BibitemShut {NoStop}%
\bibitem [{\citenamefont {Markram}\ \emph {et~al.}(2015)\citenamefont
  {Markram}, \citenamefont {Muller}, \citenamefont {Ramaswamy}, \citenamefont
  {Reimann}, \citenamefont {Abdellah}, \citenamefont {Sanchez}, \citenamefont
  {Ailamaki}, \citenamefont {Alonso-Nanclares}, \citenamefont {Antille},
  \citenamefont {Arsever}, \citenamefont {Kahou}, \citenamefont {Berger},
  \citenamefont {Bilgili}, \citenamefont {Buncic}, \citenamefont {Chalimourda},
  \citenamefont {Chindemi}, \citenamefont {Courcol}, \citenamefont
  {Delalondre}, \citenamefont {Delattre}, \citenamefont {Druckmann},
  \citenamefont {Dumusc}, \citenamefont {Dynes}, \citenamefont {Eilemann},
  \citenamefont {Gal}, \citenamefont {Gevaert}, \citenamefont {Ghobril},
  \citenamefont {Gidon}, \citenamefont {Graham}, \citenamefont {Gupta},
  \citenamefont {Haenel}, \citenamefont {Hay}, \citenamefont {Heinis},
  \citenamefont {Hernando}, \citenamefont {Hines}, \citenamefont {Kanari},
  \citenamefont {Keller}, \citenamefont {Kenyon}, \citenamefont {Khazen},
  \citenamefont {Kim}, \citenamefont {King}, \citenamefont {Kisvarday},
  \citenamefont {Kumbhar}, \citenamefont {Lasserre}, \citenamefont {B{\'{e}}},
  \citenamefont {Magalh{\~{a}}es}, \citenamefont {Merch{\'{a}}n-P{\'{e}}rez},
  \citenamefont {Meystre}, \citenamefont {Morrice}, \citenamefont {Muller},
  \citenamefont {Mu{\~{n}}oz-C{\'{e}}spedes}, \citenamefont {Muralidhar},
  \citenamefont {Muthurasa}, \citenamefont {Nachbaur}, \citenamefont {Newton},
  \citenamefont {Nolte}, \citenamefont {Ovcharenko}, \citenamefont {Palacios},
  \citenamefont {Pastor}, \citenamefont {Perin}, \citenamefont {Ranjan},
  \citenamefont {Riachi}, \citenamefont {Rodr{\'{\i}}guez}, \citenamefont
  {Riquelme}, \citenamefont {R\"{o}ssert}, \citenamefont {Sfyrakis},
  \citenamefont {Shi}, \citenamefont {Shillcock}, \citenamefont {Silberberg},
  \citenamefont {Silva}, \citenamefont {Tauheed}, \citenamefont {Telefont},
  \citenamefont {Toledo-Rodriguez}, \citenamefont {Tr\"{a}nkler}, \citenamefont
  {Geit}, \citenamefont {D{\'{\i}}az}, \citenamefont {Walker}, \citenamefont
  {Wang}, \citenamefont {Zaninetta}, \citenamefont {DeFelipe}, \citenamefont
  {Hill}, \citenamefont {Segev},\ and\ \citenamefont
  {Sch\"{u}rmann}}]{Markram2015_456}%
  \BibitemOpen
  \bibfield  {author} {\bibinfo {author} {\bibfnamefont {H.}~\bibnamefont
  {Markram}}, \bibinfo {author} {\bibfnamefont {E.}~\bibnamefont {Muller}},
  \bibinfo {author} {\bibfnamefont {S.}~\bibnamefont {Ramaswamy}}, \bibinfo
  {author} {\bibfnamefont {M.~W.}\ \bibnamefont {Reimann}}, \bibinfo {author}
  {\bibfnamefont {M.}~\bibnamefont {Abdellah}}, \bibinfo {author}
  {\bibfnamefont {C.~A.}\ \bibnamefont {Sanchez}}, \bibinfo {author}
  {\bibfnamefont {A.}~\bibnamefont {Ailamaki}}, \bibinfo {author}
  {\bibfnamefont {L.}~\bibnamefont {Alonso-Nanclares}}, \bibinfo {author}
  {\bibfnamefont {N.}~\bibnamefont {Antille}}, \bibinfo {author} {\bibfnamefont
  {S.}~\bibnamefont {Arsever}}, \bibinfo {author} {\bibfnamefont {G.~A.~A.}\
  \bibnamefont {Kahou}}, \bibinfo {author} {\bibfnamefont {T.~K.}\ \bibnamefont
  {Berger}}, \bibinfo {author} {\bibfnamefont {A.}~\bibnamefont {Bilgili}},
  \bibinfo {author} {\bibfnamefont {N.}~\bibnamefont {Buncic}}, \bibinfo
  {author} {\bibfnamefont {A.}~\bibnamefont {Chalimourda}}, \bibinfo {author}
  {\bibfnamefont {G.}~\bibnamefont {Chindemi}}, \bibinfo {author}
  {\bibfnamefont {J.-D.}\ \bibnamefont {Courcol}}, \bibinfo {author}
  {\bibfnamefont {F.}~\bibnamefont {Delalondre}}, \bibinfo {author}
  {\bibfnamefont {V.}~\bibnamefont {Delattre}}, \bibinfo {author}
  {\bibfnamefont {S.}~\bibnamefont {Druckmann}}, \bibinfo {author}
  {\bibfnamefont {R.}~\bibnamefont {Dumusc}}, \bibinfo {author} {\bibfnamefont
  {J.}~\bibnamefont {Dynes}}, \bibinfo {author} {\bibfnamefont
  {S.}~\bibnamefont {Eilemann}}, \bibinfo {author} {\bibfnamefont
  {E.}~\bibnamefont {Gal}}, \bibinfo {author} {\bibfnamefont {M.~E.}\
  \bibnamefont {Gevaert}}, \bibinfo {author} {\bibfnamefont {J.-P.}\
  \bibnamefont {Ghobril}}, \bibinfo {author} {\bibfnamefont {A.}~\bibnamefont
  {Gidon}}, \bibinfo {author} {\bibfnamefont {J.~W.}\ \bibnamefont {Graham}},
  \bibinfo {author} {\bibfnamefont {A.}~\bibnamefont {Gupta}}, \bibinfo
  {author} {\bibfnamefont {V.}~\bibnamefont {Haenel}}, \bibinfo {author}
  {\bibfnamefont {E.}~\bibnamefont {Hay}}, \bibinfo {author} {\bibfnamefont
  {T.}~\bibnamefont {Heinis}}, \bibinfo {author} {\bibfnamefont {J.~B.}\
  \bibnamefont {Hernando}}, \bibinfo {author} {\bibfnamefont {M.}~\bibnamefont
  {Hines}}, \bibinfo {author} {\bibfnamefont {L.}~\bibnamefont {Kanari}},
  \bibinfo {author} {\bibfnamefont {D.}~\bibnamefont {Keller}}, \bibinfo
  {author} {\bibfnamefont {J.}~\bibnamefont {Kenyon}}, \bibinfo {author}
  {\bibfnamefont {G.}~\bibnamefont {Khazen}}, \bibinfo {author} {\bibfnamefont
  {Y.}~\bibnamefont {Kim}}, \bibinfo {author} {\bibfnamefont {J.~G.}\
  \bibnamefont {King}}, \bibinfo {author} {\bibfnamefont {Z.}~\bibnamefont
  {Kisvarday}}, \bibinfo {author} {\bibfnamefont {P.}~\bibnamefont {Kumbhar}},
  \bibinfo {author} {\bibfnamefont {S.}~\bibnamefont {Lasserre}}, \bibinfo
  {author} {\bibfnamefont {J.-V.~L.}\ \bibnamefont {B{\'{e}}}}, \bibinfo
  {author} {\bibfnamefont {B.~R.}\ \bibnamefont {Magalh{\~{a}}es}}, \bibinfo
  {author} {\bibfnamefont {A.}~\bibnamefont {Merch{\'{a}}n-P{\'{e}}rez}},
  \bibinfo {author} {\bibfnamefont {J.}~\bibnamefont {Meystre}}, \bibinfo
  {author} {\bibfnamefont {B.~R.}\ \bibnamefont {Morrice}}, \bibinfo {author}
  {\bibfnamefont {J.}~\bibnamefont {Muller}}, \bibinfo {author} {\bibfnamefont
  {A.}~\bibnamefont {Mu{\~{n}}oz-C{\'{e}}spedes}}, \bibinfo {author}
  {\bibfnamefont {S.}~\bibnamefont {Muralidhar}}, \bibinfo {author}
  {\bibfnamefont {K.}~\bibnamefont {Muthurasa}}, \bibinfo {author}
  {\bibfnamefont {D.}~\bibnamefont {Nachbaur}}, \bibinfo {author}
  {\bibfnamefont {T.~H.}\ \bibnamefont {Newton}}, \bibinfo {author}
  {\bibfnamefont {M.}~\bibnamefont {Nolte}}, \bibinfo {author} {\bibfnamefont
  {A.}~\bibnamefont {Ovcharenko}}, \bibinfo {author} {\bibfnamefont
  {J.}~\bibnamefont {Palacios}}, \bibinfo {author} {\bibfnamefont
  {L.}~\bibnamefont {Pastor}}, \bibinfo {author} {\bibfnamefont
  {R.}~\bibnamefont {Perin}}, \bibinfo {author} {\bibfnamefont
  {R.}~\bibnamefont {Ranjan}}, \bibinfo {author} {\bibfnamefont
  {I.}~\bibnamefont {Riachi}}, \bibinfo {author} {\bibfnamefont {J.-R.}\
  \bibnamefont {Rodr{\'{\i}}guez}}, \bibinfo {author} {\bibfnamefont {J.~L.}\
  \bibnamefont {Riquelme}}, \bibinfo {author} {\bibfnamefont {C.}~\bibnamefont
  {R\"{o}ssert}}, \bibinfo {author} {\bibfnamefont {K.}~\bibnamefont
  {Sfyrakis}}, \bibinfo {author} {\bibfnamefont {Y.}~\bibnamefont {Shi}},
  \bibinfo {author} {\bibfnamefont {J.~C.}\ \bibnamefont {Shillcock}}, \bibinfo
  {author} {\bibfnamefont {G.}~\bibnamefont {Silberberg}}, \bibinfo {author}
  {\bibfnamefont {R.}~\bibnamefont {Silva}}, \bibinfo {author} {\bibfnamefont
  {F.}~\bibnamefont {Tauheed}}, \bibinfo {author} {\bibfnamefont
  {M.}~\bibnamefont {Telefont}}, \bibinfo {author} {\bibfnamefont
  {M.}~\bibnamefont {Toledo-Rodriguez}}, \bibinfo {author} {\bibfnamefont
  {T.}~\bibnamefont {Tr\"{a}nkler}}, \bibinfo {author} {\bibfnamefont {W.~V.}\
  \bibnamefont {Geit}}, \bibinfo {author} {\bibfnamefont {J.~V.}\ \bibnamefont
  {D{\'{\i}}az}}, \bibinfo {author} {\bibfnamefont {R.}~\bibnamefont {Walker}},
  \bibinfo {author} {\bibfnamefont {Y.}~\bibnamefont {Wang}}, \bibinfo {author}
  {\bibfnamefont {S.~M.}\ \bibnamefont {Zaninetta}}, \bibinfo {author}
  {\bibfnamefont {J.}~\bibnamefont {DeFelipe}}, \bibinfo {author}
  {\bibfnamefont {S.~L.}\ \bibnamefont {Hill}}, \bibinfo {author}
  {\bibfnamefont {I.}~\bibnamefont {Segev}}, \ and\ \bibinfo {author}
  {\bibfnamefont {F.}~\bibnamefont {Sch\"{u}rmann}},\ }\bibfield  {title}
  {\bibinfo {title} {Reconstruction and simulation of neocortical
  microcircuitry},\ }\href {\doibase 10.1016/j.cell.2015.09.029} {\bibfield
  {journal} {\bibinfo  {journal} {Cell}\ }\textbf {\bibinfo {volume} {163}},\
  \bibinfo {pages} {456} (\bibinfo {year} {2015})}\BibitemShut {NoStop}%
\bibitem [{\citenamefont {Ecker}\ \emph {et~al.}(2023)\citenamefont {Ecker},
  \citenamefont {Santander}, \citenamefont {Bola{\~n}os-Puchet}, \citenamefont
  {Isbister},\ and\ \citenamefont {Reimann}}]{ecker23}%
  \BibitemOpen
  \bibfield  {author} {\bibinfo {author} {\bibfnamefont {A.}~\bibnamefont
  {Ecker}}, \bibinfo {author} {\bibfnamefont {D.~E.}\ \bibnamefont
  {Santander}}, \bibinfo {author} {\bibfnamefont {S.}~\bibnamefont
  {Bola{\~n}os-Puchet}}, \bibinfo {author} {\bibfnamefont {J.~B.}\ \bibnamefont
  {Isbister}}, \ and\ \bibinfo {author} {\bibfnamefont {M.~W.}\ \bibnamefont
  {Reimann}},\ }\bibfield  {title} {\bibinfo {title} {Cortical cell assemblies
  and their underlying connectivity: an in silico study},\ }\href@noop {}
  {\bibfield  {journal} {\bibinfo  {journal} {bioRxiv}\ ,\ \bibinfo {pages}
  {2023}} (\bibinfo {year} {2023})}\BibitemShut {NoStop}%
\bibitem [{\citenamefont {Harris}\ \emph {et~al.}(2022)\citenamefont {Harris},
  \citenamefont {Meffin}, \citenamefont {Burkitt},\ and\ \citenamefont
  {Peterson}}]{harris2022}%
  \BibitemOpen
  \bibfield  {author} {\bibinfo {author} {\bibfnamefont {I.~D.}\ \bibnamefont
  {Harris}}, \bibinfo {author} {\bibfnamefont {H.}~\bibnamefont {Meffin}},
  \bibinfo {author} {\bibfnamefont {A.~N.}\ \bibnamefont {Burkitt}}, \ and\
  \bibinfo {author} {\bibfnamefont {A.~D.}\ \bibnamefont {Peterson}},\
  }\bibfield  {title} {\bibinfo {title} {Eigenvalue spectral properties of
  sparse random matrices obeying dale's law},\ }\href@noop {} {\bibfield
  {journal} {\bibinfo  {journal} {arXiv preprint arXiv:2212.01549}\ } (\bibinfo
  {year} {2022})}\BibitemShut {NoStop}%
\bibitem [{\citenamefont {Kadmon}\ and\ \citenamefont
  {Sompolinsky}(2015)}]{kadmon15}%
  \BibitemOpen
  \bibfield  {author} {\bibinfo {author} {\bibfnamefont {J.}~\bibnamefont
  {Kadmon}}\ and\ \bibinfo {author} {\bibfnamefont {H.}~\bibnamefont
  {Sompolinsky}},\ }\bibfield  {title} {\bibinfo {title} {Transition to chaos
  in random neuronal networks},\ }\href@noop {} {\bibfield  {journal} {\bibinfo
   {journal} {ArXiv}\ ,\ \bibinfo {pages} {1508.06486}} (\bibinfo {year}
  {2015})}\BibitemShut {NoStop}%
\bibitem [{\citenamefont {Helias}\ and\ \citenamefont
  {Dahmen}(2020)}]{Helias20_970}%
  \BibitemOpen
  \bibfield  {author} {\bibinfo {author} {\bibfnamefont {M.}~\bibnamefont
  {Helias}}\ and\ \bibinfo {author} {\bibfnamefont {D.}~\bibnamefont
  {Dahmen}},\ }\href {\doibase 10.1007/978-3-030-46444-8} {\emph {\bibinfo
  {title} {Statistical Field Theory for Neural Networks}}}\ (\bibinfo
  {publisher} {Springer International Publishing},\ \bibinfo {year} {2020})\
  p.\ \bibinfo {pages} {203}\BibitemShut {NoStop}%
\bibitem [{\citenamefont {Weingarten}(1978)}]{weingarten78}%
  \BibitemOpen
  \bibfield  {author} {\bibinfo {author} {\bibfnamefont {D.}~\bibnamefont
  {Weingarten}},\ }\bibfield  {title} {\bibinfo {title} {Asymptotic behavior of
  group integrals in the limit of infinite rank},\ }\href {\doibase
  10.1063/1.523807} {\bibfield  {journal} {\bibinfo  {journal} {Journal of
  Mathematical Physics}\ }\textbf {\bibinfo {volume} {19}},\ \bibinfo {pages}
  {999} (\bibinfo {year} {1978})},\ \bibinfo {note} {publisher: American
  Institute of Physics}\BibitemShut {NoStop}%
\bibitem [{\citenamefont {Collins}\ and\ \citenamefont
  {{\'S}niady}(2006)}]{collins06}%
  \BibitemOpen
  \bibfield  {author} {\bibinfo {author} {\bibfnamefont {B.}~\bibnamefont
  {Collins}}\ and\ \bibinfo {author} {\bibfnamefont {P.}~\bibnamefont
  {{\'S}niady}},\ }\bibfield  {title} {{\selectlanguage {English}\bibinfo
  {title} {Integration with {Respect} to the {Haar} {Measure} on {Unitary},
  {Orthogonal} and {Symplectic} {Group}},\ }}\href {\doibase
  10.1007/s00220-006-1554-3} {\bibfield  {journal} {\bibinfo  {journal}
  {Communications in Mathematical Physics}\ }\textbf {\bibinfo {volume}
  {264}},\ \bibinfo {pages} {773} (\bibinfo {year} {2006})}\BibitemShut
  {NoStop}%
\bibitem [{\citenamefont {Matsumoto}(2011)}]{matsumoto11}%
  \BibitemOpen
  \bibfield  {author} {\bibinfo {author} {\bibfnamefont {S.}~\bibnamefont
  {Matsumoto}},\ }\bibfield  {title} {{\selectlanguage {English}\bibinfo
  {title} {Jucys-murphy elements, orthogonal matrix integrals, and jack
  measures},\ }}\href {\doibase 10.1007/s11139-011-9317-y} {\bibfield
  {journal} {\bibinfo  {journal} {The Ramanujan Journal}\ }\textbf {\bibinfo
  {volume} {26}},\ \bibinfo {pages} {69} (\bibinfo {year} {2011})}\BibitemShut
  {NoStop}%
\end{thebibliography}%

\end{document}